\documentclass[12pt]{article}
\usepackage{amsfonts}
\usepackage{amssymb}
\usepackage{enumerate}
\usepackage{amsmath}

\begin{document}
\title{Entropic characteristics of subset of states.}
\author{M.E.Shirokov \thanks{Steklov Mathematical Institute, 119991 Moscow,
Russia, e-mail: msh@mi.ras.ru}}
\date{}
\maketitle
\date{}
\maketitle

\section{Introduction}

This paper is devoted to systematic study of the properties of the
quantum entropy and of the Holevo capacity (in what follows the
$\chi$-capacity) considered as a function of a set of quantum
states.

It is known that the quantum entropy is concave lower
semicontinuous function on the set of all quantum states with the
range $[0;+\infty]$, but it has bounded and even continuous
restrictions to some nontrivial closed subsets of states
\cite{O&P},\cite{W}. The problem of characterization of such
subsets of states arises in many applications, in particular, in
the condition of existence of an optimal measure for constrained
quantum channel \cite{H-Sh-2}. In this paper we consider this and
some other problems related to the quantum entropy.

By the HSW theorem the $\chi$-capacity of a set of states defines
the maximal rate of transmission  of classical information, which
can be achieved by using this set as an alphabet and nonentangled
encoding in the transmitter followed by entangled
measurement-decoding procedure in the receiver
\cite{H-QI},\cite{Sch-West}. Usually the notion of the
$\chi$-capacity is related to the notion of a quantum channel. But
it is easy to see that the $\chi$-capacity of a channel is
uniquely defined by the output set of this channel. So, we may
consider the $\chi$-capacity as a function of a set of states
\cite{Sch-West-1}. Despite some limitations this approach provides
a convenient way to study the $\chi$-capacity. Namely, treating
the $\chi$-capacity as a function of a set of states we obtain a
certain flexibility in studying its properties since in this case
we may speak about the $\chi$-capacity of an \textit{arbitrary}
set of states, not necessary of an output set of a particular
channel. From this point of view the $\chi$-capacity is a
nonnegative nonadditive function of a set ("nonadditive measure")
possessing  many interesting properties, which detailed
investigation seems to be useful for the development of the
infinite dimensional quantum information theory.

We begin in section 3 with considering the conditions of
boundedness and of continuity of the restriction of the quantum
entropy to subsets of quantum states as well as the conditions of
existence of the Gibbs state of these sets (propositions 1a, 3a,
4, 6a and corollaries 1,2,3). It is also shown that the quantum
entropy is continuous at a particular state with respect to the
convergence defined by the relative entropy if and only if this
state has the sufficient rate of decreasing of the spectrum
(proposition 2). The relations between several properties of sets
of states and the corresponding properties of so called "classical
projections" of these sets are considered (proposition 5). The
obtained observations show, in particular, that discontinuity and
unboundedness of the quantum entropy has purely classical nature
(the note at the end of the section).

In section 4 the definition of the $\chi$-capacity of an arbitrary
set of states and its general properties are considered.

First of all in subsection 4.1 the notion of the optimal average
state as the unique state inheriting the most important properties
of the average state of an optimal ensemble in the finite
dimensional case is introduced (theorem 1 and corollary 4).

Then in subsection 4.2 the general properties of the
$\chi$-capacity as a function of a set of states are considered
(theorem 2 and corollaries 8,9). In particular, it is shown that
every set with finite $\chi$-capacity is relatively compact and is
contained in the maximal set with the same $\chi$-capacity. This
compactness result implies many interesting observations
concerning continuity of the $\chi$-capacity with respect to
monotonous families of sets and to the problem of existence of the
minimal closed set with given $\chi$-capacity. It also implies the
following result related to quantum channels: if the
$\chi$-capacity of an infinite dimensional channel constrained by
a particular set is finite then the image of this set under this
channel is relatively compact, in particular, every unconstrained
channel with finite $\chi$-capacity has relatively compact output
(corollary 10).

The lower and the upper bounds for the $\chi$-capacity of finite
unions is obtained (proposition 7, remark 7).

It turns out that the obtained results concerning the
$\chi$-capacity imply several observations concerning general
properties of sets of states and of the quantum entropy
(corollaries 5,6,7, remark 6, the note after corollary 8).

Finally, in subsection 4.3 the notion of an optimal measure of a
set of states is considered and the generalized "maximal distance
property" (cf.\cite{Sch-West-1}) is proved (proposition 8), which
implies necessary condition of existence of an optimal measure
(corollary 11). Sufficient condition of existence of an optimal
measure is obtained (theorem 3).

The general results of sections 3 and 4 are illustrated in section
5, where different types of sets of states are considered and
their properties are explored.

The conditions of boundedness and of continuity of the restriction
of the entropy to the several sets of states as well as the
conditions of existence of the Gibbs state of these sets are
obtained (propositions 1a,3a,6a,9a,10,12  and corollary 12).

The $\chi$-capacity and the optimal average state of the several
sets of states are determined and the related properties
(existence of an optimal measure, regularity) are explored
(propositions 1b,3b,6b,9b,11,12).

The following examples of sets with finite $\chi$-capacity are
constructed (in subsections 5.1, 5.2, 5.3 and 5.5
correspondingly):
\begin{itemize}
  \item the closed countable set having no optimal measure;
  \item the closed set having no minimal closed subset with the same $\chi$-capacity;
  \item the decreasing sequence of closed sets with the same positive
$\chi$-capacity, having the intersection with zero
$\chi$-capacity;
  \item the closed set having optimal measure, but having no atomic optimal measure.
\end{itemize}

Section 6 is devoted to the "constructive" approach to the
definition of the $\chi$-capacity and of the optimal average state
for an arbitrary set of quantum states. It is shown that both
these notions can be defined by a finite dimensional construction
and a limiting procedure similarly to the case of the entropy and
of the relative entropy (theorem 4). This provides a principal
possibility of numerical approximation of the $\chi$-capacity and
of the optimal average state of a set of general quantum states.

\section{Preliminaries}

Let $\mathcal{H}$ be a separable Hilbert space,
$\mathfrak{B}(\mathcal{H})$ - the set of all bounded operators in
$\mathcal{H}$ with the cone $\mathfrak{B}_{+}(\mathcal{H})$ of all
positive operators, $\mathfrak{T}( \mathcal{H})$ - the Banach
space of all trace-class operators with the trace norm
$\Vert\cdot\Vert_{1}$ and $\mathfrak{S}(\mathcal{H})$ - the closed
convex subset of $\mathfrak{T}(\mathcal{H})$ consisting of all
density operators in $\mathcal{H}$, which is complete separable
metric space with the metric defined by the trace norm. Each
density operator uniquely defines a normal state on
$\mathfrak{B}(\mathcal{H})$ \cite{B&R}, so, in what follows we
will also for brevity use the term "state". Note that convergence
of a sequence of states to a \textit{state} in the weak operator
topology is equivalent to convergence of this sequence  to this
state in the trace norm \cite{D-A}. We will use the following
compactness criterion for subsets of states: \textit{a closed
subset $\mathcal{K}$ of states is compact if and only if for any
$\varepsilon>0$ there is a finite dimensional projector $P$ such
that $\mathrm{Tr}\rho P\geq 1-\varepsilon$ for all
$\rho\in\mathcal{K}$}. \cite{S},\cite{H-Sh-2}

In what follows $\log$ denotes the function on $[0,+\infty)$,
which coincides with the natural logarithm on $\left( 0,+\infty
\right) $ and vanishes at zero. Let $A$ and $B$ be positive trace
class operators. Let $\{|i\rangle\}$ be a complete orthonormal set
of eigenvectors of $A$. The entropy is defined by
$H(A)=-\sum_{i}\langle i|\,A\log A\,|i\rangle$ while the relative
entropy -- as $ H(A\,\|B)=\sum_{i}\langle i|\,A\log A-A\log
B+B-A\,|i\rangle$, provided
$\mathrm{ran}A\subseteq\mathrm{ran}B$,\footnote{$\mathrm{ran}$
denotes the closure of the range of an operator in $\mathcal{H}$}
and $H(A\,\Vert B)=+\infty$ otherwise (see \cite{L} for more
detailed definition). The entropy and the relative entropy are
nonnegative lower semicontinuous (in the trace-norm topology)
concave and convex functions of their arguments correspondingly
\cite{L},\cite{O&P},\cite{W}. We will use the following inequality
\begin{equation}\label{rel-entr-ineq}
H(\rho\Vert\,\sigma)\geq\textstyle\frac{1}{2}\|\rho-\sigma\|_{1}^{2},
\end{equation}
which holds for arbitrary states $\rho$ and $\sigma$ in
$\mathfrak{S}(\mathcal{H})$ \cite{O&P}.

The relative entropy $H(\rho\Vert\,\sigma)$ for two states $\rho$
and $\sigma$ can be considered as a measure of divergence of these
states which classical analog is called Kullback-Leibler
divergence. Despite the fact that this measure is not a metric it
is possible to introduce the notion of convergence of a sequence
of states $\{\rho_{n}\}$ to a particular state $\rho_{*}$ defined
by the condition
$\lim_{n\rightarrow+\infty}H(\rho_{n}\|\rho_{*})=0$. This type of
convergence plays an important role in this paper and it will be
called \textit{$H$-convergence}. By inequality
(\ref{rel-entr-ineq}) the $H$-convergence is stronger than the
convergence defined by the trace norm.

For arbitrary set $\mathcal{A}$ let $\mathrm{co}(\mathcal{A})$ and
$\overline{\mathrm{co}}(\mathcal{A})$ be the convex hull and the
convex closure of the set $\mathcal{A}$ correspondingly, let
$\mathrm{Ext}(\mathcal{A})$ be the set of all extreme points of
the set $\mathcal{A}$ \cite{J&T}.

Speaking about continuity of a particular function on some set of
states we mean continuity of the restriction of this function to
this set.

Arbitrary finite collection $\{\rho _{i}\}$ of states in
$\mathfrak{S}(\mathcal{H})$ with corresponding set of
probabilities $\{\pi _{i}\}$ is called \textit{ensemble} and is
denoted by $\{\pi _{i},\rho _{i}\}$. The state
$\bar{\rho}=\sum_{i}\pi _{i}\rho _{i}$ is called \textit{the
average state} of the ensemble. Following \cite{H-Sh-2} we treat
an arbitrary Borel probability measure $\mu$ on
$\mathfrak{S}(\mathcal{H})$ as \textit{generalized ensemble} and
the \textit{barycenter}  of the measure $\mu$ defined by the
Pettis integral
\[
\bar{\rho}(\mu )=\int\limits_{\mathfrak{S}(\mathcal{H})}\rho \mu
(d\rho )
\]
as the average state of this ensemble. In this notations the
conventional ensembles correspond to measures with finite support.
For arbitrary closed subset $\mathcal{A}$ of
$\mathfrak{S}(\mathcal{H})$ we denote by
$\mathcal{M}(\mathcal{A})$ the set of all probability measures
supported by the set $\mathcal{A}$ \cite{Par}.

In what follows an arbitrary ensemble $\{\pi _{i},\rho _{i}\}$ is
considered as a particular case of probability measure and is also
denoted by $\mu$, especially in the cases in which the specific
features of an ensemble are not essential. In particular, a convex
mixture of ensembles is defined as a convex mixture of the
corresponding probability measures.

Consider the functionals
$$
\chi(\mu)=\int
H(\rho\Vert\bar{\rho}(\mu))\mu(d\rho)\quad\mathrm{and}\quad
\hat{H}(\mu)=\int H(\rho)\mu(d\rho).
$$
In \cite{H-Sh-2} (proposition 1 and the proof of the theorem) it
is shown that both these well defined functionals are lower
semicontinuous on $\mathcal{M}(\mathfrak{S}(\mathcal{H}))$ and
\begin{equation}\label{formula}
\chi(\mu)=H(\bar{\rho}(\mu))-\hat{H}(\mu)
\end{equation}
for arbitrary $\mu$ such that $H(\bar{\rho}(\mu))<+\infty$.

If $\mu=\{\pi _{i},\rho _{i}\}$ then
$$
\chi(\{\pi_{i},\rho_{i}\})=\sum_{i=1}^{n}\pi_{i}H(\rho_{i}\Vert
\bar{\rho})\quad\mathrm{and}\quad
\hat{H}(\{\pi_{i},\rho_{i}\})=\sum_{i=1}^{n}\pi_{i}H(\rho_{i}).
$$

In analysis of the $\chi$-capacity we shall use Donald's identity
\cite{Don},\cite{O&P}
\begin{equation}\label{Donald's identity}
\sum_{i=1}^{n}\pi_{i}H(\rho_{i}\Vert
\hat{\rho})=\sum_{i=1}^{n}\pi_{i}H(\rho_{i}\Vert
\bar{\rho})+H(\bar{\rho}\Vert \hat{\rho}),
\end{equation}
which holds for arbitrary ensemble $\{\pi _{i},\rho _{i}\}$ of $n$
states with the average state $\bar{\rho}$ and arbitrary state
$\hat{\rho}$.

We shall also use the generalized integral version of Donald's
identity \cite{H-Sh-2}
\begin{equation}\label{Donald's identity-2}
\int H(\rho\Vert\hat{\rho})\mu(d\rho)=\int H(\rho\Vert
\bar{\rho}(\mu))\mu(d\rho)+H(\bar{\rho}(\mu)\Vert \hat{\rho}),
\end{equation}
which holds for arbitrary probability measure  $\mu$ with the
barycenter $\bar{\rho}(\mu)$ and arbitrary state $\hat{\rho}$.

The generalized Donald's identity (\ref{Donald's identity-2})
implies the following observation.

\textbf{Lemma 1.} \textit{Let $\{\mu_{k}\}_{k=1}^{m}$ be a finite
set of probability measures on $\mathfrak{S}(\mathcal{H})$ and
$\{\lambda_{k}\}_{k=1}^{m}$ be a probability distribution. Then
$$
\chi\left(\sum_{k=1}^{m}\lambda_{k}\mu_{k}\right)=
\sum_{k=1}^{m}\lambda_{k}\chi\left(\mu_{k}\right)
+\chi\left(\{\lambda_{k},\bar{\rho}(\mu_{k})\}_{k=1}^{m}\right).
$$}

\textit{In the case $m=2$ for arbitrary $\lambda\in [0;1]$ the
following inequality holds}
$$
\begin{array}{c}
\chi\left(\lambda\mu_{1}+ (1-\lambda)\mu_{2}\right)\geq
\lambda\chi\left(\mu_{1}\right)+
(1-\lambda)\chi\left(\mu_{2}\right)
+\frac{\lambda(1-\lambda)}{2}\|\bar{\rho}(\mu_{2})-\bar{\rho}(\mu_{1})\|_{1}^{2}.
\end{array}
$$

\textbf{Proof.} Let $\mu=\sum_{k=1}^{m}\lambda_{k}\mu_{k}$. By
definition
$$
\chi\left(\mu\right)= \sum_{k=1}^{m}\lambda_{k}\int H(\rho\Vert
\bar{\rho}(\mu))\mu_{k}(d\rho).
$$
Applying generalized Donald's identity (\ref{Donald's identity-2})
to each inner integral in the right side of the above expression
we obtain the main identity of the lemma.

To prove the inequality in the case $m=2$ it is sufficient to
apply inequality (\ref{rel-entr-ineq}) for the estimation of the
relative entropies in the main identity of the lemma:
$$
\begin{array}{c}
\lambda
H(\bar{\rho}_{1}\|\lambda\bar{\rho}_{1}+(1-\lambda)\bar{\rho}_{2})+
(1-\lambda)
H(\bar{\rho}_{2}\|\lambda\bar{\rho}_{1}+(1-\lambda)\bar{\rho}_{2})\\\\\geq
\frac{1}{2}\lambda\|(1-\lambda)(\bar{\rho}_{2}-\bar{\rho}_{1})\|_{1}^{2}+
\frac{1}{2}(1-\lambda)\|\lambda(\bar{\rho}_{2}-\bar{\rho}_{1})\|_{1}^{2}\\\\=
\frac{1}{2}\lambda(1-\lambda)\|\bar{\rho}_{2}-\bar{\rho}_{1}\|_{1}^{2}.
\square
\end{array}
$$

Note that lemma 1 implies the following inequality
\begin{equation}\label{entropy-sc}
H(\lambda\rho_{1}+(1-\lambda)\rho_{2})\geq \lambda
H(\rho_{1})+(1-\lambda)H(\rho_{2})+
\textstyle\frac{\lambda(1-\lambda)}{2}\|\rho_{2}-\rho_{1}\|_{1}^{2},
\end{equation}
valid for arbitrary states $\rho_{1}$ and $\rho_{2}$. To show this
it is sufficient to consider spectral decompositions of these
states as probability measures on $\mathfrak{S}(\mathcal{H})$.

\section{On properties of the quantum entropy}

In this section the properties of restrictions of the quantum
entropy to sets of quantum states are considered.

Let $\mathcal{A}$ be a closed set of states with finite
$\sup_{\rho\in\mathcal{A}}H(\rho)$. If this supremum is achieved
at a particular state in $\mathcal{A}$ then this state is usually
called the Gibbs state \cite{W}. We will denote it by
$\Gamma(\mathcal{A})$. Inequality (\ref{entropy-sc}) implies the
following simple observation.

\textbf{Lemma 2.} \textit{Let $\mathcal{A}$ be a closed convex
subset of states and let $\{\rho_{n}\}$ be an arbitrary sequence
of states in $\mathcal{A}$ such that
$$
\lim_{n\rightarrow+\infty}H(\rho_{n})=\sup_{\rho\in\mathcal{A}}H(\rho)<+\infty.
$$
Then this sequence converges to the uniquely defined state
$\rho_{*}(\mathcal{A})$ in $\mathcal{A}$.}\footnote{By using the
arguments from the proof of theorem 1 in section 4 it possible to
show $H$-convergence of the sequence $\{\rho_{n}\}$ to the state
$\rho_{*}(\mathcal{A})$. By using this and proposition 2 below we
conclude that $\rho_{*}(\mathcal{A})=\Gamma(\mathcal{A})$ if there
exists $\lambda<1$ such that
$\mathrm{Tr}(\rho_{*}(\mathcal{A}))^{\lambda}<+\infty$.}

\textit{If the Gibbs state $\Gamma(\mathcal{A})$ exists then it
coincides with the state $\rho_{*}(\mathcal{A})$ and the
restriction of the entropy to the set $\mathcal{A}$ is continuous
at the state $\Gamma(\mathcal{A})$}.

\textbf{Proof.} By the assumption for arbitrary $\varepsilon>0$
there exists $N_{\varepsilon}$ such that
$H(\rho_{n})>\sup_{\rho\in\mathcal{A}}H(\rho)-\varepsilon$ for all
$n\geq N_{\varepsilon}$. Inequality (\ref{entropy-sc}) with
$\lambda=1/2$ implies
$$
\begin{array}{c}
\sup_{\rho\in\mathcal{A}}H(\rho)-\varepsilon\leq
\frac{1}{2}H(\rho_{n_{1}})+\frac{1}{2}H(\rho_{n_{2}})\\\\ \leq
H\left(\frac{1}{2}\rho_{n_{1}}+\frac{1}{2}\rho_{n_{2}}\right)-
\frac{1}{8}\|\rho_{n_{2}}-\rho_{n_{1}}\|_{1}^{2}\leq
\sup_{\rho\in\mathcal{A}}H(\rho)-
\frac{1}{8}\|\rho_{n_{2}}-\rho_{n_{1}}\|_{1}^{2},
\end{array}
$$
and hence $\|\rho_{n_{2}}-\rho_{n_{1}}\|_{1}<\sqrt{8\varepsilon}$
for all $n_{1}\geq N_{\varepsilon}$ and $n_{2}\geq
N_{\varepsilon}$. Thus the sequence $\{\rho_{n}\}$ is a Cauchy
sequence and hence it converges to a particular state $\rho_{*}$
in $\mathcal{A}$. It is easy to see that this state $\rho_{*}$
does not depend on the choice of the sequence $\{\rho_{n}\}$, so,
it is determined only by the set $\mathcal{A}$. Denote this state
by $\rho_{*}(\mathcal{A})$.

If the Gibbs state $\Gamma(\mathcal{A})$ exists then by the above
observation it coincides with the state $\rho_{*}(\mathcal{A})$.
The continuity assertion follows from lower semicontinuity of the
entropy.$\square$

Following \cite{H-Sh-2} an unbounded positive operator $H$ in
$\mathcal{H}$ with discrete spectrum of finite multiplicity will
be called $\mathfrak{H}$-\textit{operator}. Let $Q_n$ be the
spectral projector of $H$ corresponding to the lowest $n$
eigenvalues. In accordance with \cite{H-c-w-c} we shall denote
\begin{equation}\label{limit}\mathrm{Tr}\rho H=\lim_{n\to\infty}\mathrm{Tr}\rho Q_n H,
\end{equation}
where the sequence on the right side is monotonously
nondecreasing. In \cite{H-c-w-c},\cite{H-Sh-2} it is shown that
any compact subset $\mathcal{K}$ of $\mathfrak{S}(\mathcal{H})$ is
contained in the convex compact set
$\mathcal{K}_{H,h}=\{\rho\in\mathfrak{S}(\mathcal{H})\,|\,\mathrm{Tr}\rho
H\leq h\}$ defined by a particular $\mathfrak{H}$-operator $H$ and
by a positive number $h$. Let $h_{\mathrm{m}}(H)$ be the minimal
eigenvalue of $H$ and $\mathcal{H}_{\mathrm{m}}(H)$ be the
corresponding (finite dimensional) eigen subspace.

Note that $\mathcal{K}_{H,h}$ is empty if $h<h_{\mathrm{m}}(H)$,
$\mathcal{K}_{H,h}=\mathfrak{S}(\mathcal{H}_{\mathrm{m}}(H))$ if
$h=h_{\mathrm{m}}(H)$ and $\mathcal{K}_{H,h}$ necessarily contains
infinite dimensional states if $h>h_{\mathrm{m}}(H)$.

As it is shown in the following proposition properties of the
restriction of the quantum entropy to the set $\mathcal{K}_{H,h}$
is determined by the \textit{increase coefficient $\mathrm{ic}(H)$
of the $\mathfrak{H}$-operator $H$} defined as
$$
\mathrm{ic}(H)=\inf\{\lambda>0\,|\,\mathrm{Tr}\exp(-\lambda
H)<+\infty\}
$$
with $\mathrm{ic}(H)=+\infty$ if $\mathrm{Tr}\exp(-\lambda
H)=+\infty$ for all $\lambda>0$.

It is known \cite{O&P},\cite{W} that under the condition
$\mathrm{ic}(H)=0$ the entropy is continuous on the compact set
$\mathcal{K}_{H,h}$ and achieves its (finite) maximum on this set at
the Gibbs state having the form $(\mathrm{Tr}\exp(-\lambda
H))^{-1}\exp(-\lambda H)$. The following proposition generalizes
this observation. It also provides necessary and sufficient
condition of existence of the Gibbs state of the set
$\mathcal{K}_{H,h}$ and reveals another sense of the term "increase
coefficient" for $\mathrm{ic}(H)$. Let
$h_{*}(H)=\frac{\displaystyle\mathrm{Tr}H\exp(-\mathrm{ic}(H)H)}{\displaystyle\mathrm{Tr}\exp(-\mathrm{ic}(H)H)}$
if $\mathrm{Tr}\exp(-\mathrm{ic}(H)H)<+\infty$ and
$h_{*}(H)=+\infty$ otherwise.\footnote{Existence of an
$\mathfrak{H}$-operator with finite $h_{*}(H)$ is verified by the
following example:
$H=\sum_{k=1}^{+\infty}\log((k+1)\log^{3}(k+1))|k\rangle\langle
k|$.}

\textbf{Proposition 1a.}\footnote{The assertions of this proposition
have classical nature and are probably obtained somewhere in
literature. The author would be grateful for any references.}
\textit{Let $H$ be a $\mathfrak{H}$-operator in the Hilbert space
$\mathcal{H}$ and $h$ be a positive number such that
$h>h_{\mathrm{m}}(H)$.}\vspace{10pt}

\textit{The entropy is bounded on the set $\mathcal{K}_{H,h}$ if
and only if $\,\mathrm{ic}(H)<+\infty$.}\vspace{10pt}

\textit{The entropy is continuous on the set $\mathcal{K}_{H,h}$
if and only if $\,\mathrm{ic}(H)=0$.}\vspace{10pt}

\textit{If $h\leq h_{*}(H)$ then
$\sup_{\rho\in\mathcal{K}_{H,h}}H(\rho)=\lambda^{*}h+\log\mathrm{Tr}\exp(-\lambda^{*}H)$,
where \break $\lambda^{*}=\lambda^{*}(H,h)\geq\mathrm{ic}(H)$ is
uniquely defined by the equation
\begin{equation}\label{main-equation}
\mathrm{Tr}H\exp(-\lambda H)=h\mathrm{Tr}\exp(-\lambda H),
\end{equation}
and there exists the Gibbs state
$\Gamma(\mathcal{K}_{H,h})=(\mathrm{Tr}\exp(-\lambda^{*}H))^{-1}\exp(-\lambda^{*}H)$
of the set $\mathcal{K}_{H,h}$.}

\textit{If $h>h_{*}(H)$ then
$\sup_{\rho\in\mathcal{K}_{H,h}}H(\rho)=\mathrm{ic}(H)h+\log\mathrm{Tr}\exp(-\mathrm{ic}(H)H)$
and there exists no state $\rho$ in $\mathcal{K}_{H,h}$ such that
$H(\rho)=\sup_{\rho\in\mathcal{K}_{H,h}}H(\rho)$.}\vspace{10pt}

\textit{In the all cases
$\displaystyle\sup_{\rho\in\mathcal{K}_{H,h}}H(\rho)=\inf_{\lambda\in(\mathrm{ic}(H);+\infty)}(\lambda
h+\log\mathrm{Tr}\exp(-\lambda H))$}.

\textit{The function
$F_{H}(h)=\sup_{\rho\in\mathcal{K}_{H,h}}H(\rho)$ has the
following properties:
\begin{itemize}
  \item the function $F_{H}(h)$ is a continuous increasing function on $[h_{\mathrm{m}};
        +\infty)$ such that $F_{H}(h_{\mathrm{m}})=\log \dim\mathcal{H}_{\mathrm{m}}(H)$ and
        $\lim\limits_{h\rightarrow+\infty}F_{H}(h)=+\infty$;
  \item the function $F_{H}(h)$ has a continuous derivative
  $$\displaystyle\frac{dF_{H}(h)}{dh}=\left\{
   \begin{array}{ll}
    \lambda^{*}(H,h),& h\in (h_{\mathrm{m}}(H),h_{*}(H))\\
    \mathrm{ic}(H), & h\in [h_{*}(H), +\infty),
   \end{array} \right.\quad such\;\; that
  $$
  $\displaystyle\frac{dF_{H}(h)}{dh}|_{h=h_{\mathrm{m}}+0}=
  \lim_{h\rightarrow h_{\mathrm{m}}(H)+0}\frac{dF_{H}(h)}{dh}=+\infty$ and
  $\displaystyle\lim_{h\rightarrow+\infty}\frac{dF_{H}(h)}{dh}=\mathrm{ic}(H)$;
  \item the function $F_{H}(h)$ is strictly concave on
  $[h_{\mathrm{m}}(H),h_{*}(H))$ and linear on $[h_{*}(H), +\infty)$ if $h_{*}(H)<+\infty$.
\end{itemize}
}

In fig.1 the result of numerical calculations of
$\sup_{\rho\in\mathcal{K}_{H,h}}H(\rho)$ as a function of $h$ for
the $\mathfrak{H}$-operator $H=-\log\sigma$ with finite $h_{*}(H)$
and $h=c$ is shown.

\textbf{Proof.} Through this proof we will assume that
$H=\sum_{k=1}^{+\infty}h_{k}|k\rangle\langle k|$, where
$\{|k\rangle\}_{k\in\mathbb{N}}$ is an orthonormal basis in the
space $\mathcal{H}$ and $\{h_{k}\}$ is a nondecreasing sequence of
positive numbers converging to the infinity. Let
$d=\dim\mathcal{H}_{\mathrm{m}}(H)$ so that
$h_{k}=h_{\mathrm{m}},\;k=\overline{1,d}$ and
$\{|k\rangle\}_{k=1}^{d}$ is a basis of
$\mathcal{H}_{\mathrm{m}}(H)$.

Begin with the proof of the first part of the proposition.

Suppose $\mathrm{ic}(H)<+\infty$. Then there exists $\lambda>0$
such that
$$\sigma=(\mathrm{Tr}\exp(-\lambda H))^{-1}\exp(-\lambda H)$$ is a state. By
using nonnegativity of relative entropy and the definition of the
set $\mathcal{K}_{H,h}$ we obtain
$$
H(\rho)=\lambda\mathrm{Tr}\rho H+\log\mathrm{Tr}\exp(-\lambda
H)-H(\rho\|\sigma)\leq\lambda h+\log\mathrm{Tr}\exp(-\lambda
H)<+\infty
$$
for all $\rho$ in $\mathcal{K}_{H,h}$, which means boundedness of
$H(\rho)$ on $\mathcal{K}_{H,h}$.

Suppose $\sup_{\rho\in\mathcal{K}_{H,h}}H(\rho)<+\infty$. Show
first that the equation
\begin{equation}\label{mlk}
\sum_{k=1}^{n}h_{k}\exp(-\lambda
h_{k})=h\sum_{k=1}^{n}\exp(-\lambda h_{k}).
\end{equation}
has the unique positive solution $\lambda_{n}$ for all
sufficiently large $n$ and that the sequence $\{\lambda_{n}\}$ is
increasing. Note that equation (\ref{mlk}) is equivalent to the
equation $f_{n}(\lambda)=0$, where
$f_{n}(\lambda)=\sum_{k=1}^{n}(h_{k}-h)\exp(-\lambda(h_{k}-h))$.
Since
$f_{n}'(\lambda)=-\sum_{k=1}^{n}(h_{k}-h)^{2}\exp(-\lambda(h_{k}-h))<0$
the function $f_{n}(\lambda)$ is strictly decreasing on $[0;
+\infty)$. It is easy to see that
$$
f_{n}(0)=\sum_{k=1}^{n}h_{k}-nh\quad\mathrm{and}\quad\lim_{\lambda\rightarrow+\infty}f_{n}(\lambda)=-\infty
\quad\mathrm{provided}\quad h>h_{\mathrm{m}}.
$$
Since the sequence $\{h_{k}\}$ is nondecreasing and unbounded
$\sum_{k=1}^{n}h_{k}>nh$ for all sufficiently large $n$ and the
above observation imply existence of the unique positive solution
$\lambda_{n}$ of the equation $f_{n}(\lambda)=0$. To show that
$\lambda_{n+1}>\lambda_{n}$ it is sufficient to note that
$f_{n+1}(\lambda)>f_{n}(\lambda)$ for all $\lambda$ in $[0;
+\infty)$ and for all $n$ such that $h_{n}>h$.

For each sufficiently large $n$ consider the state
\begin{equation}\label{rho-n}
\rho_{n}=\left(\sum_{k=1}^{n}\exp(-\lambda_{n}h_{k})\right)^{-1}
\sum_{k=1}^{n}\exp(-\lambda_{n}h_{k})|k\rangle\langle k|
\end{equation}
in $\mathcal{K}_{H,h}$. This state is the maximum point of the
entropy $H(\rho)$ on the subset $\mathcal{K}_{H,h}^{n}$ of
$\mathcal{K}_{H,h}$, consisting of states supported by the linear
hull of the vectors $\{|k\rangle\}_{k=1}^{n}$. Indeed, by using
nonnegativity of the relative entropy and definition of the state
$\rho_{n}$ it is easy to see that
$$
H(\rho)=\lambda_{n}\mathrm{Tr}\rho
H+\log\sum_{k=1}^{n}\exp(-\lambda_{n}h_{k})-
H(\rho\|\rho_{n})\leq\lambda_{n}h+\log\sum_{k=1}^{n}\exp(-\lambda_{n}h_{k})
$$
for all $\rho\in\mathcal{K}_{H,h}^{n}$ and that the equality in
this inequality takes place if and only if $\rho=\rho_{n}$. By
using this and monotonicity of logarithm we obtain
\begin{equation}\label{h-exp}
\begin{array}{c}
H(\rho_{n})=\lambda_{n}h+\log\sum\limits_{k=1}^{n}\exp(-\lambda_{n}h_{k})\geq
\lambda_{n}(h-h_{\mathrm{m}}).
\end{array}
\end{equation}

Since $h>h_{\mathrm{m}}$, the assumption
$\sup_{\rho\in\mathcal{K}_{H,h}}H(\rho)<+\infty$ implies
boundedness of the sequence $\{\lambda_{n}\}$. By this and due to
the mentioned above monotonicity of this sequence we conclude that
there exists
$\lim_{n\rightarrow+\infty}\lambda_{n}=\lambda^{*}<+\infty$. Since
$\lambda_{n}\leq\lambda^{*}$ for all $n$  the first equality in
(\ref{h-exp}) implies
\begin{equation}\label{sum-ineq}
\sum_{k=1}^{n}\exp(-\lambda^{*}h_{k})\leq\sum_{k=1}^{n}\exp(-\lambda_{n}h_{k})<
\exp\left(\sup_{\rho\in\mathcal{K}_{H,h}}H(\rho)\right)<+\infty
\end{equation}
for all $n$ and hence
\begin{equation}\label{sum-finiteness}
\sum_{k=1}^{+\infty}\exp(-\lambda^{*}h_{k})<+\infty.
\end{equation}

This shows that $\mathrm{ic}(H)\leq\lambda^{*}<+\infty$.

Since
$\mathcal{K}_{H,h}=\overline{\bigcup_{n}\mathcal{K}_{H,h}^{n}}$
and $\sup_{\rho\in\mathcal{K}_{H,h}^{n}}H(\rho)=H(\rho_{n})$ lower
semicontinuity of the entropy implies
$$
\sup_{\rho\in\mathcal{K}_{H,h}}H(\rho)=\lim_{n\rightarrow+\infty}H(\rho_{n}).
$$

By lemma 2 the sequence of states $\{\rho_{n}\}$ converges to the
state $\rho_{*}(\mathcal{K}_{H,h})$. Since
$\lim_{n\rightarrow+\infty}\lambda_{n}=\lambda^{*}$ the sequence
$\left\{A_{n}=\sum_{k=1}^{n}\exp(-\lambda_{n}h_{k})|k\rangle\langle
k|\right\}_{n}$ of operators in $\mathfrak{T}(\mathcal{H})$
converges to the operator
$A_{*}=\sum_{k=1}^{\infty}\exp(-\lambda^{*}h_{k})|k\rangle\langle
k|$ in $\mathfrak{T}(\mathcal{H})$ in the weak operator topology.
By combining these observations it is easy to see that
\begin{equation}\label{lim-exp}
\lim_{n\rightarrow+\infty}\mathrm{Tr}A_{n}=
\lim_{n\rightarrow+\infty}\sum_{k=1}^{n}\exp(-\lambda_{n}h_{k})=
\sum_{k=1}^{+\infty}\exp(-\lambda^{*}h_{k})=\mathrm{Tr}A_{*}
\end{equation}
and that
\begin{equation}\label{rho-star}
\rho_{*}(\mathcal{K}_{H,h})=\lim_{n\rightarrow+\infty}\rho_{n}=
\left(\sum_{k=1}^{+\infty}\exp(-\lambda^{*}h_{k})\right)^{-1}
\sum_{k=1}^{+\infty}\exp(-\lambda^{*}h_{k})|k\rangle\langle k|.
\end{equation}

By using (\ref{h-exp}) and (\ref{lim-exp}) we obtain
\begin{equation}\label{h-exp-2}
\sup_{\rho\in\mathcal{K}_{H,h}}H(\rho)=\lim_{n\rightarrow+\infty}H(\rho_{n})=
h\lambda^{*}+\log\sum_{k=1}^{+\infty}\exp(-\lambda^{*}h_{k}).
\end{equation}

Lower semicontinuity of the entropy implies
$$
H(\rho_{*}(\mathcal{K}_{H,h}))=\lambda^{*}\frac{\sum_{k=1}^{+\infty}h_{k}\exp(-\lambda^{*}h_{k})}
{\sum_{k=1}^{+\infty}\exp(-\lambda^{*}h_{k})}+\log\sum_{k=1}^{+\infty}\exp(-\lambda^{*}h_{k})\leq
\lim_{n\rightarrow+\infty}H(\rho_{n}).
$$
It follows from (\ref{h-exp-2}) that this inequality is equivalent
to the inequality
\begin{equation}\label{sum-ineq-2}
\sum_{k=1}^{+\infty}h_{k}\exp(-\lambda^{*}h_{k})\leq
h\sum_{k=1}^{+\infty}\exp(-\lambda^{*}h_{k}).
\end{equation}

Note that equality in this inequality implies that
$\rho_{*}(\mathcal{K}_{H,h})$ is the Gibbs state
$\Gamma(\mathcal{K}_{H,h})$. Conversely, by lemma 2 if the Gibbs
state $\Gamma(\mathcal{K}_{H,h})$ exists then it coincides with
$\rho_{*}(\mathcal{K}_{H,h})$ and hence equality holds in
(\ref{sum-ineq-2}). Thus existence of the Gibbs state
$\Gamma(\mathcal{K}_{H,h})$ is equivalent to equality in
(\ref{sum-ineq-2}). So, to complete the proof of this part of the
proposition it is sufficient to show that the inequality $h\leq
h_{*}(H)$ is equivalent to equality in (\ref{sum-ineq-2}).

Show first that $\lambda^{*}>\mathrm{ic}(H)$ implies equality in
(\ref{sum-ineq-2}). Consider the function
$$
f(\lambda)=\lim_{n\rightarrow+\infty}f_{n}(\lambda)=\sum_{k=1}^{+\infty}(h_{k}-h)\exp(-\lambda(h_{k}-h)).
$$
Since the series $\sum_{k=1}^{+\infty}h_{k}^{p}\exp(-\lambda
h_{k})$ converges uniformly on
$[\mathrm{ic}(H)+\varepsilon;+\infty)$ for arbitrary
$p\in\mathbb{N}$ and $\varepsilon>0$ the function $f(\lambda)$ has
a continuous derivative
$f'(\lambda)=-\sum_{k=1}^{+\infty}(h_{k}-h)^{2}\exp(-\lambda(h_{k}-h))<0$
on $(\mathrm{ic}(H);+\infty)$. By the construction
$f(\lambda_{n})>f_{n}(\lambda_{n})=0$ for all sufficiently large
$n$. This and continuity of the function $f(\lambda)$ at the point
$\lambda^{*}\in (\mathrm{ic}(H);+\infty)$ imply
$f(\lambda^{*})\geq 0$. Since (\ref{sum-ineq-2}) implies the
converse inequality we obtain $f(\lambda^{*})=0$, which means
equality in (\ref{sum-ineq-2}).

If $h<h_{*}(H)$ then (finite  or infinite) $f(\mathrm{ic}(H))>0$.
Since (\ref{sum-ineq-2}) implies $f(\lambda^{*})\leq 0$ this means
$\lambda^{*}>\mathrm{ic}(H)$ and by the above observation
$f(\lambda^{*})=0$.

If $h=h_{*}(H)$ then $f(\mathrm{ic}(H))=0$ and hence
$\lambda^{*}=\mathrm{ic}(H)$. Indeed, if
$\lambda^{*}>\mathrm{ic}(H)$ then by the above observation
$f(\lambda^{*})=0=f(\mathrm{ic}(H))$ contradicting to the strict
decreasing property of the function $f(\lambda)$.

If $h>h_{*}(H)$ then $f(\mathrm{ic}(H))<0$. Since the function
$f(\lambda)$ is decreasing this implies $f(\lambda^{*})<0$ and
hence equality does not hold in (\ref{sum-ineq-2}).

Let us prove the second part of the proposition. If
$\mathrm{ic}(H)=0$ then the entropy is continuous on the set
$\mathcal{K}_{H,h}$ by the observation in \cite{W}. It follows
also from the implication $(1)\Rightarrow (2)$ in the below
proposition 4.

To prove the converse implication consider the sequence of states
$$
\{\sigma_{n}=(1-q_{n})|1\rangle\langle
1|+q_{n}n^{-1}\sum_{k=2}^{n+1}|k\rangle\langle k|\},
$$
where
$\{q_{n}=(h-h_{\mathrm{m}})\left(n^{-1}\sum_{k=2}^{n+1}h_{k}-h_{\mathrm{m}}\right)^{-1}\}$
is a sequence of positive numbers obviously converging to
zero.\footnote{We assume that $n$ is sufficiently large so that
$q_{n}\leq1$.} Since the sequence $\{\sigma_{n}\}$ lies in
$\mathcal{K}_{H,h}$ and converges to the pure state
$|1\rangle\langle 1|$ continuity of the entropy on the set
$\mathcal{K}_{H,h}$ implies convergence of the sequence
$$
\{H(\sigma_{n})=h_{2}(q_{n})+q_{n}\log
n=h_{2}(q_{n})+\frac{(h-h_{\mathrm{m}})\log
n}{n^{-1}\sum_{k=2}^{n+1}h_{k}-h_{\mathrm{m}}}\}
$$
to zero. By the obvious estimation
$n^{-1}\sum_{k=2}^{n+1}h_{k}\leq h_{n+1}$ it follows that the
sequence $\{\nu_{n}=h_{n+1}^{-1}\log n\}$ converges to zero.
Therefore for arbitrary $\lambda>0$ we have
$$
\mathrm{Tr}\exp(-\lambda H)=\sum_{n=0}^{+\infty}\exp(-\lambda
h_{n+1})=\sum_{n=1}^{+\infty}n^{-\frac{\lambda}{\nu_{n}}}<+\infty
$$
and hence $\mathrm{ic}(H)=0$.

The general expression for
$\sup_{\rho\in\mathcal{K}_{H,h}}H(\rho)$ can be deduced from the
previous observation by noting that the infinum in this expression
is achieved at $\lambda^{*}$ if $h\leq h_{*}(H)$ and at
$\mathrm{ic}(H)$ if $h\geq h_{*}(H)$.

The proof of the properties of the function $F_{H}(\rho)$ is based
on the implicit function theorem and is presented in the Appendix.
$\square$

Let $\sigma$ be an arbitrary state. In what follows we will use
the \textit{decrease coefficient $\mathrm{dc}(\sigma)$ of the
state $\sigma$} defined as
$$
\mathrm{dc}(\sigma)=\inf\{\lambda>0\,|\,\mathrm{Tr}\sigma^{\lambda}<+\infty\}\in[0;1].
$$
If $\sigma$ is a full rank state then $-\log\sigma$ is an
$\mathfrak{H}$-operator and
$\mathrm{dc}(\sigma)=\mathrm{ic}(-\log\sigma)$.

It is easy to see that $\mathrm{dc}(\sigma)<1$ implies finiteness
of the entropy $H(\sigma)$ but there exist states $\sigma$ with
finite entropy such that $\mathrm{dc}(\sigma)=1$.\footnote{For
example, the state with the spectrum
$\{a((k+1)\log^{3}(k+1))^{-1}\}$, where $a$ is a coefficient.} The
special role of these states is shown in the following
proposition.

\textbf{Proposition 2.} \textit{Let $\sigma$ be a state with
finite entropy}

\textit{If $\mathrm{dc}(\sigma)<1$ then
$$
\lim_{n\rightarrow+\infty}H(\rho_{n})=H(\sigma)
$$
for arbitrary sequence $\{\rho_{n}\}$ of states $H$-converging to
the state $\sigma$.\footnote{This means that
$\lim_{n\rightarrow+\infty}H(\rho_{n}\|\sigma)=0$.}}

\textit{If $\mathrm{dc}(\sigma)=1$ then for arbitrary $h\geq
H(\sigma)$ there exists a sequence $\{\rho_{n}\}$ of states with
finite support $H$-converging to the state $\sigma$ such that
$$
\lim_{n\rightarrow+\infty}H(\rho_{n})=h.
$$}

\textbf{Remark 1.} Proposition 2 shows that the set
$\{\sigma\!\in\!\mathfrak{S}(\mathcal{H})\,|\,\mathrm{dc}(\sigma)\!<\!1\}$
is the maximal set of continuity of the entropy with respect to
the $H$-convergence.

The proof of the proposition is based on the following lemma.

\textbf{Lemma 3.} \textit{If $\sigma$ is a state with
$\mathrm{dc}(\sigma)<1$ then for arbitrary state $\rho$ such that
$H(\rho\|\sigma)<+\infty$ the entropy $H(\rho)$ is finite and for
all $\lambda>\mathrm{dc}(\sigma)$ the following identity holds
$$
H(\rho\|(\mathrm{Tr}\sigma^{\lambda})^{-1}\sigma^{\lambda})=
\lambda
H(\rho\|\sigma)+\log\mathrm{Tr}\sigma^{\lambda}-(1-\lambda)H(\rho).
$$
}\vspace{-10pt}

\textit{If $\mathrm{Tr}\sigma^{\mathrm{dc}(\sigma)}<+\infty$
then this identity holds for $\lambda=\mathrm{dc}(\sigma)$.}

\textbf{Proof.} Let $\{P_{n}\}$ be an increasing sequence of
spectral projectors of the state $\sigma$. Let $A_{n}=P_{n}\rho
P_{n}$ and $B_{n}=P_{n}\sigma$ be positive trace class operators.

By definition we have
$$
\begin{array}{c}
H(A_{n}\|B_{n}^{\lambda})=\mathrm{Tr}(A_{n}\log A_{n}-A_{n}\log
B_{n}^{\lambda}+B_{n}^{\lambda}-A_{n})\\\\=\mathrm{Tr}((\lambda+(1-\lambda))A_{n}\log
A_{n}-\lambda A_{n}\log B_{n}+B_{n}^{\lambda}-A_{n})\\\\=\lambda
H(A_{n}\|B_{n})+\mathrm{Tr}B_{n}^{\lambda}
-\lambda\mathrm{Tr}B_{n}-(1-\lambda)\mathrm{Tr}A_{n}-(1-\lambda)\mathrm{Tr}A_{n}(-\log
A_{n}).
\end{array}
$$

Since $B_{n}^{\lambda}=P_{n}\sigma^{\lambda}$ Lindblad's results
\cite{L} imply
$$
\lim_{n\rightarrow+\infty}\mathrm{Tr}A_{n}(-\log
A_{n})=H(\rho)\quad\mathrm{and}\quad
\lim_{n\rightarrow+\infty}H(A_{n}\|B_{n}^{\lambda})=H(\rho\|\sigma^{\lambda})
$$
for all $\lambda>\mathrm{dc}(\sigma)$. So, passing to the limit in
the above equality we obtain
$$
H(\rho\|\sigma^{\lambda})=\lambda
H(\rho\|\sigma)+\mathrm{Tr}\sigma^{\lambda}-1-(1-\lambda)H(\rho).
$$
Thus finiteness of $H(\rho\|\sigma)$ implies finiteness of
$H(\rho)$ and of $H(\rho\|\sigma^{\lambda})$ for all
$\lambda>\mathrm{dc}(\sigma)$. By noting that
$$
H(\rho\|(\mathrm{Tr}\sigma^{\lambda})^{-1}\sigma^{\lambda})=H(\rho\|\sigma^{\lambda})
+\log\mathrm{Tr}\sigma^{\lambda}-\mathrm{Tr}\sigma^{\lambda}+1
$$
we obtain the identity of the lemma. $\square$ \vspace{5pt}

\textbf{Proof of proposition 2.} Let $\mathrm{dc}(\sigma)<1$. Then
lemma 3 implies

\begin{equation}\label{new-ineq}
\frac{H(\rho_{n}\|(\mathrm{Tr}\sigma^{\lambda})^{-1}\sigma^{\lambda})-\lambda
H(\rho_{n}\|\sigma)}{1-\lambda}
=\frac{\log\mathrm{Tr}\sigma^{\lambda}}{1-\lambda}-H(\rho_{n})
\end{equation}
for all $\lambda>\mathrm{dc}(\sigma)$. Suppose
$\liminf_{n\rightarrow+\infty}H(\rho_{n})-H(\sigma)=\Delta>0$.
Since the first term in the right side of (\ref{new-ineq}) tends
to $H(\sigma)$ as $\lambda\rightarrow 1$ there exists $\lambda'<1$
such that the right side of (\ref{new-ineq}) is less than
$-\Delta/2$ for this $\lambda'$ and sufficiently large $n$ while
by nonegativity of the relative entropy the left side of
(\ref{new-ineq}) is greater than
$-\displaystyle\frac{\lambda'H(\rho_{n}\|\sigma)}{1-\lambda'}$,
which tends to zero as $n\rightarrow+\infty$.\vspace{5pt}

Let $\mathrm{dc}(\sigma)=1$ and let $h>H(\sigma)$. Without loss of
generality we may assume that $\sigma$ is a full rank state so
that $-\log\sigma$ is a $\mathfrak{H}$-operator such that
$\mathrm{ic}(-\log\sigma)=\mathrm{dc}(\sigma)=1$ and
$h_{*}(-\log\sigma)=H(\sigma)<+\infty$. By proposition 1a
$\sup_{\rho\in\mathcal{K}_{-\log\sigma,h}}H(\rho)=h$ for all
$h>h_{*}(-\log\sigma)$. For given $h>h_{*}(-\log\sigma)$ in the
proof of proposition 1a the sequence $\{\rho_{n}\}$ of states
defined by (\ref{rho-n}) and converging to the state
$\rho_{*}(\mathcal{K}_{-\log\sigma,h})=\sigma$ was constructed. By
this construction
$$
\lim_{n\rightarrow+\infty}H(\rho_{n})=\sup_{\rho\in\mathcal{K}_{-\log\sigma,h}}H(\rho)=h\quad
\mathrm{and}\quad
\lim_{n\rightarrow+\infty}H(\rho_{n}\|\sigma)=0.\square
$$

Consider the set
$\mathcal{V}_{\sigma,c}=\{\rho\in\mathfrak{S}(\mathcal{H})|\,H(\rho\|\sigma)\leq
c\}$ defined by a particular state $\sigma$ and by a nonnegative
number $c$. By the properties of the relative entropy the set
$\mathcal{V}_{\sigma,c}$ is a nonempty closed and convex subset of
$\mathfrak{S}(\mathcal{H})$ for arbitrary $\sigma$ and $c$. We may
consider the set $\mathcal{V}_{\sigma,c}$ as a $c$-pseudovicinity
of the state $\sigma$ with respect to the pseudometric defined by
the relative entropy. We will see in the next section that this
set plays the special role related with the notion of the
$\chi$-capacity of a set of states.

Let
$c_{*}(\sigma)=H((\mathrm{Tr}\sigma^{\mathrm{dc}(\sigma)})^{-1}\sigma^{\mathrm{dc}(\sigma)}\|\sigma)$
if $\mathrm{Tr}\sigma^{\mathrm{dc}(\sigma)}<+\infty$ and
$c_{*}(\sigma)=+\infty$ otherwise. The properties of the
restriction of the entropy to the set $\mathcal{V}_{\sigma,c}$ as
well as the necessary and sufficient condition of existence of the
Gibbs state of this set are considered in the following
proposition.\vspace{5pt}

\textbf{Proposition 3a.} \textit{Let $\sigma$ be an arbitrary
state and $c$ be a positive number.}\vspace{10pt}

\textit{The set $\mathcal{V}_{\sigma,c}$ is a compact convex
subset of $\mathfrak{S}(\mathcal{H})$.}\vspace{10pt}

\textit{The entropy is bounded on the set $\mathcal{V}_{\sigma,c}$
if and only if $\mathrm{dc}(\sigma)<1$.}\vspace{10pt}

\textit{The entropy is continuous on the set
$\mathcal{V}_{\sigma,c}$ if and only if
$\mathrm{dc}(\sigma)=0$.}\vspace{10pt}

\textit{If $\mathrm{dc}(\sigma)<1$ and $c\leq c_{*}(\sigma)$ then
$\displaystyle\sup_{\rho\in\mathcal{V}_{\sigma,c}}H(\rho)=\frac{\lambda^{*}c+
\log\mathrm{Tr}\sigma^{\lambda^{*}}}{1-\lambda^{*}}$ where
$\lambda^{*}=\lambda^{*}(\sigma,c)\geq\mathrm{dc}(\sigma)$ is
uniquely defined by the equation\footnote{This equation means that
$H((\mathrm{Tr}\sigma^{\lambda^{*}})^{-1}\sigma^{\lambda^{*}}\|\,\sigma)=c$.}
$$
(\lambda-1)\mathrm{Tr}(\sigma^{\lambda}\log
\sigma)=(c+\log\mathrm{Tr}\sigma^{\lambda})\mathrm{Tr}\sigma^{\lambda}
$$
and there exists the Gibbs state
$\Gamma(\mathcal{V}_{\sigma,c})=(\mathrm{Tr}\sigma^{\lambda^{*}})^{-1}\sigma^{\lambda^{*}}$
for the set $\mathcal{V}_{\sigma,c}$.}\vspace{10pt}

\textit{If $\mathrm{dc}(\sigma)<1$ and $c> c_{*}(\sigma)$ then
$\displaystyle\sup_{\rho\in\mathcal{V}_{\sigma,c}}H(\rho)=\frac{\mathrm{dc}(\sigma)c+
\log\mathrm{Tr}\sigma^{\mathrm{dc}(\sigma)}}{1-\mathrm{dc}(\sigma)}$
and there exists no state $\rho$ in $\mathcal{V}_{\sigma,c}$ such
that
$H(\rho)=\sup_{\rho\in\mathcal{V}_{\sigma,c}}H(\rho)$.}\vspace{10pt}

\textit{If $\mathrm{dc}(\sigma)<1$  then
$\displaystyle\sup_{\rho\in\mathcal{V}_{\sigma,c}}H(\rho)=
\inf_{\lambda\in(\mathrm{dc}(\sigma);1)}\frac{\lambda c+
\log\mathrm{Tr}\sigma^{\lambda}}{1-\lambda}$ for arbitrary
$c$.}\vspace{10pt}

In fig.1 the result of numerical calculations of
$\sup_{\rho\in\mathcal{V}_{\sigma,c}}H(\rho)$ as a function of $c$
for the state $\sigma$ with $\mathrm{dc}(\sigma)<1$ and finite
$c_{*}(\sigma)$ is shown.

\textbf{Proof.} Without loss of generality we may assume that
$\sigma$ is a full rank state so that $-\log\sigma$ is a
$\mathfrak{H}$-operator.\footnote{This assumption and infinite
dimensionality of the space $\mathcal{H}$ used in the proof imply
that $\sigma$ is a state with infinite rank. But it is possible to
show that the all assertions of proposition 3a are valid for an
arbitrary state $\sigma$ with finite rank.}

The proof of the compactness assertion is based on the compactness
criterion, described in section 2, and the inequality
\begin{equation}\label{R-H-ineq}
H(\rho\|\sigma)\geq H(P\rho P\|P\sigma P)\geq
\mathrm{Tr}(P\rho)\log\frac{\mathrm{Tr}(P\rho)}{\mathrm{Tr}(P\sigma)}+\mathrm{Tr}(P\sigma)-\mathrm{Tr}(P\rho),
\end{equation}
valid for arbitrary states $\rho,\sigma$ and arbitrary projector
$P$. This inequality follows from lemma 3 in \cite{L} and the
monotonicity property of the relative entropy \cite{L-2}, applied
to the completely positive trace preserving map
$\Phi(A)=(\mathrm{Tr}A)\tau$, where $\tau$ is an arbitrary state.

For given $\sigma$ let $\{P_{n}\}$ be a sequence of finite rank
projectors such that $\mathrm{Tr}P_{n}\sigma>1-n^{-1}$. Suppose,
$\mathcal{V}_{\sigma,c}$ is not compact. By compactness criterion
for arbitrary $n$ there exists a state $\rho_{n}$ in
$\mathcal{V}_{\sigma,c}$ such that
$\mathrm{Tr}(I_{\mathcal{H}}-P_{n})\rho_{n}>\varepsilon$ for some
positive $\varepsilon$. By this and using inequality
(\ref{R-H-ineq}) with $P=I_{\mathcal{H}}-P_{n}$ we have
$$
\begin{array}{c}
H(\rho_{n}\|\sigma)\geq
\mathrm{Tr}((I_{\mathcal{H}}-P_{n})\rho_{n})
\log\displaystyle\frac{\mathrm{Tr}((I_{\mathcal{H}}-P_{n})\rho_{n})}{\mathrm{Tr}((I_{\mathcal{H}}-P_{n})\sigma)}\\\\
+\mathrm{Tr}((I_{\mathcal{H}}-P_{n})\sigma)-\mathrm{Tr}((I_{\mathcal{H}}-P_{n})\rho_{n})\geq
\varepsilon\log(\varepsilon n)-1
\end{array}
$$
for sufficiently large $n$ and hence $H(\rho_{n}\|\sigma)$ tends
to the infinity as $n\rightarrow+\infty$, contradicting to the
definition of the set $\mathcal{V}_{\sigma,c}$.

If $\mathrm{dc}(\sigma)=1$ then by the second part of proposition
2 the entropy is unbounded on the set $\mathcal{V}_{\sigma,c}$.

If $\mathrm{dc}(\sigma)<1$ then lemma 3 implies
\begin{equation}\label{ent-bound}
H(\rho)=\frac{\lambda
H(\rho\|\sigma)+\log\mathrm{Tr}\sigma^{\lambda}-H(\rho\|\sigma_{\lambda})}{1-\lambda}\leq
\frac{c\lambda+\log\mathrm{Tr}\sigma^{\lambda}}{1-\lambda}
\end{equation}
for all $\lambda$ in $(\mathrm{dc}(\sigma);1)$ and all $\rho$ in
$\mathcal{V}_{\sigma,c}$. This implies
$\sup_{\rho\in\mathcal{V}_{\sigma,c}}H(\rho)<+\infty$.

If $\mathrm{dc}(\sigma)>0$ then by proposition 1a the entropy is
not continuous on the set $\mathcal{K}_{-\log\sigma,c}$, which is
contained in $\mathcal{V}_{\sigma,c}$.

If $\mathrm{dc}(\sigma)=0$ then by the above observation
$\sup_{\rho\in\mathcal{V}_{\sigma,c}}H(\rho)=d<+\infty$ and hence
the set $\mathcal{V}_{\sigma,c}$ is contained in
$\mathcal{K}_{-\log\sigma,c+d}$. By proposition 1a the entropy is
continuous on the set $\mathcal{K}_{-\log\sigma,c+d}$.

To prove the next part of the proposition denote the state
$(\mathrm{Tr}\sigma^{\lambda})^{-1}\sigma^{\lambda}$ by
$\sigma_{\lambda}$ and note that the continuous function
$f(\lambda)=H(\sigma_{\lambda}\|\sigma)$ is decreasing on
$(\mathrm{dc}(\sigma);1)$. Indeed, it is easy to see by direct
calculation that this function has a derivative
$$
f'(\lambda)=-(1-\lambda)\left(\mathrm{Tr}\sigma_{\lambda}\log^{2}\sigma-
(\mathrm{Tr}\sigma_{\lambda}\log\sigma)^{2}\right)<0
$$
for each $\lambda$ in $(\mathrm{dc}(\sigma);1)$. Note also that
$$
\lim_{\lambda\rightarrow\mathrm{dc}(\sigma)+0}f(\lambda)=c_{*}\leq+\infty\quad\mathrm{and}\quad
f(1)=0.
$$

Suppose that $c\leq c_{*}$. Then the above observation implies
existence of the unique solution $\lambda^{*}$ of the equation
$f(\lambda)=c$. Thus $H(\sigma_{\lambda^{*}}\|\sigma)=c$ and hence
$$
H(\sigma_{\lambda^{*}})=\frac{c\lambda^{*}+
\log\mathrm{Tr}\sigma^{\lambda^{*}}}{1-\lambda^{*}}.
$$
The inequality (\ref{ent-bound}) implies $H(\rho)\leq
H(\sigma_{\lambda^{*}})$ for all $\rho$ in
$\mathcal{V}_{\sigma,c}$.

Suppose $c_{*}$ is finite and  $c>c_{*}$. Then
$$
h=\frac{\mathrm{dc}(\sigma)c+\log\mathrm{Tr}\sigma^{\mathrm{dc}(\sigma)}}{1-\mathrm{dc}(\sigma)}>
\frac{\mathrm{dc}(\sigma)c_{*}+\log\mathrm{Tr}\sigma^{\mathrm{dc}(\sigma)}}{1-\mathrm{dc}(\sigma)}=H(\sigma_{\mathrm{dc}(\sigma)}).
$$
Since $\mathrm{dc}(\sigma_{\mathrm{dc}(\sigma)})=1$ it follows
from proposition 2 that for each sufficiently large $m$ there
exists a sequence $\{\rho_{n}^{m}\}_{n}$ of states such that
\begin{equation}\label{s-lim-exp}
\lim_{n\rightarrow+\infty}H(\rho_{n}^{m}\|\sigma_{\mathrm{dc}(\sigma)})=0\quad
\mathrm{and}\quad \lim_{n\rightarrow+\infty}H(\rho_{n}^{m})=h-1/m.
\end{equation}

By lemma 3 we have
$$
\begin{array}{c}
\displaystyle\lim_{n\rightarrow+\infty}H(\rho_{n}^{m}\|\sigma)=\lim_{n\rightarrow+\infty}
\frac{H(\rho_{n}^{m}\|\sigma_{\mathrm{dc}(\sigma)})-\log\mathrm{Tr}\sigma^{\mathrm{dc}(\sigma)}
+(1-\mathrm{dc}(\sigma))H(\rho_{n}^{m})}{\mathrm{dc}(\sigma)}\\\\
\displaystyle=\frac{(1-\mathrm{dc}(\sigma))h-\log\mathrm{Tr}\sigma^{\mathrm{dc}(\sigma)}}
{\mathrm{dc}(\sigma)}-\frac{1-\mathrm{dc}(\sigma)}{\mathrm{dc}(\sigma)m}=
c-\frac{1-\mathrm{dc}(\sigma)}{\mathrm{dc}(\sigma)m}.
\end{array}
$$

Thus for each $m$ there exists $N(m)$ such that
$\rho_{n}^{m}\in\mathcal{V}_{\sigma,c}$ for all $n\geq N(m)$. This
and (\ref{s-lim-exp}) implies possibility to extract from the
family $\{\rho_{n}^{m}\}_{n,m}$ a sequence
$\{\hat{\rho}_{n}\}_{n}$ of states in $\mathcal{V}_{\sigma,c}$
converging to the state $\sigma_{\mathrm{dc}(\sigma)}$ such that
$\lim_{n\rightarrow+\infty}H(\hat{\rho}_{n})=h$. This shows that
$\sup_{\rho\in\mathcal{V}_{\sigma,c}}H(\rho)\geq h$. Since the
converse inequality follows from (\ref{ent-bound}) we obtain
$\sup_{\rho\in\mathcal{V}_{\sigma,c}}H(\rho)=h>H(\sigma_{\mathrm{dc}(\sigma)})$,
which by lemma 2 implies nonexistence of the Gibbs state of the
set $\mathcal{V}_{\sigma,c}$ in this case.

The general expression for
$\sup_{\rho\in\mathcal{V}_{\sigma,c}}H(\rho)$ can be deduced from
the previous observation by noting that the infinum in this
expression is achieved at $\lambda^{*}$ if $c\leq c_{*}(\sigma)$
and at $\mathrm{dc}(\sigma)$ if $c\geq c_{*}(\sigma)$. $\square$

The following proposition is devoted to the question of continuity
of the entropy on arbitrary subsets of states.

\textbf{Proposition 4.} \textit{Let $\mathcal{A}$ be an arbitrary
closed subset of $\mathfrak{S}(\mathcal{H})$. The following
properties are equivalent:}
\begin{enumerate}[(i)]
  \item  \textit{$\mathcal{A}\subseteq\mathcal{K}_{H,h}$ for some positive number $h$ and $\mathfrak{H}$-operator
         $H$ with $\mathrm{ic}(H)=0$;}
  \item \textit{The entropy is continuous on the set $\mathcal{A}$ and there exists a state $\sigma$ in
        $\mathfrak{S}(\mathcal{H})$ such that the relative entropy $H(\rho\|\sigma)$ is continuous and bounded
        on the set $\mathcal{A}$;}
  \item \textit{There exists a $\mathfrak{H}$-operator
         $\widetilde{H}$ with $\mathrm{ic}(\widetilde{H})<+\infty$ such that
        the linear function $\mathrm{Tr}\rho\widetilde{H}$ is continuous and bounded
        on the set $\mathcal{A}$.}
\end{enumerate}
\textit{If equivalent properties $(i)-(iii)$ hold then the
$\mathfrak{H}$-operators $H$,$\widetilde{H}$  and the state
$\sigma$ can be chosen in such a way that $\mathrm{Tr}\,\sigma
H<+\infty$, $\widetilde{H}=-\log\sigma$ and $H(\sigma)<+\infty$.}

\textbf{Remark 2.} The last assertion of this proposition implies
that the properties $(i)-(iii)$ remain valid for the set
$\overline{\mathrm{co}}\{\mathcal{A}, \sigma\}$.$\square$

\textbf{Proof.} $(ii)\Rightarrow(iii)\;$ Since every continuous
function is finite we have
\begin{equation}\label{imp-exp}
 H(\rho\|\sigma)=-H(\rho)+\mathrm{Tr}\rho(-\log\sigma),\quad\forall\rho\in\mathcal{A}.
\end{equation}

By proposition 3a the set $\mathcal{A}$ is compact and hence the
entropy is bounded on $\mathcal{A}$. Thus the conditions of $(ii)$
and (\ref{imp-exp}) imply continuity and boundedness of the
function $\mathrm{Tr}\rho(-\log\sigma)$ on the set $\mathcal{A}$.
Hence $(iii)$ holds with $\widetilde{H}=-\log\sigma$.

$(iii)\Rightarrow(ii)\;$ For given
$\lambda>\mathrm{ic}(\widetilde{H})$ let
$\sigma=(\mathrm{Tr}\exp(-\lambda
\widetilde{H}))^{-1}\exp(-\lambda \widetilde{H})$ be a state in
$\mathfrak{S}(\mathcal{H})$ with finite entropy. Then $(iii)$
means continuity and boundedness of the function
$\mathrm{Tr}\rho(-\log\sigma)$ on the set $\mathcal{A}$. By lower
semicontinuity of the entropy and of the relative entropy this and
(\ref{imp-exp}) imply  continuity and boundedness of the functions
$H(\rho)$ and $H(\rho\|\sigma)$ on the set $\mathcal{A}$.

$(i)\Rightarrow(iii)\;$ By the assumption $\sum_{k}\exp(-\lambda
h_{k})<+\infty$ for all $\lambda>0$ and hence
$\sum_{k}h_{k}\exp(-\lambda h_{k})<+\infty$ for all $\lambda>0$.
This implies existence of a sequence $\{\lambda_{k}\}$ of positive
numbers monotonously converging to zero and such that
$\sum_{k}h_{k}\exp(-\lambda_{k}h_{k})<+\infty$. This sequence can
be constructed as follows. For arbitrary natural $m$ let $N(m)$ be
the minimal number such that
$\sum_{k=N(m)}^{+\infty}h_{k}\exp(-h_{k}/m)<2^{-m}$. Consider a
sequence
$$
    \lambda_{k}=\left\{\begin{array}{lc}
    1,& k<N(2)\\
     1/m, & N(m)\leq k < N(m+1), m\geq 2.
    \end{array}\right.
$$

It is easy to see that this sequence satisfies the above
condition. Since $\mathrm{Tr}\,\rho H=\sum_{k}h_{k}\langle
k|\rho|k\rangle\leq h$ for all $\rho$ in $\mathcal{A}$ the series
$\sum_{k}\lambda_{k}h_{k}\langle k|\rho|k\rangle$ converges
uniformly on $\mathcal{A}$. This implies continuity of the
function $\mathrm{Tr}\rho(-\log\sigma)$, where
$\sigma=\left(\sum_{k}\exp(-\lambda_{k}h_{k})\right)^{-1}\sum_{k}\exp(-\lambda_{k}h_{k})|k\rangle
\langle k|$. Note that the condition
$\sum_{k}h_{k}\exp(-\lambda_{k}h_{k})<+\infty$ implies
$\mathrm{Tr}\,\sigma H<+\infty$ and $H(\sigma)<+\infty$. Thus
$(iii)$ holds with $\widetilde{H}=-\log\sigma$

$(iii)\Rightarrow(i)\;$ Let
$\widetilde{H}=\sum_{k}\tilde{h}_{k}|k\rangle\langle k|$, where
$\{|k\rangle\}$ is an orthonormal basis in $\mathcal{H}$. Since
$(iii)$ means $(ii)$ proposition 3a implies compactness of the set
$\mathcal{A}$. By the assumption the series
$\sum_{k}\tilde{h}_{k}\langle k|\rho|k\rangle$ converges on the
compact set $\mathcal{A}$ to the continuous function
$\mathrm{Tr}\rho\widetilde{H}$. By Dini's lemma it converges
uniformly on $\mathcal{A}$. This implies existence of a sequence
$\{\lambda_{k}\}$ of positive numbers monotonously converging to
the infinity and such that
$\sum_{k}\lambda_{k}\tilde{h}_{k}\langle k|\rho|k\rangle\leq
h<+\infty$ for all $\rho$ in $\mathcal{A}$. It is easy to see that
the $\mathfrak{H}$-operator
$H=\sum_{k}\lambda_{k}\tilde{h}_{k}|k\rangle\langle k|$ has the
all properties stated in $(i)$.

The last assertion of the proposition follows from the above
construction.  $\square$

Propositions 1a and 4 imply the following observation.

\textbf{Corollary 1.} \textit{If $H$ is a $\mathfrak{H}$-operator
with $\mathrm{ic}(H)=0$ then there exist a state $\sigma$ in
$\mathfrak{S}(\mathcal{H})$ and a $\mathfrak{H}$-operator
$\widetilde{H}$ with $\mathrm{ic}(\widetilde{H})<+\infty$ such
that the relative entropy $H(\rho\|\sigma)$ and the linear
functional $\mathrm{Tr}\rho\widetilde{H}$ are continuous on the
set $\mathcal{K}_{H,h}$.}

Since the set $\mathcal{K}_{H,h}$ is convex by definition
propositions 1a and 4 also provide the following result.

\textbf{Corollary 2.} \textit{If the entropy is continuous on the
closed set $\mathcal{A}$ and there exists a state $\sigma$ in
$\mathfrak{S}(\mathcal{H})$ such that the relative entropy
$H(\rho\|\sigma)$ is continuous and bounded on the set
$\mathcal{A}$ then the entropy is continuous on the set
$\overline{\mathrm{co}}(\mathcal{A})$.}

\textbf{Remark 3.} The assumption of existence of the state
$\sigma$ in the statement $(ii)$ of proposition 4 and in corollary
2 is essential. Indeed, let $\mathcal{A}$ be the closed subset of
all pure states in $\mathfrak{S}(\mathcal{H})$. Then the entropy
is trivially continuous on this set $\mathcal{A}$, but it is not
continuous on
$\overline{\mathrm{co}}(\mathcal{A})=\mathfrak{S}(\mathcal{H})$.
There exists \textit{compact countable} set $\mathcal{A}$ of pure
states such that the entropy is unbounded on the set
$\overline{\mathrm{co}}(\mathcal{A})$ (see the example in
subsection 5.1 below). $\square$

The implication $(iii)\Rightarrow(ii)$ in proposition 4 makes
possible to show continuity of the entropy on some nontrivial
subsets of states, which will be used in subsection 5.5.

\textbf{Corollary 3.} \textit{Let $\lambda\mapsto U_{\lambda}$ be
a continuous mapping from some compact set $\Lambda$ into the set
of all unitaries (antiunitaries) in $\mathcal{H}$ and let $\omega$
be a state in $\mathfrak{S}(\mathcal{H})$ such that
$U_{\lambda}\omega U_{\lambda}^{*}=\omega$ for all
$\lambda\in\Lambda$. Then for arbitrary state $\sigma$ such that
$\mathrm{Tr}\sigma(-\log\omega)<+\infty$ the functions $H(\rho)$
and $H(\rho\|\omega)$ are continuous on the set
$\overline{\mathrm{co}}(\{U_{\lambda}\sigma
U_{\lambda}^{*}\}_{\lambda\in\Lambda})$. }\vspace{10pt}

For an arbitrary orthonormal basis
$\{|k\rangle\}\subset\mathcal{H}$ consider the expectation
$$
\Pi_{\{|k\rangle\}}:\rho\mapsto\sum_{k}\langle
k|\rho|k\rangle|k\rangle\langle k|.
$$
Note that the output states of $\Pi_{\{|k\rangle\}}$ can be
considered as classical states (probability distributions). So, we
may call the set $\Pi_{\{|k\rangle\}}(\mathcal{A})$
\textit{classical projection} of the set $\mathcal{A}$,
corresponding to the basis $\{|k\rangle\}$.

The following proposition shows, roughly speaking, that properties
of sets of quantum states are closely related to the properties of
classical projections of these sets.

\textbf{Proposition 5.} \textit{Let $\mathcal{A}$ be an arbitrary
closed subset of $\mathfrak{S}(\mathcal{H})$.}

\begin{enumerate}[A)]
  \item \textit{The set $\mathcal{A}$ is compact if the set
$\Pi_{\{|k\rangle\}}(\mathcal{A})$ is compact for at least one
basis $\{|k\rangle\}$.}
  \item \textit{If the set $\mathcal{A}$ is compact then the set
$\Pi_{\{|k\rangle\}}(\mathcal{A})$ is compact for arbitrary basis
$\{|k\rangle\}$.}
 \item \textit{The entropy is bounded on the set $\mathcal{A}$ if it is
bounded on the set $\Pi_{\{|k\rangle\}}(\mathcal{A})$ for at least
one basis $\{|k\rangle\}$. }

\item \textit{If the entropy is bounded on the set $\mathcal{A}$ and the
set $\mathcal{A}$ is convex then it is bounded on the set
$\Pi_{\{|k\rangle\}}(\mathcal{A})$ for at least one basis
$\{|k\rangle\}$. }

\item \textit{The entropy is continuous on the set $\mathcal{A}$ if it
is continuous on the set $\Pi_{\{|k\rangle\}}(\mathcal{A})$ for at
least one basis $\{|k\rangle\}$.}

\item \textit{If the entropy is continuous on the set $\mathcal{A}$ and
there exists a state $\sigma$ in $\mathfrak{S}(\mathcal{H})$ such
that the relative entropy $H(\rho\|\sigma)$ is continuous and
bounded on the set $\mathcal{A}$ then the entropy is continuous on
the set $\Pi_{\{|k\rangle\}}(\mathcal{A})$ for at least one basis
$\{|k\rangle\}$. }

\end{enumerate}

\textbf{Proof.} If the set $\Pi_{\{|k\rangle\}}(\mathcal{A})$ is
compact then by the compactness criterion for subsets of classical
states for arbitrary $\varepsilon>0$ there exists
$N_{\varepsilon}$ such that
$$\mathrm{Tr}P_{\varepsilon}\rho=\sum_{k=1}^{N_{\varepsilon}}\langle
k|\rho|k\rangle\geq 1-\varepsilon,\quad \forall
\rho\in\mathcal{A},
$$
where
$P_{\varepsilon}=\sum_{k=1}^{N_{\varepsilon}}|k\rangle\langle k|$
is a finite rank projector. By the compactness criterion for
subsets of $\mathfrak{S}(\mathcal{H})$ this implies compactness of
the set $\mathcal{A}$.

If the set $\mathcal{A}$ is compact then for arbitrary basis
$\{|k\rangle\}$ the set $\Pi_{\{|k\rangle\}}(\mathcal{A})$ is
compact as an image of a compact set under a continuous mapping.

In the proof of the following statements we will use the following
identity
\begin{equation}\label{H-H-identity}
H(\rho\|\Pi_{\{|k\rangle\}}(\rho))=H(\Pi_{\{|k\rangle\}}(\rho))-H(\rho),
\end{equation}
valid for arbitrary state $\rho$ in $\mathfrak{S}(\mathcal{H})$
with finite $H(\Pi_{\{|k\rangle\}}(\rho))$.

If the entropy is bounded on the set
$\Pi_{\{|k\rangle\}}(\mathcal{A})$ then  it is bounded on the set
$\mathcal{A}$ since identity (\ref{H-H-identity}) and
nonnegativity of the relative entropy implies $H(\rho)\leq
H(\Pi_{\{|k\rangle\}}(\rho))$ for arbitrary $\rho$ in
$\mathcal{A}$.

If the entropy is bounded on the convex set $\mathcal{A}$ then by
corollary 5 below this set $\mathcal{A}$ is contained in the set
$\mathcal{K}_{H,h}$ defined by a particular
$\mathfrak{H}$-operator $H$  with $\mathrm{ic}(H)<+\infty$. Let
$\{|k\rangle\}$ be the basis of eigenvectors for the
$\mathfrak{H}$-operator $H$. Then
$\Pi_{\{|k\rangle\}}(\mathcal{A})$ also is contained in the set
$\mathcal{K}_{H,h}$ and hence the entropy is bounded on the set
$\Pi_{\{|k\rangle\}}(\mathcal{A})$ by proposition 1a.

Suppose the entropy is continuous on the set
$\Pi_{\{|k\rangle\}}(\mathcal{A})$. Then the entropy is finite on
this set and by (\ref{H-H-identity}) it is finite on the set
$\mathcal{A}$. Let $\rho$ be a state in $\mathcal{A}$ and
$\{\rho_{n}\}$ be a sequence of states in $\mathcal{A}$ converging
to the state $\rho$. By the assumption, lower semicontinuity of
the relative entropy and (\ref{H-H-identity}) we have
$$
\begin{array}{c}
\limsup\limits_{n\rightarrow+\infty}H(\rho_{n})=\lim\limits_{n\rightarrow+\infty}H(\Pi_{\{|k\rangle\}}(\rho_{n}))
-\liminf\limits_{n\rightarrow+\infty}
H(\rho_{n}\|\Pi_{\{|k\rangle\}}(\rho_{n}))\\\\\leq
H(\Pi_{\{|k\rangle\}}(\rho))-H(\rho\|\Pi_{\{|k\rangle\}}(\rho))=H(\rho).
\end{array}
$$
This and lower semicontinuity of the entropy imply
$\lim\limits_{n\rightarrow+\infty}H(\rho_{n})=H(\rho)$.

If the entropy is continuous on the set $\mathcal{A}$ and there
exists a state $\sigma$ in $\mathfrak{S}(\mathcal{H})$ such that
the relative entropy $H(\rho\|\sigma)$ is continuous and bounded
on the set $\mathcal{A}$ then by proposition 4 the set
$\mathcal{A}$ is contained in the set $\mathcal{K}_{H,h}$ defined
by a particular $\mathfrak{H}$-operator $H$ with
$\mathrm{ic}(H)=0$. Let $\{|k\rangle\}$ be the basis of
eigenvectors for $H$. Then $\Pi_{\{|k\rangle\}}(\mathcal{A})$ also
is contained in the set $\mathcal{K}_{H,h}$ and hence the entropy
is continuous on the set $\Pi_{\{|k\rangle\}}(\mathcal{A})$ by
proposition 1a. $\square$

\textbf{Remark 4.} Note that  the expression "for at least one" in
the statements D and F of the proposition 5 can not be changed to
"for arbitrary" in contrast to the statement B. Indeed, it is easy
to find a pure state $\rho$ and a basis $\{|k\rangle\}$ such that
$H(\Pi_{\{|k\rangle\}}(\rho))=+\infty$.$\square$

Let $\sigma$ be a  state with the basis of eigenvectors
$\{|k\rangle\}$. The set $\Pi_{\{|k\rangle\}}^{-1}(\sigma)$ of all
states having the same diagonal values in the basis
$\{|k\rangle\}$ as the state $\sigma$ will be called
\textit{layer, corresponding to the state $\sigma$} and denoted by
$\mathcal{L}(\sigma)$.\footnote{If the state has different
eigenvalues then the basis $\{|k\rangle\}$ is (essentially) unique
and the set $\mathcal{L}(\sigma)$ depends only on the state
$\sigma$. If there are multiple eigenvalues then the set
$\mathcal{L}(\sigma)$ depends also on the choice of the basis
$\{|k\rangle\}$. Since in the last case all "variants" of the set
$\mathcal{L}(\sigma)$ are isomorphic to each other we will assume
that one of them is chosen.} In a sense a layer can be considered
as the simplest purely quantum subset of states.

By (\ref{H-H-identity}) we have
\begin{equation}\label{layer-ineq}
H(\rho)\leq H(\sigma),\quad \forall\rho\in \mathcal{L}(\sigma)
\end{equation}
and hence the quantum entropy is bounded on the layer
corresponding to the state $\sigma$ if and only if
$H(\sigma)<+\infty$. The above proposition implies that
boundedness of the entropy on a layer means its
continuity.\vspace{5pt}

\textbf{Proposition 6a.}  \textit{Let $\sigma$ be an arbitrary
state.}\vspace{5pt}

\textit{The set $\mathcal{L}(\sigma)$ is a compact convex subset
of $\mathfrak{S}(\mathcal{H})$.}\vspace{5pt}

\textit{The entropy $H(\rho)$ is continuous on the set
$\mathcal{L}(\sigma)$ if and only if
$$
\sup_{\rho\in\mathcal{L}(\sigma)}H(\rho)=H(\sigma)<+\infty
$$}\vspace{-5pt}

\textit{If $H(\sigma)<+\infty$ then
$H(\rho\|\sigma)=H(\sigma)-H(\rho)$ for arbitrary state $\rho$ in
$\mathcal{L}(\sigma)$.}\vspace{5pt}

\textit{If $H(\sigma)=+\infty$ then $H(\rho\|\sigma)=+\infty$ for
arbitrary pure state $\rho$ in $\mathcal{L}(\sigma)$.}\vspace{5pt}

\textbf{Proof.} The first and the second assertions follows from
the statements A and E of proposition 5 correspondingly since
$\Pi_{\{|k\rangle\}}(\mathcal{L}(\sigma))=\{\sigma\}$ if
$\{|k\rangle\}$ is the basis of eigenvectors for the state
$\sigma$.

The expression for the relative entropy in the case
$H(\sigma)<+\infty$ is a reformulation of (\ref{H-H-identity}).

Let $H(\sigma)=+\infty$ and $\rho$ be an arbitrary pure state in
$\mathcal{L}(\sigma)$. Consider the sequences of states
$\{\sigma_{n}=(\mathrm{Tr}P_{n}\sigma)^{-1}P_{n}\sigma\}$ and
$\{\rho_{n}=(\mathrm{Tr}P_{n}\rho)^{-1}P_{n}\rho P_{n}\}$, where
$P_{n}$ be a spectral projector of the state $\sigma$
corresponding to its $n$ maximal eigen values.

Since for each $n$ pure state $\rho_{n}$ lies in
$\mathcal{L}(\sigma_{n})$ by using (\ref{H-H-identity}) we obtain
\begin{equation*}
H(\rho_{n}\|\sigma_{n})=H(\sigma_{n})-H(\rho_{n})=H(\sigma_{n}).
\end{equation*}

By Lindblad's results \cite{L} the left and right sides of this
equality tends to $H(\rho\|\sigma)$ and to $H(\sigma)=+\infty$
correspondingly as $n\rightarrow+\infty$.$\square$

Propositions 5 and 6a imply the following observation:
\textit{Absence of such properties of the quantum entropy as
finiteness and continuity in the infinite dimensional case has
purely classical nature.} Indeed, the set of all quantum states
can be considered as a union of the layers corresponding to all
states diagonizable in a particular basis. The set of these states
can be identified with the set of all classical states -
probability distributions while a single layer - with a set of
purely quantum states. Proposition 6a shows that the entropy is
continuous on the whole layer if it is finite on the corresponding
classical state. By proposition 5 possible discontinuity of the
quantum entropy is connected with transitions between layers
corresponding to a set of classical states, on which the entropy
is not continuous.

\section{The $\chi$-capacity}

\subsection{The optimal average state}

Let $\mathcal{A}$ be an arbitrary subset of
$\mathfrak{S}(\mathcal{H})$. Consider the $\chi$-\textit{capacity}
of the set $\mathcal{A}$ defined by
\begin{equation}\label{ccap-1}
\bar{C}(\mathcal{A})=\sup_{\{\pi_{i},\rho_{i}\}}\chi(\{\pi_{i},\rho_{i}\}).
\end{equation}
where the supremum is over all ensembles $\{\pi_{i},\rho_{i}\}$ of
states in $\mathcal{A}$.

If the entropy is bounded on the set
$\overline{\mathrm{co}}(\mathcal{A})$ then
\begin{equation*}
\bar{C}(\mathcal{A})=\sup_{\{\pi_{i},\rho_{i}\}}\left(H\left(\sum_{i}\pi_{i}\rho_{i}\right)
-\sum_{i}\pi_{i}H(\rho_{i})\right)\leq
\sup_{\rho\in\overline{\mathrm{co}}(\mathcal{A})}H(\rho)<+\infty.
\end{equation*}

But boundedness of the entropy is not a necessary condition for
finiteness of the $\chi$-capacity, as it follows from the examples
in the next section.

In accordance to \cite{Sh-2} a sequence of ensembles
$\{\{\pi_{i}^{n},\rho _{i}^{n}\}\}_{n}$ of states in $\mathcal{A}$
such that
$$
\lim_{n\rightarrow+\infty}\chi(\{\pi_{i}^{n},\rho_{i}^{n}\})=\bar{C}(\mathcal{A})
$$
is called \textit{approximating sequence} for the set
$\mathcal{A}$.

If $\mathcal{A}$ is a set of states in finite dimensional Hilbert
space then three exists ensemble $\{\pi_{i},\rho_{i}\}$ - optimal
ensemble for the set $\mathcal{A}$ - at which the supremum in the
definition (\ref{ccap-1}) of the $\chi$-capacity is achieved
\cite{Sch-West-1}. If $\mathcal{A}$ is a set of states in infinite
dimensional Hilbert space then we can not assert existence of
optimal ensemble but we can assert existence of the unique state,
possessing the properties of the average state of the optimal
ensemble in the finite dimensional case.

\textbf{Theorem 1.} \textit{Let $\mathcal{A}$ be a set with finite
$\chi$-capacity $\bar{C}(\mathcal{A})$. Then there exists the
unique state $\Omega(\mathcal{A})$ in $\mathfrak{S}(\mathcal{H})$
such that
$$
H(\rho\|\Omega(\mathcal{A}))\leq \bar{C}(\mathcal{A})\quad for\;
all\; \rho\; in\;\mathcal{A}.
$$
The state $\Omega(\mathcal{A})$ lies in
$\overline{\mathrm{co}}(\mathcal{A})$ and for arbitrary
approximating sequence of ensembles $\{\{\pi_{i}^{n},\rho
_{i}^{n}\}\}_{n}$ for the set $\mathcal{A}$ the corresponding
sequence $\{\bar{\rho}_{n}\}$ of their average states
$H$-converges to the state $\Omega(\mathcal{A})$.}\footnote{This
means that
$\lim_{n\rightarrow+\infty}H(\bar{\rho}_{n}\|\Omega(\mathcal{A}))=0$.}

\textit{The $\chi$-capacity $\bar{C}(\mathcal{A})$ can be defined
by the expression
\begin{equation}\label{chi-cap-exp}
\bar{C}(\mathcal{A})=\inf_{\sigma\in\mathfrak{S}(\mathcal{H})}\sup_{\rho\in
\mathcal{A}}H(\rho\|\sigma)=\inf_{\sigma\in\overline{\mathrm{co}}(\mathcal{A})}\sup_{\rho\in
\mathcal{A}}H(\rho\|\sigma)=\sup_{\rho\in\mathcal{A}}H(\rho\|\Omega(\mathcal{A})),
\end{equation}
in which the first two equalities remain valid in the case
$\bar{C}(\mathcal{A})=+\infty$.}

\textbf{Proof.} Show first that for arbitrary approximating
sequence of ensembles $\left\{\mu_{n}=\{\pi_{i}^{n},\rho
_{i}^{n}\}_{i=1}^{N(n)}\right\}$ for the set $\mathcal{A}$ the
corresponding sequence of the average states $\{\bar{\rho}_{n}\}$
converges to a particular state in $\mathfrak{S}(\mathcal{H})$. By
definition of an approximating sequence for arbitrary
$\varepsilon>0$ there exists $N_{\varepsilon}$ such that
$\chi(\mu_{n})>\bar{C}(\mathcal{A})-\varepsilon$ for all $n\geq
N_{\varepsilon}$. By lemma 1 with $m=2$ and $\lambda=1/2$  we have
$$
\begin{array}{c}
\bar{C}(\mathcal{A})-\varepsilon\leq
\frac{1}{2}\chi(\mu_{n_{1}})+\frac{1}{2}\chi(\mu_{n_{2}})\\\\
\!\!\leq\chi\left(\frac{1}{2}\mu_{n_{1}}+\frac{1}{2}\mu_{n_{2}}\right)-
\frac{1}{8}\|\bar{\rho}_{n_{2}}-\bar{\rho}_{n_{1}}\|_{1}^{2}\leq
\bar{C}(\mathcal{A})-
\frac{1}{8}\|\bar{\rho}_{n_{2}}-\bar{\rho}_{n_{1}}\|_{1}^{2},
\end{array}
$$
and hence
$\|\bar{\rho}_{n_{2}}-\bar{\rho}_{n_{1}}\|_{1}<\sqrt{8\varepsilon}$
for all $n_{1}\geq N_{\varepsilon}$ and $n_{2}\geq
N_{\varepsilon}$. Thus the sequence $\{\bar{\rho}_{n}\}$ is a
Cauchy sequence and hence it converges to a particular state
$\rho_{*}$ in $\mathfrak{S}(\mathcal{H})$.

Let $\sigma$ be an arbitrary state in $\mathcal{A}$. For each $n$
consider the ensemble
\[
\mu_{n}^{\eta} =\{(1-\eta )\pi^{n} _1\rho^{n} _1,...,(1-\eta
)\pi^{n}_{N(n)}\rho^{n}_{N(n)},\eta\sigma\},\quad \eta\in [0,1]
\]
obtained from the ensemble $\mu_{n}=\{\pi _{i}^{n},\rho
_{i}^{n}\}=\mu_{n}^{0}$ of the approximating sequence by adding
the state $\sigma$ with probability $\eta$.\footnote{This trick
was originally used in \cite{Sch-West-1} in the finite dimensional
case.} We obtain the sequence of ensembles $\{\mu_{n}^{\eta}\}$
with the corresponding sequence of the average states
$\{\bar{\rho}^{\eta}_{n} =(1-\eta)\bar{\rho}_{n}+\eta\sigma\}_{n}$
converging to the state $\bar{\rho}_{\eta}=(1-\eta
)\rho_{*}+\eta\sigma$ as $n\rightarrow+\infty$.

For arbitrary $n$ we have
\begin{equation}
\chi\left( \mu^{\eta}_{n} \right) =(1-\eta
)\sum_{i}\pi^{n}_{i}H(\rho^{n}_{i}\Vert\bar{\rho}^{\eta}_{n}
)+\eta H(\sigma\Vert\bar{\rho}^{\eta}_{n}). \label{m-chi}
\end{equation}
By the assumption $\bar{C}(\mathcal{A})<+\infty$ both sums in the
right side of the above expression are finite. Applying Donald's
identity (\ref{Donald's identity}) to the first sum in the right
side we obtain
\[
\sum_{i}\pi^{n}_{i}H(\rho^{n}_{i}\Vert\bar{\rho}^{\eta}_{n})=\chi(\mu^{0}_{n})+H(\bar{\rho}_{n}\Vert
\bar{\rho}^{\eta}_{n}).
\]
Substitution of the above expression into (\ref{m-chi}) gives
$$
\chi\left( \mu^{\eta}_{n} \right)
=\chi(\mu^{0}_{n})+(1-\eta)H(\bar{\rho}_{n}\Vert\bar{\rho}^{\eta}_{n})
+\eta\left( H(\sigma\Vert
\bar{\rho}^{\eta}_{n})-\chi(\mu^{0}_{n})\right).
$$
Due to nonnegativity of the relative entropy it follows that
\begin{equation}
H(\sigma\Vert\bar{\rho}^{\eta}_{n})\leq\eta^{-1}\left(\chi\left(\mu^{\eta}_{n}\right)-\chi\left(
\mu_{n}^{0}\right)\right)+\chi\left(\mu^{0}_{n}\right),\;\;
\eta\neq 0. \label{chi-ineq-1}
\end{equation}
By definition of the approximating sequence we have
\begin{equation}\label{a-p-exp}
\lim_{n\rightarrow+\infty}\chi\left(
\mu^{0}_{n}\right)=\bar{C}(\mathcal{A})\geq
\chi\left(\mu^{\eta}_{n}\right)
\end{equation}
for all $n$ and $\eta>0$. It follows that
\begin{equation}
\liminf_{\eta\rightarrow+0}\,\liminf_{n\rightarrow+\infty}\,\eta^{-1}\left[\chi\left(\mu^{\eta}_{n}\right)
-\chi\left(\mu_{n}^{0}\right)\right]\leq 0 \label{d-lim-exp}
\end{equation}

Lower semicontinuity of the relative entropy with
(\ref{chi-ineq-1}),(\ref{a-p-exp}) and (\ref{d-lim-exp}) implies
$$
H(\sigma\Vert\rho_{*})\leq
\liminf_{\eta\rightarrow+0}\,\liminf_{n\rightarrow+\infty}
H(\sigma\Vert \bar{\rho}^{\eta}_{n})\leq \bar{C}(\mathcal{A}).
$$

This proves that
\begin{equation}\label{C-exp}
\sup_{\sigma\in\mathcal{A}}H(\sigma\Vert
\rho_{*})\leq\bar{C}(\mathcal{A}),
\end{equation}

Let $\{\{\lambda_{j}^{n},\sigma_{j}^{n}\}\}_{n}$ be an arbitrary
approximating sequence of ensembles. By inequality (\ref{C-exp})
we have
\[
\sum_{j}\lambda_{j}^{n}H(\sigma_{j}^{n}\Vert\,\rho_{*})\leq
\bar{C}(\mathcal{A}).
\]
Applying Donald's identity (\ref{Donald's identity}) to the left
side we obtain
\begin{equation}\label{D-decomp}
\sum_{j}\lambda_{j}^{n}H(\sigma_{j}^{n}\Vert\,\rho_{*})=\sum_{j}\lambda_{j}^{n}H(\sigma_{j}^{n}\Vert\,\bar{\sigma}_{n})
+H(\bar{\sigma}_{n}\Vert\,\rho_{*})
\end{equation}
From the two above expressions we have
$$
H(\bar{\sigma}_{n}\Vert\,\rho_{*})\leq
\bar{C}(\mathcal{A})-\sum_{j}\lambda_{j}^{n}H(\sigma_{j}^{n}\Vert\,\bar{\sigma}_{n}).
$$
The right side of this inequality tends to zero as
$n\rightarrow+\infty$ due to the approximating property of the
sequence $\{\{\lambda_{j}^{n},\sigma_{j}^{n}\}\}_{n}$. Thus the
sequence $\{\bar{\sigma}_{n}\}_{n}$ $H$-converges to the state
$\rho_{*}$ and hence it converges to this state in the trace norm
topology. Hence this state $\rho_{*}$ does not depend on the
choice of an approximating sequence, so, it is determined only by
the set $\mathcal{A}$. Denote this state by $\Omega(\mathcal{A})$.
The above observation implies also that
$\rho_{*}=\Omega(\mathcal{A})$ is the unique state in
$\mathfrak{S}(\mathcal{H})$ for which inequality (\ref{C-exp})
holds.

To prove expression (\ref{chi-cap-exp}) show first that inequality
(\ref{C-exp}) is in fact equality. Indeed expression
(\ref{D-decomp}), valid for an approximating sequence
$\{\{\lambda_{j}^{n},\sigma_{j}^{n}\}\}_{n}$, and nonnegativity of
the relative entropy imply
$$
\sum_{j}\lambda_{j}^{n}H(\sigma_{j}^{n}\Vert\,\bar{\sigma}_{n})\leq
\sum_{j}\lambda_{j}^{n}H(\sigma_{j}^{n}\Vert\,\rho_{*})\leq
\sup_{\sigma\in\mathcal{A}}H(\sigma\Vert \rho_{*}).
$$
By the approximating property of the sequence
$\{\{\lambda_{j}^{n},\sigma_{j}^{n}\}\}_{n}$ the left side in the
above inequality tends to $\bar{C}(\mathcal{A})$ as
$n\rightarrow+\infty$. This proves $"="$ in (\ref{C-exp}).

Consider the function
$F(\sigma)=\sup_{\rho\in\mathcal{A}}H(\rho\|\sigma)$ on
$\mathfrak{S}(\mathcal{H})$. By the equality in (\ref{C-exp}) we
have $F(\Omega(\mathcal{A}))=\bar{C}(\mathcal{A})$. It follows
that the state $\Omega(\mathcal{A})$ is the unique minimal point
of the function $F(\sigma)$ on $\mathfrak{S}(\mathcal{H})$.
Indeed, let $\sigma_{0}$ be a state in $\mathfrak{S}(\mathcal{H})$
such that
$$
\sup_{\rho\in\mathcal{A}}H(\rho\Vert\,\sigma_{0})=F(\sigma_{0})\leq
F(\Omega(\mathcal{A}))=\bar{C}(\mathcal{A})
$$
By the first part of the theorem this implies
$\sigma_{0}=\Omega(\mathcal{A})$.

If $\bar{C}(\mathcal{A})=+\infty$ then the right side of
expression (\ref{chi-cap-exp}) is equal to $+\infty$ as well.
Indeed, if $\sigma'$ is a state in $\mathfrak{S}(\mathcal{H})$
such that
$\sup_{\rho\in\mathcal{A}}H(\rho\Vert\,\sigma')=c<+\infty$ then by
using Donald's identity and nonnegativity of the relative entropy
we have
\[
\sum_{i}\pi_{i}H(\rho
_{i}\Vert\,\bar{\rho})\leq\sum_{i}\pi_{i}H(\rho
_{i}\Vert\,\sigma')-H(\bar{\rho}\|\sigma')\leq c.
\]
for arbitrary ensemble $\{\pi_{i},\rho _{i}\}$ of states in
$\mathcal{A}$. This implies $\bar{C}(\mathcal{A})\leq c<+\infty$.
$\square$

\textbf{Definition 1.} \textit{The state $\Omega(\mathcal{A})$
described in theorem 1 is called the optimal average state of the
set $\mathcal{A}$. }

Theorem 1, Donald identity (\ref{Donald's identity}) and
inequality (\ref{rel-entr-ineq}) imply the following useful
result.

\textbf{Corollary 4.} \textit{Let $\mathcal{A}$ be a set with
finite $\chi$-capacity. For arbitrary ensemble
$\{\pi_{i},\rho_{i}\}$ of states in $\mathcal{A}$ with the average
state $\bar{\rho}$ the following inequality holds}
$$
\bar{C}(\mathcal{A})-\chi(\{\pi_{i},\rho_{i}\})\geq
H(\bar{\rho}\Vert\Omega(\mathcal{A}))\geq\textstyle\frac{1}{2}\|\bar{\rho}-\Omega(\mathcal{A})\|_{1}^{2}.
$$

Theorem 1 and proposition 1a provide the following observation on
the properties of the entropy.

\textbf{Corollary 5.} \textit{The entropy is bounded on a convex
set $\mathcal{A}$ if and only if this set $\mathcal{A}$ is
relatively compact and is contained in the set $\mathcal{K}_{H,h}$
defined by a particular $\mathfrak{H}$-operator $H$ with
$\mathrm{ic}(H)<+\infty$ and positive $h$.}

\textbf{Proof.} If the set $\mathcal{A}$ is contained in the set
$\mathcal{K}_{H,h}$ with $\mathrm{ic}(H)<+\infty$ then by
proposition 1a $\sup_{\rho\in\mathcal{A}}H(\rho)<+\infty$.

If $\sup_{\rho\in\mathcal{A}}H(\rho)<+\infty$ then
$\bar{C}(\mathcal{A})<+\infty$ and by theorem 1
$$
H(\rho\|\Omega(\mathcal{A}))=\mathrm{Tr}\rho(-\log\Omega(\mathcal{A}))-H(\rho)\leq
\bar{C}(\mathcal{A})
$$
for all $\rho$ in $\mathcal{A}$. It follows
$$
\mathrm{Tr}\rho(-\log\Omega(\mathcal{A}))\leq
\bar{C}(\mathcal{A})+\sup_{\rho\in\mathcal{A}}H(\rho)
$$
for all $\rho$ in $\mathcal{A}$ and hence
$\mathcal{A}\subseteq\mathcal{K}_{H,h}$, where
$H=-\log\Omega(\mathcal{A})$ and
$h=\bar{C}(\mathcal{A})+\sup_{\rho\in\mathcal{A}}H(\rho)$.
$\square$

By corollary 5 boundedness of the entropy on a convex set
$\mathcal{A}$ means that this set $\mathcal{A}$ is contained in
the set $\mathcal{K}_{H,h}$ defined by a particular
$\mathfrak{H}$-operator $H$ with finite $\mathrm{ic}(H)$. By
theorem 1 finiteness of the $\chi$-capacity  of an arbitrary set
$\mathcal{A}$ means that this set $\mathcal{A}$ is contained in
the set $\mathcal{V}_{\Omega(\mathcal{A}),\bar{C}(\mathcal{A})}$,
having the same $\chi$-capacity and the same optimal average
state.

\subsection{General properties}

In this section we consider general properties of the
$\chi$-capacity as a function of a set. We show also the special
role of the optimal average state introduced in the previous
subsection. It turns out that many properties of sets of states
related to the $\chi$-capacity depend on validity for these sets
of one of the two special continuity properties. So, it is
convenient to introduce the following definition.

\textbf{Definition 2.} \textit{An arbitrary set $\mathcal{A}$ with
finite $\chi$-capacity is called regular if one of the two
following conditions holds:}
\begin{itemize}
\item \textit{$H(\Omega(\mathcal{A}))$ is finite and $\lim_{n\rightarrow+\infty}H(\rho_{n})=H(\Omega(\mathcal{A}))$
       for arbitrary sequence $\{\rho_{n}\}$ of states in
       $\mathrm{co}(\mathcal{A})$ $H$-converging to the state
       $\Omega(\mathcal{A})$;}\footnote{This means that $\lim_{n\rightarrow+\infty}H(\rho_{n}\|\Omega(\mathcal{A}))=0$.}
  \item \textit{the relative entropy $H(\rho\|\Omega(\mathcal{A}))$
                is  continuous on the set $\overline{\mathcal{A}}$.}
\end{itemize}

Note that continuity of the entropy on the set
$\overline{\mathrm{co}}(\mathcal{A})$ is a sufficient condition
for regularity of the  set $\mathcal{A}$, but it is very
restrictive requirement. In a sense the conditions in the above
definition are the minimal continuity requirements which
guarantees the "good" properties of the $\chi$-capacity. These
conditions do not imply each other: there exist sets, for which
the first condition holds but the second one is not valid and vise
versa. The most of the examples of sets with finite
$\chi$-capacity presented in section 5 are regular. The examples
of the nonregular sets with finite $\chi$-capacity and
consequences of this nonregularity are considered in subsections
5.1,5.2 and 5.3.

In the following theorem we summarize the properties of the
$\chi$-capacity and of the optimal average state, which will be
used later. These properties shows that $\chi$-capacity can be
considered as a specific nonadditive measure of a set of quantum
states.

\textbf{Theorem 2.} \textit{The following properties
hold\footnote{In all statements concerning the optimal average
state of a particular set it is assumed that this set has finite
$\chi$-capacity.}}

\begin{enumerate} [A)]

  \item \textit{$\bar{C}(\mathcal{A})\geq 0$ for arbitrary set $\mathcal{A}$
  and equality here takes place if and only if the set
  $\mathcal{A}$ consists of a single point;}

  \item \textit{$\bar{C}(\mathcal{A})=\bar{C}(\overline{\mathrm{co}}(\mathcal{A}))$
         and $\Omega(\mathcal{A})=\Omega(\overline{\mathrm{co}}(\mathcal{A}))$
         for arbitrary set $\mathcal{A}$;}
  \item \textit{if $\mathcal{A}\subseteq\mathcal{B}$ then
  $\bar{C}(\mathcal{A})\leq\bar{C}(\mathcal{B})$ and equality here
  implies $\Omega(\mathcal{A})=\Omega(\mathcal{B})$;}\footnote{Note that $\mathcal{A}\varsubsetneq\mathcal{B}$
  does not imply
  $\bar{C}(\mathcal{A})<\bar{C}(\mathcal{B})$ even in the case of convex and closed sets $\mathcal{A}$ and $\mathcal{B}$
  (see the examples in section 5).}
  \item \textit{if $\bar{C}(\mathcal{A})<+\infty$ then
  $\mathcal{A}$ is relatively compact and hence
  $\bar{C}(\mathcal{A})=\bar{C}(\mathrm{Ext}\overline{\mathcal{A}})$;}

 \item \textit{if
 $\mathrm{dc}(\Omega(\mathcal{A}))<1$ then the set $\mathcal{A}$
 is regular and the entropy is bounded on the set $\overline{\mathrm{co}}\mathcal{A}$,\\ if
 $\mathrm{dc}(\Omega(\mathcal{A}))=0$ then the entropy is
 continuous on the set $\overline{\mathrm{co}}\mathcal{A}$;}

\item\textit{let $\{\mathcal{A}_{n}\}$ be a
sequence of sets such that $\mathcal{A}_{n}\subseteq
\mathcal{A}_{n+1}$ for all $n$ then
$$
\lim_{n\rightarrow+\infty}\bar{C}(\mathcal{A}_{n})=\bar{C}\left(\bigcup_{n}\mathcal{A}_{n}\right)
\quad and \quad
\lim_{n\rightarrow+\infty}\Omega(\mathcal{A}_{n})=\Omega\left(\bigcup_{n}\mathcal{A}_{n}\right);
$$}

\item \textit{let $\{\mathcal{A}_{n}\}$ be a
sequence of closed sets such that $\mathcal{A}_{n}\supseteq
\mathcal{A}_{n+1}$ for all $n$ then
$$
\lim_{n\rightarrow+\infty}\bar{C}(\mathcal{A}_{n})=\bar{C}\left(\bigcap_{n}\mathcal{A}_{n}\right)
\quad and \quad
\lim_{n\rightarrow+\infty}\Omega(\mathcal{A}_{n})=\Omega\left(\bigcap_{n}\mathcal{A}_{n}\right)
$$
take place if one of the following conditions
holds:}\footnote{These condition are essential (see remark 5
below).}
\begin{itemize}

\item \textit{the set $\mathcal{A}_{1}$ is regular and
   $\Omega(\mathcal{A}_{n})=\Omega(\mathcal{A}_{1})$ for all  $n$;}
  \item \textit{the restriction of the entropy $H(\rho)$ to the set $\mathcal{A}_{1}$
               is continuous at some limit point $\omega$ of the sequence
  $\{\Omega(\mathcal{A}_{n})\}$;}\footnote{By the assertion D the
  set of limit points of the sequence $\{\Omega(\mathcal{A}_{n})\}$
  is nonempty.}
  \item \textit{the relative entropy $H(\rho\|\omega)$ is continuous on the set $\mathcal{A}_{1}$
   for some limit point $\omega$ of the sequence
  $\{\Omega(\mathcal{A}_{n})\}$;}

\end{itemize}

\item \textit{each set $\mathcal{A}$ with finite $\chi$-capacity is contained in the maximal
       set $\mathcal{V}_{\Omega(\mathcal{A}),\bar{C}(\mathcal{A})}$ with the same $\chi$-capacity;}
       \footnote{A set is called  maximal set with given
       $\chi$-capacity if it is not a proper subset of a set with the same $\chi$-capacity.}

\item \textit{each regular closed set $\mathcal{A}$ with finite $\chi$-capacity contains the minimal
       closed set with the same $\chi$-capacity;}\footnote{A set is called  minimal closed set with given
       $\chi$-capacity if it has no proper closed subsets with the same $\chi$-capacity.}

\item \textit{ if $\bar{C}(\mathcal{A})<+\infty$ and $\bar{C}(\mathcal{B})<+\infty$ then
$\bar{C}(\mathcal{A}\cup\mathcal{B})<+\infty$, in particular, the
coincidence $\Omega(\mathcal{A})=\Omega(\mathcal{B})$ implies
$\bar{C}(\mathcal{A}\cup\mathcal{B})=\max(\bar{C}(\mathcal{A}),\bar{C}(\mathcal{B}))$;}

\item \textit{if $\Phi:\mathfrak{S}(\mathcal{H})\mapsto\mathfrak{S}(\mathcal{H}')$ is an arbitrary channel
 then $\bar{C}(\Phi(\mathcal{A}))\leq\bar{C}(\mathcal{A})$ and equality here
 implies $\;\Omega(\Phi(\mathcal{A}))=\Phi(\Omega(\mathcal{A}))$;}

  \item \textit{if $\{\Phi_{t}\}_{t\in\mathbb{R}_{+}}$ is an arbitrary family of channels from
   $\mathfrak{S}(\mathcal{H})$ into itself such that
$\lim_{t\rightarrow+0}\Phi_{t}(\rho)=\rho$ for all states $\rho$
in $\mathcal{A}$ then\footnote{This assertion can be considered as
a stability property of the $\chi$-capacity and of the optimal
average state with respect to a quantum noise.}
$$
\lim_{t\rightarrow+0}\bar{C}(\Phi_{t}(\mathcal{A}))=\bar{C}(\mathcal{A})\quad
and \quad
\lim_{t\rightarrow+0}\Omega(\Phi_{t}(\mathcal{A}))=\Omega(\mathcal{A}).
$$}

\end{enumerate}

\textbf{Remark 5.} The regularity and the continuity requirements
in the assertions G and I are essential. Moreover, nonregularity
of a particular set with finite $\chi$-capacity can be shown by
finding a decreasing family of subset of this set for which the
assertion G does not hold. This possibility is used in the proof
of proposition 3b in subsection 5.3. The example of a closed set
with finite $\chi$-capacity having no minimal closed subset with
the same $\chi$-capacity is considered in subsection 5.2.$\square$

\textbf{Proof.} The assertions A, B and the first part of C
directly follows from the definition of the $\chi$-capacity due to
lower semicontinuity and convexity of the relative entropy. The
second part of C is proved as follows. Let
$\mathcal{A}\subseteq\mathcal{B}$ and
$\bar{C}(\mathcal{A})=\bar{C}(\mathcal{B})$. Then by theorem 1
$H(\rho\|\Omega(\mathcal{B}))\leq
\bar{C}(\mathcal{B})=\bar{C}(\mathcal{A})$ for all states $\rho$
in $\mathcal{B}$. Since $\mathcal{A}\subseteq\mathcal{B}$ this
inequality holds for all states $\rho$ in $\mathcal{A}$. Thus the
uniqueness assertion of theorem 1 implies
$\Omega(\mathcal{A})=\Omega(\mathcal{B})$.

The first part of D follows from proposition 3a since by theorem 1
each set $\mathcal{A}$ with finite $\chi$-capacity is contained in
the set $\mathcal{V}_{\Omega(\mathcal{A}),\bar{C}(\mathcal{A})}$.
The second part of D is a corollary of B and the Krein-Milman
theorem.

Since theorem 1 implies
$\overline{\mathrm{co}}(\mathcal{A})\subseteq\mathcal{V}_{\Omega(\mathcal{A}),\bar{C}(\mathcal{A})}$
the assertion E follows from propositions 2 and 3a.

To prove F note that C implies existence of the limit and the
inequality
\begin{equation}\label{one}
\lim_{n\rightarrow+\infty}\bar{C}(\mathcal{A}_{n})\leq\bar{C}\left(\bigcup_{n}\mathcal{A}_{n}\right).
\end{equation}

Let $\{\{\pi_{i}^{k}, \rho_{i}^{k}\}\}_{k}$ be an arbitrary
approximating sequence of ensembles for the set
$\bigcup_{n}\mathcal{A}_{n}$, so that
\begin{equation}\label{two}
\lim_{k\rightarrow+\infty}\chi(\{\pi_{i}^{k},
\rho_{i}^{k}\})=\bar{C}\left(\bigcup\mathcal{A}_{n}\right).
\end{equation}

Since an ensemble is a \textit{finite} collection of states for
each $k$ there exists $n(k)$ such that $\rho_{i}^{k}\in
\mathcal{A}_{n(k)}$ for all $i$ and hence
$\bar{C}\left(\mathcal{A}_{n(k)}\right)\geq \chi(\{\pi_{i}^{k},
\rho_{i}^{k}\})$. This and (\ref{two}) imply $"="$ in (\ref{one}).

Suppose
$\bar{C}\left(\bigcup_{n}\mathcal{A}_{n}\right)=
\bar{C}\left(\overline{\mathrm{co}}\left(\bigcup_{n}\mathcal{A}_{n}\right)\right)<+\infty$.
By the assertion D the set
$\overline{\mathrm{co}}\left(\bigcup_{n}\mathcal{A}_{n}\right)$ is
compact. It follows that the sequence
$\{\Omega(\mathcal{A}_{n})\}$ has partial limits. Let
$\omega=\lim_{k\rightarrow+\infty}\Omega(\mathcal{A}_{n(k)})$ for
a particular subsequence $n(k)$.

By theorem 1 for each $n$ there exists ensemble $\{\pi_{i}^{n},
\rho_{i}^{n}\}$ of states in $\mathcal{A}_{n}$ with the average
state $\bar{\rho}_{n}$ such that
\begin{equation}\label{three}
\chi(\{\pi_{i}^{n}, \rho_{i}^{n}\})\geq
\bar{C}(\mathcal{A}_{n})-1/n \quad\mathrm{and}\quad
\left\|\bar{\rho}_{n} - \Omega(\mathcal{A}_{n})\right \|_{1}\leq
1/n
\end{equation}

By the proved equality in (\ref{one}) the sequence
$\{\{\pi_{i}^{n}, \rho_{i}^{n}\}\}_{n}$ is approximating for the
set $\bigcup_{n}\mathcal{A}_{n}$ and hence, by theorem 1, the
sequence $\{\bar{\rho}_{n}\}_{n}$ converges to the state
$\Omega(\bigcup_{n}\mathcal{A}_{n})$ as $n\rightarrow+\infty$. By
(\ref{three}) the subsequence $\{\bar{\rho}_{n(k)}\}_{k}$
converges to the state $\omega$. So, we have
$\omega=\Omega(\bigcup_{n}\mathcal{A}_{n})$. Thus each partial
limit of the sequence $\{\Omega(\mathcal{A}_{n})\}$ coincides with
the state $\Omega(\bigcup_{n}\mathcal{A}_{n})$.

To prove G note that C implies existence of the above limit and
the inequality
\begin{equation}\label{one-2}
\lim_{n\rightarrow+\infty}\bar{C}(\mathcal{A}_{n})\geq\bar{C}\left(\bigcap_{n}\mathcal{A}_{n}\right).
\end{equation}

The additional conditions in G provide  different ways of proving
the equality in this inequality.

Consider first the second and the third conditions. Without loss
of generality we may assume that
\begin{equation}\label{four}
\lim_{n\rightarrow+\infty}\Omega(\mathcal{A}_{n})=\omega.
\end{equation}

By theorem 1 for each natural $n$ there exists a measure $\mu_{n}$
(finitely) supported by the set $\mathcal{A}_{n}$ such that
\begin{equation}\label{five}
\chi(\mu_{n})\geq
\bar{C}(\mathcal{A}_{n})-1/n\quad\mathrm{and}\quad
\|\bar{\rho}(\mu_{n})-\Omega(\mathcal{A}_{n})\|_{1}\leq 1/n
\end{equation}

The supports of all measures in the sequence $\{\mu_{n}\}$ lie in
the set $\mathcal{A}_{1}$, which is compact by the assertion D.
Hence this sequence is compact in the weak topology and contains
subsequence $\{\mu_{n(k)}\}$ weakly converging to a particular
measure $\mu_{*}$. Continuity of the mapping
$\mu\mapsto\bar{\rho}(\mu)$, (\ref{four}) and (\ref{five}) imply
$\omega=\bar{\rho}(\mu_{*})=\lim_{k\rightarrow+\infty}\bar{\rho}(\mu_{n_{k}})$
By theorem 6.3 in \cite{Par} $\mathrm{supp}\mu_{*}\subseteq
\bigcap_{n}\mathcal{A}_{n}$.

Suppose the second condition in G is valid. Then there exists
\begin{equation}\label{six}
\lim_{k\rightarrow+\infty}H(\bar{\rho}(\mu_{n_{k}}))=H(\bar{\rho}(\mu_{*}))=H(\omega)<+\infty
\end{equation}
and by using (\ref{formula}) we have
$$
\chi(\mu_{n_{k}})=H(\bar{\rho}(\mu_{n_{k}}))-\hat{H}(\mu_{n_{k}})
$$
for sufficiently large $k$. By using (\ref{six}) and lower
semicontinuity of the functional $\hat{H}(\mu)$ we obtain
$$
\begin{array}{c}
\lim\limits_{n\rightarrow+\infty}\bar{C}(\mathcal{A}_{n})=
\limsup\limits_{k\rightarrow+\infty}\chi(\mu_{n_{k}})=\lim\limits_{k\rightarrow+\infty}H(\bar{\rho}(\mu_{n_{k}}))-
\liminf\limits_{k\rightarrow+\infty}\hat{H}(\mu_{n_{k}})\\\\\leq
H(\bar{\rho}(\mu_{*}))-\hat{H}(\mu_{*})=\chi(\mu_{*})\leq
\bar{C}\left(\bigcap\limits_{n}\mathcal{A}_{n}\right),
\end{array}
$$
which implies equality in (\ref{one-2}).

Suppose the third condition in G is valid.  Since this means
continuity of the function $H(\rho\|\omega)$ on the compact set
$\mathcal{A}_{1}$ the definition of the weak convergence implies
$$
\lim_{k\rightarrow+\infty}\int
H(\rho\|\omega)\mu_{n_{k}}(d\rho)=\int
H(\rho\|\omega)\mu_{*}(d\rho)=\chi(\mu_{*})\leq
\bar{C}\left(\bigcap_{n}\mathcal{A}_{n}\right).
$$

By generalized Donald's identity (\ref{Donald's identity-2}) we
have
$$
\int
H(\rho\|\omega)\mu_{n_{k}}(d\rho)=\chi(\mu_{n_{k}})+H(\bar{\rho}(\mu_{n_{k}})\|\omega)\geq
\chi(\mu_{n_{k}})
$$
and by the above inequality we obtain
$$
\bar{C}\left(\bigcap_{n}\mathcal{A}_{n}\right)\geq\lim_{k\rightarrow+\infty}\int
H(\rho\|\omega)\mu_{n_{k}}(d\rho)\geq\lim_{k\rightarrow+\infty}\chi(\mu_{n_{k}})=
\lim_{n\rightarrow+\infty}\bar{C}(\mathcal{A}_{n}),
$$
which implies equality in (\ref{one-2}).

To complete the consideration of the second and the third
conditions in  G it is sufficient to show that the limit state
$\omega$ in (\ref{four}) is the optimal average state of the set
$\bigcap_{n}\mathcal{A}_{n}$. By theorem 1
$H(\rho\|\Omega(\mathcal{A}_{n}))\leq \bar{C}(\mathcal{A}_{n})$
for arbitrary state $\rho$ in $\bigcap_{n}\mathcal{A}_{n}$ and for
arbitrary $n$. By using (\ref{four}), the proved equality in
(\ref{one-2}) and lower semicontinuity of the relative entropy we
obtain that
$$
H(\rho\|\omega)\leq\liminf_{n\rightarrow+\infty}H(\rho\|\Omega(\mathcal{A}_{n}))\leq
\liminf_{n\rightarrow+\infty}\bar{C}(\mathcal{A}_{n})=\bar{C}\left(\bigcap_{n}\mathcal{A}_{n}\right)
$$
for all such $\rho$. Theorem 1 implies that
$\omega=\Omega\left(\bigcap_{n}\mathcal{A}_{n}\right)$.

Now consider the first condition in G. Note that the assumed
regularity of the set $\mathcal{A}_{1}$ and the condition
$\Omega(\mathcal{A}_{n})=\Omega(\mathcal{A}_{1})$ for all $n$
implies regularity of the sets $\mathcal{A}_{n}$  for all $n$. By
theorem 3 in the next section for each $n$ there exists an optimal
measure $\mu_{n}$ supported by the set $\mathcal{A}_{n}$ such that
(\ref{five}) holds with $0$ instead of $1/n$.  If the first
condition of regularity is valid then relation (\ref{six}) in this
case holds trivially and by repeating arguments in the proof of
the second condition we complete the proof. If the second
condition of regularity is valid then the arguments in the proof
of the third condition are applied immediately.

The assertion H immediately follows from theorem 1.

To prove I consider the nonempty set $\mathfrak{A}$ of all closed
subsets of $\mathcal{A}$ having the same $\chi$-capacity endowed
with the partial order $"\prec"$ defined by
$$
\mathcal{B}\prec\mathcal{C}\quad\Leftrightarrow\quad
\mathcal{B}\supseteq\mathcal{C}.
$$

It is clear that I means existence of a maximal element in
$\mathfrak{A}$. By the Zorn lemma to show this it is sufficient to
show that an arbitrary chain in $\mathfrak{A}$ has maximal
element. The role of this maximal element for a given chain can be
plaid by the intersection of all elements of the chain provided
that this intersection is an element of $\mathfrak{A}$. Since D
implies compactness of the set $\mathcal{A}$ the intersection of
an \textit{arbitrary} decreasing family of subsets of the set
$\mathcal{A}$ coincides with the intersection of its particular
\textit{countable} subfamily. So, it is sufficient to show that
$$
\bar{C}\left(\bigcap_{n}\mathcal{B}_{n}\right)=\bar{C}(\mathcal{A})
$$
for arbitrary monotonuosly decreasing sequence
$\{\mathcal{B}_{n}\}$ of closed subsets of $\mathcal{A}$ such that
$\bar{C}(\mathcal{B}_{n})=\bar{C}(\mathcal{A})$. But this follows
from regularity of the set $\mathcal{A}$ and G with the first
condition since C implies
$\Omega(\mathcal{B}_{n})=\Omega(\mathcal{A})$ for all $n$.

The first part of J follows from proposition 4 below. The second
is a corollary of C and theorem 1 since it implies
$$
H(\rho\|\Omega(\mathcal{A})=\Omega(\mathcal{B}))\leq
\max(\bar{C}(\mathcal{A}),\bar{C}(\mathcal{B}))
$$
for all $\rho$ in $\mathcal{A}\cup\mathcal{B}$.

The first part of K is a direct corollary of the definition of the
$\chi$-capacity and the monotonicity property of the relative
entropy.

To prove the second suppose
$\bar{C}(\Phi(\mathcal{A}))=\bar{C}(\mathcal{A})$. By using
monotonicity of the relative and theorem 1 we obtain
$$
H(\Phi(\rho)\|\Phi(\Omega(\mathcal{A})))\leq
H(\rho\|\Omega(\mathcal{A}))\leq
\bar{C}(\mathcal{A})=\bar{C}(\Phi(\mathcal{A}))
$$
for arbitrary state $\rho$ in $\mathcal{A}$. By theorem 1 this
implies $\Omega(\Phi(\mathcal{A}))=\Phi(\Omega(\mathcal{A}))$.

The assertion L follows the first part of K and lemma 4 below.
$\square$

Theorem 2E implies the following observation.

\textbf{Corollary 6.} \textit{Let $\mathcal{A}$ be a closed convex
set with finite $\chi$-capacity.}

\textit{If $\mathrm{dc}(\rho)<1$ for all $\rho$ in $\mathcal{A}$
then the set $\mathcal{A}$ is regular and the entropy is bounded
on the set $\mathcal{A}$.}

\textit{If $\mathrm{dc}(\rho)=0$ for all $\rho$ in $\mathcal{A}$
then the entropy is continuous on the set $\mathcal{A}$.}

\textbf{Remark 6.} Corollary 6 implies, in particular, that
boundedness of the entropy on a particular closed convex set of
states with zero decrease coefficient (for example, Gaussian
states) implies continuity of the entropy on this set. $\square$

The assertions D and F in theorem 2 provide a sufficient condition
for compactness of unions.

\textbf{Corollary 7.} \textit{If $\{\mathcal{A}_{n}\}$ be a
sequence of sets such that $\mathcal{A}_{n}\subseteq
\mathcal{A}_{n+1}$ and $\bar{C}(\mathcal{A}_{n})\leq M<+\infty$
for all $n$ then the set $\bigcup_{n}\mathcal{A}_{n}$ is
relatively compact.}

Theorem 2K implies the following observation.

\textbf{Corollary 8.} \textit{Let $\mathcal{A}$ be a set with
finite $\chi$-capacity  $\bar{C}(\mathcal{A})$. Then
$\Omega(\mathcal{A})$ is an invariant state for arbitrary channel
$\Phi$ such that
$\Phi(\mathcal{A})\subseteq\overline{\mathrm{co}}(\mathcal{A})$
and $\bar{C}(\Phi(\mathcal{A}))=\bar{C}(\mathcal{A})$. In
particular, $\Omega(\mathcal{A})$ is an invariant state for
arbitrary automorphism $\alpha$ of
$\mathfrak{S}(\mathcal{H})$\footnote{By Wigner's theorem each
automorphism of $\mathfrak{S}(\mathcal{H})$ has the form
$U(\cdot)U^{*}$, where $U$ is either unitary or antiunitary
operator in $\mathcal{H}$.} such that
$\alpha(\mathcal{A})\subseteq\overline{\mathrm{co}}(\mathcal{A})$.}

Let $\mathfrak{F}(\mathcal{A})$ be the set of all channels $\Phi$
from $\mathfrak{S}(\mathcal{H})$ into itself such that
$\Phi(\mathcal{A})\subseteq\overline{\mathrm{co}}(\mathcal{A})$
and $\bar{C}(\Phi(\mathcal{A}))=\bar{C}(\mathcal{A})$. This set is
nonempty and contains all automorphisms $\alpha$ of
$\mathfrak{S}(\mathcal{H})$ such that
$\alpha(\mathcal{A})\subseteq\overline{\mathrm{co}}(\mathcal{A})$.

Corollary 8 implies the following observation (in the spirit of
the Markov-Kakutany theorem): \textit{For arbitrary set
$\mathcal{A}$ with finite $\chi$-capacity the set
$\overline{\mathrm{co}}(\mathcal{A})$ contains at least one common
invariant state for all channels from
$\mathfrak{F}(\mathcal{A})$.}

Theorem 1 and corollary 8 provide the following result.

\textbf{Corollary 9.} \textit{Let $\mathcal{A}$ be an arbitrary
set of states and $\mathfrak{F}_{0}$ be an arbitrary subset of
$\mathfrak{F}(\mathcal{A})$. Let $\mathrm{Inv}\mathfrak{F}_{0}$ be
a set of all common invariant states for all channels from
$\mathfrak{F}_{0}$.}

\textit{The $\chi$- capacity of the set $\mathcal{A}$ can be
defined by the expression
$$
\bar{C}(\mathcal{A})=\inf_{\sigma\in\mathrm{Inv}\mathfrak{F}_{0}\cap\overline{\mathrm{co}}(\mathcal{A})}
\sup_{\rho\in \mathcal{A}}H(\rho\|\sigma),
$$
keeping in mind that $\bar{C}(\mathcal{A})=+\infty$ if
$\,\mathrm{Inv}\mathfrak{F}_{0}\cap\overline{\mathrm{co}}(\mathcal{A})=\emptyset$.}

\textit{In particular, if there exists the unique invariant state
$\sigma_{0}\in\overline{\mathrm{co}}(\mathcal{A})$ for all
channels from $\mathfrak{F}_{0}$ then
$\bar{C}(\mathcal{A})=\sup_{\rho\in
\mathcal{A}}H(\rho\|\sigma_{0})$ and if
$\bar{C}(\mathcal{A})<+\infty$ then
$\Omega(\mathcal{A})=\sigma_{0}$.}

Corollaries 8 an 9 provide a possibility to determine the optimal
average state and to calculate the $\chi$-capacity of a particular
set of states by finding a sufficient family $\mathfrak{F}_{0}$ of
channels from $\mathfrak{F}(\mathcal{A})$. We will use this
possibility in the next section.

Theorem 2D implies the following observation concerning the
$\chi$-capacity of constrained quantum channels
\cite{H-Sh-2},\cite{Sh-2}.

\textbf{Corollary 10.} \textit{Let
$\Phi:\mathfrak{S}(\mathcal{H})\mapsto\mathfrak{S}(\mathcal{H}')$
be an arbitrary quantum channel and $\mathcal{A}$ be a subset of
$\mathfrak{S}(\mathcal{H})$. If
$\bar{C}(\Phi,\mathcal{A})<+\infty$ then $\Phi(\mathcal{A})$ is a
relatively compact subset of $\mathfrak{S}(\mathcal{H}')$.}

\textbf{Proof.} It is easy to see by the definitions that
$$
\bar{C}(\Phi(\mathcal{A})) \leq\bar{C}(\Phi,\mathcal{A}).\square
$$
By this corollary the $\chi$-capacity of an unconstrained quantum
channel can be finite only if the output set of this channel is
relatively compact.

Now we consider the bounds for the $\chi$-capacity of finite union
of sets.

\textbf{Proposition 7.} \textit{If $\{\mathcal{A}_{k}\}_{k=1}^{n}$
is a finite collection of sets then
$$
\max_{\{\lambda_{k}\}}\left(\sum_{k=1}^{n}\lambda_{k}\bar{C}(\mathcal{A}_{k})
+\chi(\{\lambda_{k},\Omega(\mathcal{A}_{k})\})\right)\leq
\bar{C}\left(\bigcup_{k=1}^{n}\mathcal{A}_{k}\right)\leq
\max_{1\leq k\leq n }\bar{C}(\mathcal{A}_{k})+\log n,
$$
where the first maximum is over all probability distributions with
$n$ outcomes.}

\textit{In the case $\bar{C}(\mathcal{A}_{k})=C$ for all
$k=\overline{1,n}$ this implies
$$
C+\bar{C}(\{\Omega(\mathcal{A}_{1}),...,\Omega(\mathcal{A}_{n})\})\leq
\bar{C}\left(\bigcup_{k=1}^{n}\mathcal{A}_{k}\right)\leq C+\log n.
$$}

\textbf{Proof.} By theorem 1 for each natural $m$ and each
$k=\overline{1,n}$ there exists ensemble $\mu^{m}_{k}$ such that

\begin{equation}\label{23}
\chi(\mu^{m}_{k})\geq \bar{C}(\mathcal{A}_{k})-1/m
\quad\mathrm{and}\quad\|\bar{\rho}(\mu^{m}_{k})-\Omega(\mathcal{A}_{k})\|_{1}\leq
1/m.
\end{equation}

Taking arbitrary probability distribution
$\{\lambda_{k}\}_{k=1}^{n}$ consider the ensemble
$\mu_{m}=\sum_{k=1}^{n}\lambda_{k}\mu^{m}_{k}$ of states in
$\bigcup\limits_{k=1}^{n}\mathcal{A}_{k}$. By using lemma 1, lower
semicontinuity of the relative entropy and (\ref{23}) we obtain
$$
\begin{array}{c}
 \bar{C}\left(\bigcup\limits_{k=1}^{n}\mathcal{A}_{k}\right)\geq
\liminf\limits_{m\rightarrow+\infty}\chi(\mu_{m})=\liminf\limits_{m\rightarrow+\infty}
\left(\sum\limits_{k=1}^{n}\lambda_{k}\chi(\mu^{m}_{k})+\chi(\{\lambda_{k},
\bar{\rho}(\mu^{m}_{k})\})\right)\\\\=\sum\limits_{k=1}^{n}\lambda_{k}\bar{C}(\mathcal{A}_{k})
+\liminf\limits_{m\rightarrow+\infty}\chi(\{\lambda_{k},
\bar{\rho}(\mu^{m}_{k})\})\geq\sum\limits_{k=1}^{n}\lambda_{k}\bar{C}(\mathcal{A}_{k})
+\chi(\{\lambda_{k},\Omega(\mathcal{A}_{k})\}),
\end{array}
$$
which implies the lower bound of the proposition.

To prove the upper bound note that arbitrary ensemble $\mu$ of
states in $\bigcup\limits_{k=1}^{n}\mathcal{A}_{k}$ can be
represented as a convex combination
$\sum_{k=1}^{n}\lambda_{k}\mu_{k}$, where $\mu_{k}$ is a ensemble
of states in $\mathcal{A}_{k}$ for $k=\overline{1,n}$ and
$\{\lambda_{k}\}_{k=1}^{n}$ is a probability distribution. By
using lemma 1 and proposition 9b below we obtain
$$
\chi(\mu)=
\sum_{k=1}^{n}\lambda_{k}\chi(\mu_{k})+\chi(\{\lambda_{k},
\bar{\rho}(\mu_{k})\})\leq \max_{1\leq k\leq n
}\bar{C}(\mathcal{A}_{k})+\log n. \square
$$

\textbf{Remark 7.} Proposition 7 shows that the $\chi$-capacity of
a union of sets with given $\chi$-capacities depends on relative
positions of their optimal average states. By theorem 2J if all
the optimal average states coincide with each other then the
$\chi$-capacity of the union is minimal and is equal to the
maximal $\chi$-capacity of the united sets. The greater diversity
of the optimal average states the higher the $\chi$-capacity of
the union. This is obvious in the case of union of two sets for
which the lower bound in proposition 7 and inequality
(\ref{rel-entr-ineq}) imply
$$
\bar{C}(\mathcal{A}\cup\mathcal{B})\geq\max_{\lambda\in[0,1]}\left(\lambda\bar{C}(\mathcal{A})+
(1-\lambda)\bar{C}(\mathcal{B})+
\textstyle\frac{1}{2}\lambda(1-\lambda)\|\Omega(\mathcal{A})-\Omega(\mathcal{B})\|_{1}^{2}\right).
$$
Note also that the lower and the upper bounds in proposition 7
coincides if and only if
$$
\bar{C}(\mathcal{A}_{i})=\bar{C}(\mathcal{A}_{j})\quad\mathrm{and}\quad
\bigcup_{\rho\in\mathcal{A}_{i}}\mathrm{supp}\rho\; \perp \;
\bigcup_{\rho\in\mathcal{A}_{j}}\mathrm{supp}\rho\quad
\mathrm{for}\; \mathrm{all}\quad i\neq j.
$$

To complete the proof of theorem 2 we obtain the following result,
which will be also used in section 6.

\textbf{Lemma 4.} \textit{Let
$\{\Psi_{\lambda}\}_{\lambda\in\Lambda}$ be a family of continuous
mappings from $\mathfrak{S}(\mathcal{H})$ into itself indexed by
some ordered set $\Lambda$ and such that
$\lim_{\lambda}\Psi_{\lambda}(\rho)=\rho$ for all states $\rho$ in
a particular subset $\mathcal{A}$ of $\mathfrak{S}(\mathcal{H})$.
Then
$$
\liminf_{\lambda}\bar{C}(\Psi_{\lambda}(\mathcal{A}))\geq\bar{C}(\mathcal{A}).
$$
If there exists
$\displaystyle\lim_{\lambda}\bar{C}(\Psi_{\lambda}(\mathcal{A}))=\bar{C}(\mathcal{A})$
then there exists
$\displaystyle\lim_{\lambda}\Omega(\Psi_{\lambda}(\mathcal{A}))=\Omega(\mathcal{A})$.}

\textbf{Proof.} The first assertion of the lemma easily follows
from lower semicontinuity of the relative entropy. Indeed, for
arbitrary $\varepsilon>0$ there exists an ensemble $\{\pi_{i},
\rho_{i}\}$ such that
$$
\chi(\{\pi_{i}, \rho_{i}\})\geq C(\varepsilon)=\left\{
   \begin{array}{lr}
    \bar{C}(\mathcal{A})-\varepsilon, &
    \bar{C}(\mathcal{A})<+\infty\\
    \varepsilon, &
    \bar{C}(\mathcal{A})=+\infty
    \end{array}\right.
$$
By the assumption and due to lower semicontinuity of the relative
entropy we obtain
$$
\liminf_{\lambda}\bar{C}(\Psi_{\lambda}(\mathcal{A}))\geq
\liminf_{\lambda}\chi(\{\pi_{i}, \Psi_{\lambda}(\rho_{i})\})\geq
\chi(\{\pi_{i}, \rho_{i}\})\geq C(\varepsilon).
$$
Since $\varepsilon$ can be arbitrary this implies the first
assertion on the lemma.

Let
$\lim_{\lambda}\bar{C}(\Psi_{\lambda}(\mathcal{A}))=\bar{C}(\mathcal{A})<+\infty$.
By theorem 1 for arbitrary $\varepsilon>0$ there exists an
ensemble $\{\pi_{i}, \rho_{i}\}$ such that
\begin{equation}\label{1}
\chi(\{\pi_{i}, \rho_{i}\})\geq \bar{C}(\mathcal{A})-\varepsilon
\quad\mathrm{and}\quad
\|\textstyle\sum_{i}\pi_{i}\rho_{i}-\Omega(\mathcal{A})\|_{1}<\varepsilon
\end{equation}

Applying the arguments from the first part of the proof we obtain
that there exists $\lambda^{1}_{\varepsilon}$ such that
$$
\chi(\{\pi_{i}, \Psi_{\lambda}(\rho_{i})\})\geq \chi(\{\pi_{i},
\rho_{i}\})-\varepsilon,\quad\forall\lambda\geq\lambda^{1}_{\varepsilon}.
$$

By the assumption there exists $\lambda^{2}_{\varepsilon}$ such
that
$$
\bar{C}(\Psi_{\lambda}(\mathcal{A}))\leq\bar{C}(\mathcal{A})+\varepsilon,
\quad\forall\lambda\geq\lambda^{2}_{\varepsilon}.
$$

Thus for all $\lambda\geq
\max(\lambda^{1}_{\varepsilon},\lambda^{2}_{\varepsilon})$ we have
$$
0\leq \bar{C}(\Psi_{\lambda}(\mathcal{A}))-\chi(\{\pi_{i},
\Psi_{\lambda}(\rho_{i})\})\leq
\bar{C}(\mathcal{A})-\chi(\{\pi_{i}, \rho_{i}\})+2\varepsilon\leq
3\varepsilon
$$
and by using corollary 4 we obtain
\begin{equation}\label{2}
\begin{array}{c}
\frac{1}{2}\|\sum_{i}\pi_{i}\Psi_{\lambda}(\rho_{i})-\Omega(\Psi_{\lambda}(\mathcal{A}))\|_{1}^{2}\leq
H(\sum_{i}\pi_{i}\Psi_{\lambda}(\rho_{i})\|\Omega(\Psi_{\lambda}(\mathcal{A})))\\\\\leq
\bar{C}(\Psi_{\lambda}(\mathcal{A}))-\chi(\{\pi_{i},
\Psi_{\lambda}(\rho_{i})\})\leq 3\varepsilon
\end{array}
\end{equation}

The continuity property of the family $\{\Psi_{\lambda}\}$ implies
existence of $\lambda^{3}_{\varepsilon}$ such that
\begin{equation}\label{3}
\|\textstyle\sum_{i}\pi_{i}\Psi_{\lambda}(\rho_{i})
-\textstyle\sum_{i}\pi_{i}\rho_{i}\|_{1}\leq\varepsilon,
\quad\forall\lambda\geq\lambda^{3}_{\varepsilon}.
\end{equation}

By using (\ref{1}),(\ref{2}) and (\ref{3}) we obtain
$$
\begin{array}{c}
\|\Omega(\Psi_{\lambda}(\mathcal{A}))-\Omega(\mathcal{A})\|_{1}\leq
\|\Omega(\Psi_{\lambda}(\mathcal{A}))-\sum_{i}\pi_{i}\Psi_{\lambda}(\rho_{i})\|_{1}\\\\+
\|\sum_{i}\pi_{i}\Psi_{\lambda}(\rho_{i})
-\sum_{i}\pi_{i}\rho_{i}\|_{1}+\|\sum_{i}\pi_{i}\rho_{i}-\Omega(\mathcal{A})\|_{1}\leq
2\varepsilon+\sqrt{6\varepsilon}
\end{array}
$$
for all $\lambda\geq
\max(\lambda^{1}_{\varepsilon},\lambda^{2}_{\varepsilon},
\lambda^{3}_{\varepsilon})$. Since $\varepsilon$ is arbitrary this
implies the second statement of the lemma. $\square$

\subsection{The optimal measure}

Let $\mathcal{A}$ be a closed set with finite $\chi$-capacity. By
theorem 2D the set $\mathcal{A}$ is compact. Hence the set
$\mathcal{M}(\mathcal{A})$ of all probability measures supported
by the set $\mathcal{A}$ is compact in the topology of weak
convergence (Prokhorov's topology). Since an arbitrary measure in
$\mathcal{M}(\mathcal{A})$ can be weakly approximated by a
sequence of measures with finite support, lower semicontinuity of
the functional $\chi(\mu)$ implies
\begin{equation}\label{ccap-2}
\bar{C}(\mathcal{A})=\sup_{\mu\in\mathcal{M}(\mathcal{A})}\chi(\mu),
\end{equation}
which means that the supremum over all measures coincides with the
supremum over all measures with finite support.

\textbf{Definition 3.} \textit{A measure $\mu_{*}$ supported by
the set $\mathcal{A}$ and such that
$$
\bar{C}(\mathcal{A})=\chi(\mu_{*})=\int\limits_{\mathcal{A}}H(\rho\Vert\bar{\rho}(\mu_{*}))\mu_{*}(d\rho)
$$
is called the optimal measure for the set $\mathcal{A}$.}

By using the arguments from the proof of proposition 1 in
\cite{H-Sh-2} it is easy to see that the functional
$\mu\mapsto\int H(\rho\Vert\Omega(\mathcal{A}))\mu(d\rho)$ is
lower semicontinuous on $\mathcal{M}(\mathcal{A})$. This, the
above mentioned weak density of measures with finite support in
$\mathcal{M}(\mathcal{A})$ and generalized Donald's identity
(\ref{Donald's identity-2}) imply the following generalization of
theorem 1 and corollary 4: \textit{For arbitrary closed set
$\mathcal{A}$ with finite $\chi$-capacity and arbitrary measure
$\mu$ from $\mathcal{M}(\mathcal{A})$ the following inequalities
hold
$$
\begin{array}{c}
\int\limits_{\mathcal{A}}H(\rho\Vert\Omega(\mathcal{A}))\mu(d\rho)\leq
\bar{C}(\mathcal{A}),\\\\ \bar{C}(\mathcal{A})-\chi(\mu)\geq
H(\bar{\rho}(\mu)\Vert\Omega(\mathcal{A}))\geq\textstyle\frac{1}{2}\|\bar{\rho}(\mu)-\Omega(\mathcal{A})\|_{1}^{2}.
\end{array}
$$}

This provides the following generalization of the "maximal
distance property" \cite{Sch-West-1} of an optimal ensemble to the
infinite dimensional case.

\textbf{Proposition 8.} \textit{Let $\mu_{*}$ be an optimal
measure for the closed set $\mathcal{A}$ with finite
$\chi$-capacity. Then its barycenter $\bar{\rho}(\mu_{*})$
coincides with the optimal average state $\Omega(\mathcal{A})$ and
$H(\rho\|\Omega(\mathcal{A}))=\bar{C}(\mathcal{A})$ for
$\mu_{*}$-almost all $\rho$.}

In particular, if there exists finite or countable ensemble
$\{\pi_{i}, \rho_{i}\}$ on which the supremum in definition
(\ref{ccap-1}) of the $\chi$-capacity is achieved - an optimal
ensemble for the set $\mathcal{A}$ - then its the average state
$\bar{\rho}$ coincides with the optimal average state
$\Omega(\mathcal{A})$ and
$H(\rho_{i}\|\Omega(\mathcal{A}))=\bar{C}(\mathcal{A})$ for all
$i$ such that $\pi_{i}>0$.

\textbf{Corollary 11.} \textit{Let $\mathcal{A}$ be a closed set
with finite $\chi$-capacity. Existence of an optimal measure for
the set $\mathcal{A}$ implies $\bar{C}(\mathcal{A})\leq
H(\Omega(\mathcal{A}))$.}

\textbf{Proof.} It is sufficient to consider the case
$H(\Omega(\mathcal{A}))<+\infty$ for which (\ref{formula}), the
definition of an optimal measure $\mu_{*}$ and proposition 8 imply
$$
\bar{C}(\mathcal{A})=\chi(\mu_{*})=H(\bar{\rho}(\mu_{*}))-\hat{H}(\mu_{*})\leq
H(\bar{\rho}(\mu_{*}))=H(\Omega(\mathcal{A})).\square
$$

This corollary provides the simple way to show nonexistence of an
optimal measure for a particular set of states, which will be used
in the proof of proposition 1b and 3b in section 5 below.

The following theorem provides the sufficient condition for
existence of an optimal measure.

\textbf{Theorem 3.} \textit{Let $\mathcal{A}$ be a convex closed
set with finite $\chi$-capacity. If $\mathrm{Ext}(\mathcal{A})$ is
a regular set then there exists an optimal measure for the set
$\mathcal{A}$ supported by the set
$\overline{\mathrm{Ext}(\mathcal{A})}$.}

The main ingredient of the proof of this theorem is the following
lemma.

\textbf{Lemma 5.} \textit{Let $\mathcal{A}$ be a convex closed set
with finite the $\chi$-capacity. There exists a sequence of
measures $\{\mu_{n}\}$ supported by the set
$\overline{\mathrm{Ext}(\mathcal{A})}$ weakly converging to some
measure ${\mu_{*}}$ supported by the set
$\overline{\mathrm{Ext}(\mathcal{A})}$ with the barycenter
$\Omega(\mathcal{A})$ such that
$$
\lim_{n\rightarrow+\infty}H(\overline{\rho}(\mu_{n})\|\Omega(\mathcal{A}))=0\quad
and
\quad\lim_{n\rightarrow+\infty}\chi(\mu_{n})=\bar{C}(\mathcal{A}).
$$}

\textbf{Proof.} Let $\{\{\pi^{n}_{i}, \rho^{n}_{i}\}\}_{n}$ be an
approximating sequence of ensembles for the set $\mathcal{A}$ with
the corresponding sequence of the average states
$\{\bar{\rho}_{n}\}$. Theorem 1 implies
$\lim_{n\rightarrow+\infty}H(\bar{\rho}_{n}\|\Omega(\mathcal{A}))=0$.
Since by theorem 2D the set $\mathcal{A}$ is compact the theory of
barycentric decomposition \cite{Alf},\cite{B&R} implies existence
for each $n$ and $i$ of a measure $\mu_{i}^{n}$ supported by
$\overline{\mathrm{Ext}(\mathcal{A})}$ such that
$\bar{\rho}(\mu_{i}^{n})=\rho^{n}_{i}$. Convexity of the relative
entropy and Jensen's inequality\footnote{Application of Jensen's
inequality in this case is valid since the relative entropy can be
represented as a pointwise limit of a monotonously increasing
sequence of continuous convex functions \cite{L}.} imply
$$
H(\rho^{n}_{i}\|\bar{\rho}_{n})=H\left(\int \rho
\mu^{n}_{i}(d\rho)\|\bar{\rho}_{n}\right)\leq \int
H(\rho\|\bar{\rho}_{n})\mu^{n}_{i}(d\rho).
$$
By using this and (\ref{ccap-2}) we obtain
$$
\sum_{i}\pi^{n}_{i}H(\rho^{n}_{i}\|\bar{\rho}_{n})\leq
\sum_{i}\pi^{n}_{i}\int
H(\rho\|\bar{\rho}_{n})\mu^{n}_{i}(d\rho)=\chi\left(\sum_{i}\pi^{n}_{i}\mu^{n}_{i}\right)\leq
\bar{C}(\mathcal{A}).
$$
Let $\mu_{n}=\sum_{i}\pi^{n}_{i}\mu^{n}_{i}$ be a measure with the
barycenter $\bar{\rho}_{n}$ for each $n$. It follows from the
approximating property of the sequence $\{\{\pi^{n}_{i},
\rho^{n}_{i}\}\}_{n}$ and from the above inequality that
$\lim_{n\rightarrow+\infty}\chi(\mu_{n})=\bar{C}(\mathcal{A})$.
Compactness of the set $\mathcal{A}$ implies compactness of the
set $\mathcal{M}(\mathcal{A})$ in the weak topology and hence
existence of a subsequence of the sequence $\{\mu_{n}\}$
converging to a particular measure $\mu_{*}$, supported by
$\overline{\mathrm{Ext}(\mathcal{A})}$ due to theorem 6.1 in
\cite{Par}. Continuity of the mapping $\mu\mapsto \bar{\rho}(\mu)$
and theorem 1 imply $\bar{\rho}(\mu_{*})=\Omega(\mathcal{A})$.
Thus this subsequence has the all properties stated in the lemma.
$\square$

\textbf{Proof of theorem 3.} The two regularity conditions provide
two different ways to show that the limit measure $\mu_{*}$
involved in the above lemma is an optimal measure for the set
$\mathcal{A}$.

Let $\{\mu_{n}\}$ be a sequence provided by lemma 5.

By the first regularity condition
$$
\lim_{n\rightarrow+\infty}H(\bar{\rho}(\mu_{n}))=H(\bar{\rho}(\mu_{*}))=H(\Omega(\mathcal{A}))<+\infty.
$$
Hence expression (\ref{formula}) and lower semicontinuity of the
functional $\hat{H}(\mu)$ imply
$$
\limsup_{n\rightarrow+\infty}\chi(\mu_{n})=
\limsup_{n\rightarrow+\infty}(H(\bar{\rho}(\mu_{n}))-\hat{H}(\mu_{n}))\leq
H(\bar{\rho}(\mu_{*}))-\hat{H}(\mu_{*})=\chi(\mu_{*}).
$$
Since
$\lim_{n\rightarrow+\infty}\chi(\mu_{n})=\bar{C}(\mathcal{A})$ and
$\chi(\mu_{*})\leq\bar{C}(\mathcal{A})$ this inequality  implies
$\chi(\mu_{*})=\bar{C}(\mathcal{A})$, which means optimality of
the measure $\mu_{*}$.

The second regularity condition, compactness of the set
$\mathcal{A}$ and the definition of the weak convergence imply
$$
\chi(\mu_{*})=\int H(\rho\|\Omega(\mathcal{A}))\mu_{*}(d\rho)=
\lim_{n\rightarrow+\infty}\int
H(\rho\|\Omega(\mathcal{A}))\mu_{n}(d\rho).
$$
By generalized Donald's identity (\ref{Donald's identity-2}) and
nonegativity of the relative entropy we have
$$
\begin{array}{c}
\int
H(\rho\|\Omega(\mathcal{A}))\mu_{n}(d\rho)=\chi(\mu_{n})+H(\bar{\rho}(\mu_{n})\|\Omega(\mathcal{A}))\geq
\chi(\mu_{n}).
\end{array}
$$
Since
$\lim_{n\rightarrow+\infty}\chi(\mu_{n})=\bar{C}(\mathcal{A})$ the
above expressions imply $\chi(\mu_{*})=\bar{C}(\mathcal{A})$,
which means optimality of the measure $\mu_{*}$. $\square $

\textbf{Remark 8.} The regularity condition in theorem 3 is
essential but is not necessary. There exist nonregular sets with
finite $\chi$-capacity having no optimal measure (see propositions
1b and 3b in subsections 5.2 and 5.3 correspondingly). It is
surprising that there exist converging sequences of states with
finite $\chi$-capacity having no optimal measure (see the example
in subsection 5.1). There also exists nonregular sets having an
optimal measure (see the note before lemma 6 in subsections 5.2).

\section{Examples}

The general results of the previous section are illustrated in
this section by considering several examples of sets of states.

\subsection{Finite set of states and converging sequences}

By theorem 2D each set of states with finite $\chi$-capacity is
relatively compact. In this subsection we consider the following
simplest examples of relatively compact sets:
\begin{itemize}
  \item a finite collection of states $\{\rho_{n}\}_{n=1}^{N}$;
  \item a sequence of states $\{\rho_{n}\}_{n=1}^{+\infty}$ converging to a particular state $\rho_{*}$;
  \item a sequence of states $\{\rho_{n}\}_{n=1}^{+\infty}$ $H$-converging to a particular state
  $\rho_{*}$.\footnote{This means that
$\lim_{n\rightarrow+\infty}H(\rho_{n}\|\rho_{*})=0$.}
\end{itemize}

The properties of the restriction of the entropy to the convex
closure of the above sets are considered in the following
proposition.

\textbf{Proposition 9a.} A) \textit{Let $\{\rho_{n}\}_{n=1}^{N}$
be a finite collection of states in $\mathfrak{S}(\mathcal{H})$.}

\textit{The entropy is continuous on the (closed) set
$\mathrm{co}\left(\{\rho_{n}\}_{n=1}^{N}\right)$ if and only if
$$
H(\rho_{n})<+\infty\quad for \quad all \quad n=1,2,...N.
$$}
B) \textit{Let $\{\rho_{n}\}_{n=1}^{+\infty}$ be a sequence of
states converging to a state $\rho_{*}$. }

\textit{The entropy is bounded on the set
$\overline{\mathrm{co}}\left(\{\rho_{n}\}_{n=1}^{+\infty}\right)$
if and only if there exists $\mathfrak{H}$-operator $H$ with
$\mathrm{ic}(H)<+\infty$ such that
$$
\sup_{n}\mathrm{Tr}\rho_{n}H<+\infty.
$$}
\textit{The entropy is continuous on the set
$\overline{\mathrm{co}}\left(\{\rho_{n}\}_{n=1}^{+\infty}\right)$
if one of the following equivalent conditions holds: }
\begin{itemize}
  \item \textit{$H(\rho_{n})<+\infty$ for all $n$, $\lim\limits_{n\rightarrow+\infty}H(\rho_{n})=H(\rho_{*})<+\infty$
  and there exists a state $\sigma$ such that
}  $$
  \lim\limits_{n\rightarrow+\infty}H(\rho_{n}\|\sigma)=H(\rho_{*}\|\sigma)<+\infty;
  $$\vspace{-20pt}
\item \textit{there exists a $\mathfrak{H}$-operator $H$ with
$\mathrm{ic}(H)=0$ such that }
$$
\sup\limits_{n}\mathrm{Tr}\rho_{n}H<+\infty;
$$\vspace{-20pt}
\item \textit{there exists $\mathfrak{H}$-operator $H$ with
$\mathrm{ic}(H)<+\infty$ such that }
$$
\mathrm{Tr}\rho_{n}H<+\infty\;\, for\; all\;\, n\quad
and\quad\lim\limits_{n\rightarrow+\infty}\mathrm{Tr}\rho_{n}H=\mathrm{Tr}\rho_{*}H<+\infty.
$$
\end{itemize}\vspace{-10pt}

C) \textit{Let $\{\rho_{n}\}_{n=1}^{+\infty}$ be a sequence of
states $H$-converging to a  state $\rho_{*}$. }

\textit{The entropy is bounded on the set
$\overline{\mathrm{co}}\left(\{\rho_{n}\}_{n=1}^{+\infty}\right)$
if and only if
$$
\sup_{n}H(\rho_{n})<+\infty.
$$}\vspace{-10pt}

\textit{The entropy is continuous on the set
$\overline{\mathrm{co}}\left(\{\rho_{n}\}_{n=1}^{+\infty}\right)$
if and only if
$$
H(\rho_{n})<+\infty\;\, for\; all\;\, n\quad
and\quad\lim\limits_{n\rightarrow+\infty}H(\rho_{n})=H(\rho_{*})<+\infty.
$$}

\textbf{Remark 9.} It is interesting to compare the boundedness and
the continuity conditions for converging and for $H$-converging
sequences. The conditions for $H$-converging sequence look like
natural generalizations of the corresponding conditions for finite
set of states while the conditions for converging sequence include
some additional requirements. These requirements are essential -
there exists a converging sequence $\{\rho_{n}\}_{n=1}^{+\infty}$ of
states for which $H(\rho_{n})$ is finite for all $n$ and
$$
\lim\limits_{n\rightarrow+\infty}H(\rho_{n})=H\left(\lim\limits_{n\rightarrow+\infty}\rho_{n}\right)<+\infty.
$$
but the entropy is unbounded on the set
$\overline{\mathrm{co}}\left(\{\rho_{n}\}_{n=1}^{+\infty}\right)$
(see the example below). $\square$

\textbf{Proof.} A) Let $\mathcal{A}=\{\rho_{i}\}_{i=1}^{N}$.
Necessity of the continuity condition is obvious. To show its
sufficiency note that this condition and general properties of
quantum entropy \cite{W} implies its boundedness on the closed set
$\mathrm{co}(\mathcal{A})$ and hence finiteness of the
$\chi$-capacity of this set. By theorem 1 there exists the unique
state $\Omega(\mathcal{A})$ such that
$$
H(\rho_{n}\|\Omega(\mathcal{A}))=
\mathrm{Tr}\rho_{n}(-\log\Omega(\mathcal{A}))-
H(\rho_{n})\leq\bar{C}(\mathcal{A})<+\infty
$$
and hence $\mathrm{Tr}\rho_{n}(-\log\Omega(\mathcal{A}))\leq
\bar{C}(\mathcal{A})+ \max_{n}H(\rho_{n})<+\infty$ for all
$n=\overline{1,N}$. Thus the linear functional
$\mathrm{Tr}\rho(-\log\Omega(\mathcal{A}))$ is finite and hence
continuous on the \textit{finite} set $\mathcal{A}$. By
proposition 4 this means continuity of the entropy on the set
$\overline{\mathrm{co}}(\mathcal{A})$.

B) The boundedness condition for this case follows from
proposition 1a while the continuity condition - from proposition
4.

C) Let $\mathcal{A}=\{\rho_{i}\}_{i=1}^{+\infty}$. Necessity of
the boundedness and of the continuity conditions for this case is
obvious. To show sufficiency of the boundedness condition note
that the $\chi$-capacity of the set $\mathcal{A}$ is finite (see
proposition 9b below). By theorem 1 there exists the unique state
$\Omega(\mathcal{A})$ such that
$$
H(\rho_{n}\|\Omega(\mathcal{A})=
\mathrm{Tr}\rho_{n}(-\log\Omega(\mathcal{A}))-
H(\rho_{n})\leq\bar{C}(\mathcal{A})<+\infty\
$$
for all $n$ and hence
$\sup_{n}\mathrm{Tr}\rho_{n}(-\log\Omega(\mathcal{A}))\leq
\bar{C}(\mathcal{A})+ \sup_{n}H(\rho_{n})<+\infty$. By proposition
1a this implies boundedness of the entropy on the set
$\overline{\mathrm{co}}\mathcal{A}$. Sufficiency of the continuity
condition follows from the first continuity condition for the case
B with $\sigma=\rho_{*}$. $\square$

The questions concerning the $\chi$-capacity of finite sets of
states and of converging sequences are considered in the following
proposition.

\textbf{Proposition 9b.} A) \textit{Let $\{\rho_{n}\}_{n=1}^{N}$
be a finite collection of states in $\mathfrak{S}(\mathcal{H})$.}

\textit{The set $\{\rho_{n}\}_{n=1}^{N}$ is regular and
$$
\bar{C}(\{\rho_{n}\}_{n=1}^{N})\leq\log N
$$}\vspace{-10pt}

\textit{There exists optimal ensemble
$\mu_{*}=\{\pi_{n},\rho_{n}\}_{n=1}^{N}$ for the set
$\{\rho_{n}\}_{n=1}^{N}$.}

B) \textit{Let $\{\rho_{n}\}_{n=1}^{+\infty}$ be a sequence of
states converging to a state $\rho_{*}$. }

\textit{The $\chi$-capacity of the set
$\{\rho_{n}\}_{n=1}^{+\infty}$ is finite if and only if there
exists a state $\sigma$ such that}\footnote{The below example
shows that the $\chi$-capacity of a converging sequence can be
infinite}
$$
\sup_{n}H(\rho_{n}\|\sigma)<+\infty.
$$
C) \textit{Let $\{\rho_{n}\}_{n=1}^{+\infty}$ be a sequence of
states $H$-converging to a state $\rho_{*}$. }

\textit{The $\chi$-capacity of the set
$\{\rho_{n}\}_{n=1}^{+\infty}$ is finite and}
$$
\bar{C}\left(\{\rho_{n}\}_{n=1}^{+\infty}\right)\leq
\inf_{m}\max\left(\sup_{n>m}H(\rho_{n}\|\rho_{*});\log
m\right)+\log 2.
$$

\textit{In the cases A,B,C existence of an optimal measure
$\mu_{*}=\{\pi_{n},\rho_{n}\}$ for the set $\{\rho_{n}\}$ is
equivalent to existence of a probability distribution
$\{\pi_{n}\}$ and of a positive number $C$ satisfying to the
following system}
\begin{equation}\label{opt-ens-sys}
   \left\{
   \begin{array}{lr}
    H(\rho_{n}\|\sum_{k}\pi_{k}\rho_{k})=C,& \pi_{n}>0\\
    H(\rho_{n}\|\sum_{k}\pi_{k}\rho_{k})\leq C,& \pi_{n}=0
   \end{array}\right.
\end{equation}

\textit{If this system has a solution then
$\bar{C}\left(\{\rho_{n}\}\right)=C$ and
$\Omega\left(\{\rho_{n}\}\right)=\sum_{n}\pi_{n}\rho_{n}$.}

\textbf{Proof.} A) To prove the upper bound for the
$\chi$-capacity of the set $\mathcal{A}=\{\rho_{n}\}_{n=1}^{N}$ it
is sufficient to note that $\mathrm{co}(\mathcal{A})$ is an output
set for the channel $\sigma\mapsto\sum_{n=1}^{N}\langle
n|\sigma|n\rangle\rho_{n}$ from $N$-dimensional Hilbert space with
orthonormal basis $\{|n\rangle\}_{n=1}^{N}$ and to use the
monotonicity property of the relative entropy. Finiteness of the
$\chi$-capacity and theorem 1 imply finiteness of
$H(\rho_{n}\|\Omega(\mathcal{A}))$ for all $n$ and hence
regularity of the set $\mathcal{A}$. Existence of optimal
measure=optimal ensemble follows from theorem 3.

B) This directly follows from theorem 1.

C) To prove the upper bound for the $\chi$-capacity of the set
$\mathcal{A}=\{\rho_{n}\}_{n=1}^{+\infty}$ consider this set as
the  union of the finite set
$\mathcal{A}_{1}=\{\rho_{n}\}_{n=1}^{m}$ and the "tail"
$\mathcal{A}_{2}=\{\rho_{n}\}_{n=m+1}^{+\infty}$. Proposition 7,
theorem 1 and the part A of this proposition imply
$$
\begin{array}{c}
\bar{C}(\mathcal{A})=\bar{C}(\mathcal{A}_{1}\cup\mathcal{A}_{2})\leq
\max(\bar{C}(\mathcal{A}_{1}),\bar{C}(\mathcal{A}_{2}))+\log 2\\\\
\leq \max\left(\sup_{n>m}H(\rho_{n}\|\rho_{*}),\;\log m\right)
+\log 2.
\end{array}
$$\vspace{5pt}

If $\{\pi_{n}\}$ is an optimal probability distribution then  by
proposition 8 it satisfies system (\ref{opt-ens-sys}) with
$C=\bar{C}(\{\rho_{n}\})$. Conversely, if $(\{\pi_{n}\}, C)$  is a
solution of this system then by using the second part of theorem 1
it is easy to see that the ensemble $\{\pi_{n},\rho_{n}\}$ is
optimal for the set $\{\rho_{n}\}$ and
$C=\bar{C}\left(\{\rho_{n}\}\right)$. $\square$

Consider the case of finite set of states.

If $N=2$ we have
$\Omega(\{\rho_{1},\rho_{2}\})=\pi\rho_{1}+(1-\pi)\rho_{2}$, where
$\pi$ is uniquely defined by the equation
$$
H(\rho_{1}\|\pi\rho_{1}+(1-\pi)\rho_{2})=H(\rho_{2}\|\pi\rho_{1}+(1-\pi)\rho_{2})
$$
and both sides of this equality are equal to
$\bar{C}(\{\rho_{1},\rho_{2}\})$. In the case $N>2$ the situation
is more difficult in general. It may happen that there exists
proper subset $\{\rho_{n_{1}},...\rho_{n_{N'}}\}$, $N'<N$, of the
set $\{\rho_{1},...\rho_{N}\}$ such that
$\bar{C}(\{\rho_{n_{1}},...\rho_{n_{N'}}\})=\bar{C}(\{\rho_{1},...\rho_{N}\})$.
This means that some "weights" in the above optimal probability
distribution $\{\pi_{n}\}$ are equal to zero. Indeed, this
situation takes place if we add to the set $\{\rho_{1},\rho_{2}\}$
arbitrary state $\rho_{3}$ such that
$H(\rho_{3}\|\Omega(\{\rho_{1},\rho_{2}\}))\leq\bar{C}(\{\rho_{1},\rho_{2}\})$.
By using theorem 1 it is easy to see that
$\Omega(\{\rho_{1},\rho_{2}\})=\Omega(\{\rho_{1},\rho_{2},\rho_{3}\})$
and
$\bar{C}(\{\rho_{1},\rho_{2}\})=\bar{C}(\{\rho_{1},\rho_{2},\rho_{3}\})$
in this case. This provides the simplest example showing that
$\mathcal{A}\varsubsetneq\mathcal{B}$ does not imply
$\bar{C}(\mathcal{A})<\bar{C}(\mathcal{B})$ in general.

There are two cases in which the optimal average state can be
easily determined as the uniform average:
$\Omega(\{\rho_{n}\}_{n=1}^{N})=N^{-1}\sum_{i=1}^{N}\rho_{n}$. The
first one is the case when the states $\rho_{1},...\rho_{N}$ form
an orbit of some group of automorphisms of
$\mathfrak{S}(\mathcal{H})$ (see subsection 5.5). The second one
is the case when the supports of the states $\rho_{1},...\rho_{N}$
are orthogonal to each other. It is this case in which the
$\chi$-capacity achieves its maximal value $\log N$ independently
of types of the states $\rho_{1},...\rho_{N}$ and of values of
their entropies. Indeed, this follows from the equality
$$
H\left(\rho_{n}\|N^{-1}\sum_{k=1}^{N}\rho_{k}\right)=H\left(\rho_{n}\|N^{-1}\rho_{n}\right)+1-N^{-1}=\log
N,\quad n=\overline{1,N},
$$
obtained by using properties of relative entropy
\cite{O&P},\cite{W}.

The case of converging sequence is illustrated by the following
example, which shows in particular that system (\ref{opt-ens-sys})
defining the optimal probability distribution and the value of the
$\chi$-capacity can be solved directly in some nontrivial cases.

\textbf{Example of a converging sequence of states.} Let
$\{|n\rangle\}$ be an orthonormal basis in $\mathcal{H}$ and let
$\{q_{n}\}$ be a sequence of numbers in $[0;1]$ converging to
zero. For given $\varepsilon\in [0;1]$ consider the set
$\mathcal{S}^{\varepsilon}_{\{q_{n}\}}=\{\rho^{\pm}_{n}\}$ of
states
$$
\rho^{\pm}_{n}=(1-q_{n})|1\rangle\langle 1|+q_{n}|n\rangle\langle
n|\pm\eta_{n}(q_{n},\varepsilon)\sqrt{(1-q_{n})q_{n}}(|1\rangle\langle
n|+|n\rangle\langle 1|),\;n\geq 2,
$$
where the parameter $\eta_{n}(q_{n},\varepsilon)\in [0;1]$ is
defined by the condition
$$
H(\rho^{\pm}_{n})=(1-\varepsilon)
h_{2}(q_{n})=-(1-\varepsilon)((1-q_{n})\log(1-q_{n})+q_{n}\log
q_{n}).
$$
Thus $\varepsilon$ can be considered as a purity parameter. If
$\varepsilon=0$ then $\eta_{n}(q_{n},\varepsilon)=0$ and the all
states $\rho^{+}_{n}=\rho^{-}_{n}$ are diagonizable in the basis
$\{|n\rangle\}$ and have maximal entropy, if $\varepsilon=1$ then
$\eta_{n}(q_{n},\varepsilon)=1$ and the all states $\rho^{\pm}_{n}$
are pure.

The set $\mathcal{S}^{\varepsilon}_{\{q_{n}\}}$ can be considered
as a sequence converging to the state $\rho_{1}=|1\rangle\langle
1|$. We will establish that:

\textit{The $\chi$-capacity of the set
$\mathcal{S}^{\varepsilon}_{\{q_{n}\}}$ is finite if and only if
there exists positive $\lambda$ such that}
  \begin{equation}\label{fsc}
  \sum_{n}\exp\left(-\frac{\lambda}{q_{n}}\right)<+\infty
  \end{equation}

\textit{If condition (\ref{fsc}) holds then the necessary and
sufficient condition of existence of optimal measure = optimal
ensemble $\mu_{*}=\{\pi^{\pm}_{n}, \rho^{\pm}_{n}\}$ for the set
$\mathcal{S}^{\varepsilon}_{\{q_{n}\}}$ is given by the inequality
\begin{equation}\label{fsc+}
\sum_{n>1}q_{n}^{-\varepsilon}(1-q_{n})^{1+\frac{(1-q_{n})(1-\varepsilon)}{q_{n}}}
\exp\left(-\frac{\lambda^{*}_{\{q_{n}\}}}{q_{n}}\right)\geq 1,
\end{equation}
where
$$
\lambda^{*}_{\{q_{n}\}}=\inf\left\{\lambda:\;\sum_{n}\exp\left(-\frac{\lambda}{q_{n}}\right)<+\infty\right\}.
$$}

\textit{If conditions (\ref{fsc}) and (\ref{fsc+}) with given
$\varepsilon$ hold for the sequence $\{q_{n}\}$ then}

\begin{itemize}

\item \textit{the $\chi$-capacity of the set
$\mathcal{S}^{\varepsilon}_{\{q_{n}\}}$ is expressed by}
$$
\bar{C}\left(\mathcal{S}^{\varepsilon}_{\{q_{n}\}}\right)
=\lambda_{\{q_{n}\}}^{\varepsilon}-\log\pi_{\{q_{n}\}}^{\varepsilon},
$$\vspace{-10pt}
\item \textit{the optimal average state $\Omega(\mathcal{S}^{\varepsilon}_{\{q_{n}\}})$ of the set
$\mathcal{S}^{\varepsilon}_{\{q_{n}\}}$ has the form}
$$
\pi_{\{q_{n}\}}^{\varepsilon}|1\rangle\langle
1|+\pi_{\{q_{n}\}}^{\varepsilon}\sum_{n>1}\left(q_{n}(1-q_{n})^{\frac{(1-q_{n})}{q_{n}}}\right)^{(1-\varepsilon)}
   \exp\left(-\frac{\lambda_{\{q_{n}\}}^{\varepsilon}}{q_{n}}\right)|n\rangle\langle n|,
$$\vspace{-10pt}

\item \textit{the optimal probability
  distribution $\{\pi^{\pm}_{n}\}$ is defined as follows}
  $$
  \!\!\pi^{\pm}_{1}=0,\quad
  \pi^{\pm}_{n}=\textstyle\frac{1}{2}\displaystyle\pi_{\{q_{n}\}}^{\varepsilon}
  q_{n}^{-\varepsilon}(1-q_{n})^{\frac{(1-q_{n})(1-\varepsilon)}{q_{n}}}
  \exp\left(-\frac{\lambda_{\{q_{n}\}}^{\varepsilon}}{q_{n}}\right),n\geq 2,
  $$
\end{itemize}
\textit{where $\lambda_{\{q_{n}\}}^{\varepsilon}$ is the unique
solution of the equation
$$
\sum_{n>1}q_{n}^{-\varepsilon}(1-q_{n})^{1+\frac{(1-q_{n})(1-\varepsilon)}{q_{n}}}
\exp\left(-\frac{\lambda}{q_{n}}\right)=1
$$
and
$\pi_{\{q_{n}\}}^{\varepsilon}=\left(\sum_{n>1}q_{n}^{-\varepsilon}(1-q_{n})^{\frac{(1-q_{n})(1-\varepsilon)}{q_{n}}}
\exp\left(-\frac{\lambda_{\{q_{n}\}}^{\varepsilon}}{q_{n}}\right)\right)^{-1}\in[0;1]$.}\vspace{5pt}

\textit{Condition (\ref{fsc}) means boundedness of the entropy on
the set
$\overline{\mathrm{co}}(\mathcal{S}^{\varepsilon}_{\{q_{n}\}})$
for arbitrary $\varepsilon$.}

\textit{Existence of the Gibbs state
$\Gamma(\overline{\mathrm{co}}(\mathcal{S}^{\varepsilon}_{\{q_{n}\}}))$
of the set
$\overline{\mathrm{co}}(\mathcal{S}^{\varepsilon}_{\{q_{n}\}})$
for some and hence for arbitrary $\varepsilon$ is equivalent to
validity of conditions (\ref{fsc}) and (\ref{fsc+}) with
$\varepsilon=1$ for the sequence $\{q_{n}\}$. If these conditions
hold then
$$
\Gamma(\overline{\mathrm{co}}(\mathcal{S}^{\varepsilon}_{\{q_{n}\}}))=\pi_{\{q_{n}\}}^{1}|1\rangle\langle
1|+\pi_{\{q_{n}\}}^{1}\sum_{n>1}\exp\left(-\frac{\lambda_{\{q_{n}\}}^{1}}{q_{n}}\right)|n\rangle\langle
n|,
$$
for arbitrary $\varepsilon$, where $\pi_{\{q_{n}\}}^{1}$ and
$\lambda_{\{q_{n}\}}^{1}$ are the above defined
parameters.}\footnote{It is interesting to compare this
observation with the results of proposition 1a with the
$\mathfrak{H}$-operator
$H=\sum_{n=2}^{+\infty}q_{n}^{-1}|n\rangle\langle n|$.}

\textit{If condition (\ref{fsc}) holds for arbitrary $\lambda>0$
then the entropy is continuous on the set
$\overline{\mathrm{co}}(\mathcal{S}^{\varepsilon}_{\{q_{n}\}})$
for arbitrary $\varepsilon$.}\vspace{5pt}

In fig.2 the results of numerical calculation of the
$\chi$-capacity of the set $\mathcal{S}^{\varepsilon}_{\{q_{n}\}}$
as a function of $\varepsilon$ for different sequences $\{q_{n}\}$
are presented.\vspace{5pt}

By theorem 1 finiteness of the $\chi$-capacity of the set
$\mathcal{S}^{\varepsilon}_{\{q_{n}\}}$ means existence of the
optimal average state
$\Omega(\mathcal{S}^{\varepsilon}_{\{q_{n}\}})$ in
$\overline{\mathrm{co}}(\mathcal{S}^{\varepsilon}_{\{q_{n}\}})$
such that
\begin{equation}\label{f-cond}
\sup_{n\geq
1}H(\rho^{\pm}_{n}\|\Omega(\mathcal{S}^{\varepsilon}_{\{q_{n}\}}))<+\infty.
\end{equation}

By lemma 1 in \cite{H-Sh-W} the optimal average state can be
represented as follows
\begin{equation}\label{o-a-s}
\Omega(\mathcal{S}^{\varepsilon}_{\{q_{n}\}})=\pi_{1}\rho_{1}+\sum_{n>1,\pm}\pi^{\pm}_{n}\rho^{\pm}_{n}.
\end{equation}
Since the set $\mathcal{S}^{\varepsilon}_{\{q_{n}\}}$ is invariant
under action of the automorphism $U(\cdot)U^{*}$, where $U$ is a
unitary operator diagonizable in the basis $\{|n\rangle\}$ and
having eigen values $\pm 1$, corollary 8 implies that the state
$\Omega(\mathcal{S}^{\varepsilon}_{\{q_{n}\}})$ is invariant under
the action of the above automorphism and hence it is diagonizable
in the basis $\{|n\rangle\}$. This means that
$\pi^{+}_{n}=\pi^{-}_{n}=\frac{1}{2}\pi_{n}$ for all $n>1$ in
(\ref{o-a-s}), where $\{\pi_{n}\}_{n=1}^{+\infty}$ is a
probability distribution. So we have
\begin{equation}\label{o-a-s-2}
\Omega(\mathcal{S}^{\varepsilon}_{\{q_{n}\}})=\pi|1\rangle\langle
1|+\sum_{n>1}\pi_{n}q_{n}|n\rangle\langle n|,
\end{equation}
where $\pi=\pi_{1}+\sum_{n>1}(1-q_{n})\pi_{n}$. Thus
\begin{equation}\label{r-e-exp-1}
H(\rho_{1}\|\Omega(\mathcal{S}^{\varepsilon}_{\{q_{n}\}}))=-\log\pi
\end{equation}
and \begin{equation}\label{r-e-exp}
\begin{array}{c}
H(\rho^{\pm}_{n}\|\Omega(\mathcal{S}^{\varepsilon}_{\{q_{n}\}}))=-(1-q_{n})\log\pi-q_{n}\log(\pi_{n}q_{n})
\\\\+(1-\varepsilon)((1-q_{n})\log(1-q_{n})+q_{n}\log
q_{n})=-(1-q_{n})\log\pi\\\\-q_{n}\log\pi_{n}-\varepsilon
q_{n}\log q_{n}+(1-\varepsilon)(1-q_{n})\log(1-q_{n}),\quad n>1.
\end{array}
\end{equation}

Since $q_{n}\rightarrow 0$ as $n\rightarrow+\infty$ condition
(\ref{f-cond}) means that $\sup_{n>1}q_{n}(-\log\pi_{n})$ is
finite. It is easy to see that existence of a probability
distribution $\{\pi_{n}\}$ satisfying this condition is equivalent
to existence of positive $\lambda$ such that the series
$\sum_{n}\exp\left(-\frac{\lambda}{q_{n}}\right)$ is finite.

Note that (\ref{f-cond}) and (\ref{r-e-exp}) imply $\pi_{n}>0$ for
all $n>1$. By using this (\ref{r-e-exp-1}) and (\ref{r-e-exp})
system (\ref{opt-ens-sys}) can be rewritten in the form
\begin{equation}\label{opt-ens-sys-2}
   \left\{
   \begin{array}{lr}
   -\log\pi\leq C,\quad \pi_{1}(C+\log\pi)=0\\\\
   (1-q_{n})((1-\varepsilon)\log(1-q_{n})-\log\pi)-q_{n}\log\pi_{n}-\varepsilon q_{n}\log
   q_{n}=C.
   \end{array}\right.
\end{equation}

The second part of this system  implies
  \begin{equation}\label{opt-pi-2}
  \pi_{n}=\pi q_{n}^{-\varepsilon}(1-q_{n})^{\frac{(1-q_{n})(1-\varepsilon)}{q_{n}}}
  \exp\left(-\frac{C+\log\pi}{q_{n}}\right),\;n\geq
  2.
  \end{equation}
Since $\pi_{n}$ must be arbitrary small for large $n$ we conclude
that $-\log\pi<C$ and the first part of the above system implies
$\pi_{1}=0$.

It is easy to see that if there exists a probability distribution
$\{\pi_{n}\}$ satisfying system (\ref{opt-ens-sys-2}) then
$\pi=\sum_{n>1}(1-q_{n})\pi_{n}$ and $C$ forms a solution of the
system
  \begin{equation}\label{system}
  \left\{
   \displaystyle\begin{array}{lr}
   \sum_{n>1}q_{n}^{-\varepsilon}(1-q_{n})^{1+\frac{(1-q_{n})(1-\varepsilon)}{q_{n}}}
   \exp\left(-\frac{C+\log\pi}{q_{n}}\right)=1\\
   \sum_{n>1}q_{n}^{-\varepsilon}(1-q_{n})^{\frac{(1-q_{n})(1-\varepsilon)}{q_{n}}}
   \exp\left(-\frac{C+\log\pi}{q_{n}}\right)=\pi^{-1}.
   \end{array}\right.
  \end{equation}
and vise versa by means of (\ref{opt-pi-2}) any solution $(\pi,
C)$ of system (\ref{system}) provides a probability distribution
$\{\pi_{n}\}$ satisfying system (\ref{opt-ens-sys-2}).

Now will show that system (\ref{system}) has a solution $(\pi, C)$
if and only if inequality (\ref{fsc+}) holds. Consider the
functions
$$
F(x)=\sum_{n>1}q_{n}^{-\varepsilon}(1-q_{n})^{1+\frac{(1-q_{n})(1-\varepsilon)}{q_{n}}}
\exp\left(-\frac{x}{q_{n}}\right)
$$
and
$$G(x)=\sum_{n>1}q_{n}^{-\varepsilon}(1-q_{n})^{\frac{(1-q_{n})(1-\varepsilon)}{q_{n}}}
\exp\left(-\frac{x}{q_{n}}\right).
$$
It is easy to see that these functions are continuous and strictly
decreasing on $(\lambda^{*}_{\{q_{n}\}};+\infty)$ such that
$F(x)\leq G(x)$. Hence there exist the converse functions
$F^{-1}(y)$ and $G^{-1}(y)$, which are continuous and strictly
decreasing on $F((\lambda^{*}_{\{q_{n}\}};+\infty))$ and on
$G((\lambda^{*}_{\{q_{n}\}};+\infty))$ correspondingly. By means of
these functions system (\ref{system}) can be rewritten in the form
$$
 \left\{
 \displaystyle\begin{array}{lr}
   F(C+\log\pi)=1\\
   G(C+\log\pi)=\pi^{-1}.
   \end{array}\right.
$$
It is easy to see that inequality (\ref{fsc+}) is equivalent to the
following one $\lim_{x\rightarrow\lambda^{*}_{\{q_{n}\}}+0}F(x)\geq
1$, which by the previous observation  means that $F^{-1}(1)$ is
well defined. So, if inequality (\ref{fsc+}) holds then
$C+\log\pi=F^{-1}(1)$. Hence
$\pi=(G(F^{-1}(1)))^{-1}\leq(F(F^{-1}(1)))^{-1}=1$ and
$C=F^{-1}(1)+\log G(F^{-1}(1))$ form the unique solution of system
(\ref{system}). Denoting $F^{-1}(1)$ and $\pi$ by
$\lambda_{\{q_{n}\}}^{\varepsilon}$ and
$\pi_{\{q_{n}\}}^{\varepsilon}$ correspondingly we obtain the all
statements, concerning the $\chi$-capacity of the set
$\mathcal{S}^{\varepsilon}_{\{q_{n}\}}$. If inequality (\ref{fsc+})
does not hold then there exists no solution of system (\ref{system})
and hence there exists no optimal probability distribution
$\{\pi_{n}\}$.\footnote{It is easy to construct a sequence
$\{q_{n}\}$ for which (\ref{fsc}) holds while (\ref{fsc+}) does not
hold (see the example at the end of this subsection).} Thus the set
$\mathcal{S}^{\varepsilon}_{\{q_{n}\}}$ is not reqular in this case.

Since boundedness of the entropy on the set
$\overline{\mathrm{co}}(\mathcal{S}^{\varepsilon}_{\{q_{n}\}})$
implies finiteness of the $\chi$-capacity of the set
$\mathcal{S}^{\varepsilon}_{\{q_{n}\}}$ it implies, by the above
observation, validity of condition (\ref{fsc}). But the
boundedness condition in the part B of proposition 9a with the
$\mathfrak{H}$-operator
$\sum_{n=2}^{+\infty}q_{n}^{-1}|n\rangle\langle n|$ provides the
converse implication. Thus condition (\ref{fsc}) means boundedness
of the entropy on the set
$\overline{\mathrm{co}}(\mathcal{S}^{\varepsilon}_{\{q_{n}\}})$.

Suppose condition (\ref{fsc+}) with $\varepsilon=1$ holds for the
sequence $\{q_{n}\}$. Since the closed set
$\mathcal{S}^{1}_{\{q_{n}\}}$ consists of pure states existence of
the optimal measure for this set provided by the above condition
implies that the optimal average state
$\Omega(\mathcal{S}^{1}_{\{q_{n}\}})$ coincides with the Gibbs
state
$\Gamma(\overline{\mathrm{co}}(\mathcal{S}^{1}_{\{q_{n}\}}))$. By
noting that $\Omega(\mathcal{S}^{1}_{\{q_{n}\}})$ lies in
$\overline{\mathrm{co}}(\mathcal{S}^{0}_{\{q_{n}\}})$ and that
$\overline{\mathrm{co}}(\mathcal{S}^{0}_{\{q_{n}\}})\subseteq\overline{\mathrm{co}}(\mathcal{S}^{\varepsilon}_{\{q_{n}\}})$
for arbitrary $\varepsilon$ we conclude that
$$
\Gamma(\overline{\mathrm{co}}(\mathcal{S}^{\varepsilon}_{\{q_{n}\}}))=\Omega(\mathcal{S}^{1}_{\{q_{n}\}})
$$
for arbitrary $\varepsilon$ in this case.

Suppose there exists the Gibbs state
$\Gamma(\overline{\mathrm{co}}(\mathcal{S}^{\varepsilon}_{\{q_{n}\}}))$
for some $\varepsilon$. By using the observations in the end of
section 3 it is easy to see that this implies existence the Gibbs
state
$\Gamma(\overline{\mathrm{co}}(\mathcal{S}^{\varepsilon}_{\{q_{n}\}}))$
for arbitrary $\varepsilon$, in particular, for $\varepsilon=1$.
Since the closed set $\mathcal{S}^{1}_{\{q_{n}\}}$ consists of
pure states the Gibbs state
$\Gamma(\overline{\mathrm{co}}(\mathcal{S}^{1}_{\{q_{n}\}}))$
coincides with the optimal average state
$\Omega(\mathcal{S}^{1}_{\{q_{n}\}})$. By lemma 2 the restriction
of the entropy to the set
$\overline{\mathrm{co}}(\mathcal{S}^{1}_{\{q_{n}\}})$ is
continuous at the state
$\Omega(\mathcal{S}^{1}_{\{q_{n}\}})=\Gamma(\overline{\mathrm{co}}(\mathcal{S}^{1}_{\{q_{n}\}}))$,
which implies reqularity of the set $\mathcal{S}^{1}_{\{q_{n}\}}$.
By theorem 3 there  exists an optimal measure for the set
$\mathcal{S}^{1}_{\{q_{n}\}}$ and hence, by the above observation,
condition (\ref{fsc+}) with $\varepsilon=1$ holds for the sequence
$\{q_{n}\}$.

By the second continuity condition in the part B of proposition 9a
with the $\mathfrak{H}$-operator
$\sum_{n=2}^{+\infty}q_{n}^{-1}|n\rangle\langle n|$ finiteness of
the series in (\ref{fsc}) for arbitrary $\lambda$ implies
continuity of the entropy on the set
$\overline{\mathrm{co}}(\mathcal{S}^{\varepsilon}_{\{q_{n}\}})$
for arbitrary $\varepsilon$.

We complete this subsection with the example of the sequence
$\{q_{n}\}$ for which condition (\ref{fsc}) holds while condition
(\ref{fsc+}) with arbitrary $\varepsilon$ does not hold. Let
$q_{n}=1/\log(n\log^{3}(2n+1))$ for $n\geq 2$. Then
$\lambda^{*}_{\{q_{n}\}}=1$ and the left side of (\ref{fsc+}) with
$\varepsilon=1$ is approximately equal to $0.89$. It follows that
condition (\ref{fsc+}) does not hold with arbitrary $\varepsilon$.
By the above observation for arbitrary $\varepsilon$ the entropy
is bounded on the set
$\overline{\mathrm{co}}(\mathcal{S}^{\varepsilon}_{\{q_{n}\}})$
and the $\chi$-capacity of the set
$\mathcal{S}^{\varepsilon}_{\{q_{n}\}}$ is finite  but the Gibbs
state
$\Gamma(\overline{\mathrm{co}}(\mathcal{S}^{\varepsilon}_{\{q_{n}\}}))$
of the set
$\overline{\mathrm{co}}(\mathcal{S}^{\varepsilon}_{\{q_{n}\}})$
and the optimal measure $\mu_{*}=\{\pi^{\pm}_{n},
\rho^{\pm}_{n}\}$ for the set
$\mathcal{S}^{\varepsilon}_{\{q_{n}\}}$ do not exist.

\subsection{The sets $\mathcal{L}(\sigma)$ and
$\mathcal{K}_{H,h}$}

Let $\sigma=\sum_{k}\lambda_{k}|k\rangle\langle k|$ be an
arbitrary state. The layer $\mathcal{L}(\sigma)$ is defined in
section 3 as the set consisting of all states, having the same
diagonal values as the state $\sigma$ in the basis
$\{|k\rangle\}$. By proposition 6a the entropy is continuous on
the set $\mathcal{L}(\sigma)$ if and only if $H(\sigma)<+\infty$
and $\sup_{\rho\in\mathcal{L}(\sigma)}H(\rho)=H(\sigma)$. The
questions concerning the $\chi$-capacity of the set
$\mathcal{L}(\sigma)$ are considered in the following proposition.

\textbf{Proposition 6b.} \textit{Let $\sigma$ be an arbitrary
state.}

\textit{The $\chi$-capacity of the set $\mathcal{L}(\sigma)$ is
equal to $H(\sigma)$.}

\textit{The set $\mathcal{L}(\sigma)$ is regular if and only if
$H(\sigma)<+\infty$. If this condition holds then there exists an
optimal measure for the set $\mathcal{L}(\sigma)$ with the
barycenter $\Omega(\mathcal{L}(\sigma))=\sigma$ supported by pure
states in $\mathcal{L}(\sigma)$.}

\textbf{Proof.} Suppose $\bar{C}(\mathcal{L}(\sigma))$ is finite.
Let $G$ be the group of all unitaries in
$\mathfrak{B}(\mathcal{H})$ diagonizable in the basis
$\{|k\rangle\}$. Since the set $\mathcal{L}(\sigma)$ is invariant
under the action of the automorphism $U(\cdot)U^{*}$ for each
$U\in G$ corollary 9 implies $\Omega(\mathcal{L}(\sigma))=\sigma$.
Let $\rho$ be an arbitrary pure state in $\mathcal{L}(\sigma)$,
for example, the state, corresponding to the vector
$\sum_{k}\sqrt{\lambda_{k}}|k\rangle$. By theorem 1 and
proposition 6a we have
$$
\bar{C}(\mathcal{L}(\sigma))\geq H(\rho\|\sigma)=H(\sigma).
$$
Since obviously
$\bar{C}(\mathcal{L}(\sigma))\leq\sup_{\rho\in\mathcal{L}(\sigma)}H(\rho)=H(\sigma)$
there is equality here. To complete the proof of the first
assertion of the proposition note that by the last inequality
$\bar{C}(\mathcal{L}(\sigma))=+\infty$  implies
$H(\sigma)=+\infty$.

The regularity assertion follows from proposition 6a.

Since
$\bar{C}(\mathcal{L}(\sigma))=H(\Omega(\mathcal{L}(\sigma)))$ the
assertion concerning existence of optimal measure follows from
theorem 3, propositions 6a and 8. $\square$

The set $\mathcal{K}_{H,h}$ is introduced in section 3 as the set
defined by the inequality $\mathrm{Tr}\rho H\leq h$, where $H$ is
a $\mathfrak{H}$-operator and $h$ is a positive number.
Proposition 1a gives necessary and sufficient conditions of
boundedness and of continuity of the entropy on the set
$\mathcal{K}_{H,h}$ in terms of the increase coefficient
$\mathrm{ic}(H)$ of the $\mathfrak{H}$-operator $H$. This
proposition also shows that existence of the Gibbs state of the
set $\mathcal{K}_{H,h}$ is equivalent to the inequality $h\leq
h_{*}(H)$.\footnote{The parameters $\mathrm{ic}(H)$ and $h_{*}(H)$
are defined before proposition 1a.} The questions concerning the
$\chi$-capacity of the set $\mathcal{K}_{H,h}$ are considered in
the following proposition.

\textbf{Proposition 1b.} \textit{Let $H$ be a
$\mathfrak{H}$-operator on the Hilbert space $\mathcal{H}$ and $h$
be a positive number such that $h\geq
h_{\mathrm{m}}(H)$.}\vspace{5pt}

\textit{The $\chi$-capacity of the set $\mathcal{K}_{H,h}$
coincides with $\sup_{\rho\in\mathcal{K}_{H,h}}H(\rho)$ and hence
it is finite if and only if $\mathrm{ic}(H)<+\infty$. If this
condition holds then
$$
\Omega(\mathcal{K}_{H,h})=\left\{
   \begin{array}{ll}
    \Gamma(\mathcal{K}_{H,h})=(\mathrm{Tr}\exp(-\lambda^{*}H))^{-1}\exp(-\lambda^{*}H),&
    h\leq h_{*}(H) \\
    (\mathrm{Tr}\exp(-\mathrm{ic}(H)H))^{-1}\exp(-\mathrm{ic}(H)H),&
    h> h_{*}(H),
    \end{array}\right.
$$
where $\lambda^{*}$ is uniquely defined by equation
(\ref{main-equation}).}\vspace{5pt}

\textit{The following statements are equivalent}
\begin{enumerate}[i)]
  \item \textit{the inequality $h\leq h_{*}(H)$ holds;}
  \item \textit{the set $\mathcal{K}_{H,h}$ is regular;}
  \item
  \textit{$\bar{C}(\mathcal{K}_{H,h})\leq
  H(\Omega(\mathcal{K}_{H,h}))$;}\footnote{This inequality implies equality.}
  \item
  \textit{$\bar{C}(\mathcal{K}_{H,h})=\bar{C}(\mathcal{K}_{H,h}\cap\mathcal{L}(\Omega(\mathcal{K}_{H,h})))$;}
  \item \textit{there exists an optimal measure for the set  $\mathcal{K}_{H,h}$.}
\end{enumerate}

\textbf{Proof.} Let $H=\sum_{k}h_{k}|k\rangle\langle k|$ and
$\mathcal{K}_{H,h}^{c}$ be the subset of $\mathcal{K}_{H,h}$
consisting of states diagonizable in the basis $\{|k\rangle\}$.
Then
$\mathcal{K}_{H,h}=\bigcup_{\rho\in\mathcal{K}_{H,h}^{c}}\mathcal{L}(\rho)$
and hence
$$
\bar{C}(\mathcal{K}_{H,h})\geq
\sup_{\rho\in\mathcal{K}_{H,h}^{c}}\bar{C}(\mathcal{L}(\rho))=\sup_{\rho\in\mathcal{K}_{H,h}^{c}}H(\rho)=
\sup_{\rho\in\mathcal{K}_{H,h}}H(\rho),
$$
where the last equality follows from inequality
(\ref{layer-ineq}). Since the converse inequality is obvious the
first statement of the proposition is proved.

In the proof of proposition 1a the sequence $\{\rho_{n}\}$ of
states in $\mathcal{K}_{H,h}^{c}$ such that
$\lim_{n\rightarrow\infty}H(\rho_{n})=\sup_{\rho\in\mathcal{K}_{H,h}}H(\rho)$
and
$\lim_{n\rightarrow\infty}\rho_{n}=\rho_{*}(\mathcal{K}_{H,h})$
was constructed. By proposition 6b for each $n$ there exists
optimal measure $\mu_{n}$ for the set $\mathcal{L}(\rho_{n})$ such
that $\bar{\rho}(\mu_{n})=\rho_{n}$ and
$\chi(\mu_{n})=H(\rho_{n})$. By the first part of the proposition
the sequence of measures $\{\mu_{n}\}$ is an approximating
sequence for the set $\mathcal{K}_{H,h}$. By theorem 1 the limit
$\rho_{*}(\mathcal{K}_{H,h})$ of the corresponding sequence of
barycenters $\{\rho_{n}\}$ is the optimal average state of the set
$\mathcal{K}_{H,h}$.

The asserted equivalence of statements $(i)-(iv)$ will be proved
in the following order
$(i)\Rightarrow(ii)\Rightarrow(v)\Rightarrow(iii)\Rightarrow(iv)\Rightarrow(i)$.

$(i)\Rightarrow(ii)$. By proposition 1a $(i)$ means that
$\Omega(\mathcal{K}_{H,h})$ is the Gibbs state
$\Gamma(\mathcal{K}_{H,h})$ of the set $\mathcal{K}_{H,h}$. By
lemma 2 the restriction of the entropy to the set
$\mathcal{K}_{H,h}$ is continuous at the state
$\Omega(\mathcal{K}_{H,h})$, which implies regularity of the set
$\mathcal{K}_{H,h}$.

$(ii)\Rightarrow(v)$. This directly follows from theorem 3.

$(v)\Rightarrow(iii)$. This directly follows from corollary 11.

$(iii)\Rightarrow(iv)$. This follows from proposition 6b and the
first part of this proposition.

$(iv)\Rightarrow(i)$. If $h>h_{*}(H)$ then by propositions 1a and
6b
$$
\bar{C}(\mathcal{L}(\Omega(\mathcal{K}_{H,h})))=H(\Omega(\mathcal{K}_{H,h}))<
\sup_{\rho\in\mathcal{K}_{H,h}}H(\rho)=\bar{C}(\mathcal{K}_{H,h}).\square
$$

The observations in the proof of propositions 1a and 1b provide
the following example, showing that the regularity condition in
theorem 2I is essential.

\textbf{Example of a closed set with finite $\chi$-capacity having
no minimal closed subset with the same $\chi$-capacity.}

Let $H$ be a $\mathfrak{H}$-operator such that
$h_{*}(H)=\frac{\displaystyle\mathrm{Tr}H\exp(-\mathrm{ic}(H)H)}{\displaystyle\mathrm{Tr}\exp(-\mathrm{ic}(H)H)}<+\infty$.
For example,
$H=\sum_{k=1}^{+\infty}\log((k+1)\log^{3}(k+1))|k\rangle\langle k
|$. By the observation in the proof of proposition 1a for given
$h>h_{*}(H)$ there exists natural $n_{0}$ such that the state
$\rho_{n}$ is well defined by (\ref{rho-n}) for all $n\geq n_{0}$
and the sequence $\{\rho_{n}\}_{n\geq n_{0}}$ converges to the
state $\rho_{*}(\mathcal{K}_{H,h})$ defined by (\ref{rho-star}).
Let $\mathcal{A}_{0}=\bigcup_{n\geq n_{0}}\mathcal{L}(\rho_{n})$
and
$\mathcal{A}=\overline{\mathcal{A}_{0}}=\mathcal{A}_{0}\cup\mathcal{L}(\rho_{*}(\mathcal{K}_{H,h}))$.
By the observation in the proof of proposition 1a  and proposition
6a
$$
\bar{C}(\mathcal{A})=\lim_{n\rightarrow+\infty}H(\rho_{n})>H(\rho_{*}(\mathcal{K}_{H,h}))=
\sup_{\rho\in\mathcal{L}(\rho_{*}(\mathcal{K}_{H,h}))}H(\rho).
$$
We assert that the closed set $\mathcal{A}$ has no minimal closed
subsets with the same $\chi$-capacity. Suppose, $\mathcal{B}$ is
the minimal subset of $\mathcal{A}$. Since
$\bar{C}(\mathcal{L}(\rho_{*}(\mathcal{K}_{H,h})))$ is less than
$\bar{C}(\mathcal{A})=\bar{C}(\mathcal{B})$ the set $\mathcal{B}$
has nonempty intersection with the set $\mathcal{L}(\rho_{n_{*}})$
for some $n_{*}\geq n_{0}$. We will show that the closed set
$\mathcal{B}\backslash\mathcal{L}(\rho_{n_{*}})\varsubsetneq
\mathcal{B}$ has the same $\chi$-capacity as the set $\mathcal{B}$
contradicting to the assumed minimality of this set.

Since
$\sup_{\rho\in\mathcal{L}(\rho_{*}(\mathcal{K}_{H,h}))}H(\rho)<\bar{C}(\mathcal{B})$
there exists approximating sequence $\{\{\pi_{i}^{k},
\rho_{i}^{k}\}\}_{k}$ for the set $\mathcal{B}$ such that the
corresponding sequence of the average states
$\{\bar{\rho}_{k}\}_{k}$ has no intersection with
$\mathcal{L}(\rho_{*}(\mathcal{K}_{H,h}))$ and hence
$\Pi_{\{|k\rangle\}}(\bar{\rho}_{k})=\rho_{n(k)}$ for some
sequence of natural numbers $\{n(k)\}_{k}$. Since
$\sup_{\rho\in\mathcal{L}(\rho_{n})}H(\rho)<\bar{C}(\mathcal{B})$
for each $n\geq n_{0}$ the sequence $\{n(k)\}_{k}$ tends to
$+\infty$. Since for arbitrary state $\rho$ in $\mathcal{B}$ the
state $\Pi_{\{|k\rangle\}}(\rho)$ is either
$\rho_{*}(\mathcal{K}_{H,h})$ or $\rho_{n}$ for a particular $n$
and since $\bar{\rho}_{k}=\sum_{i}\pi_{i}^{k}\rho_{i}^{k}$ implies
$\Pi_{\{|k\rangle\}}(\bar{\rho}_{k})=\sum_{i}\pi_{i}^{k}\Pi_{|k\rangle}(\rho_{i}^{k})$
for each $k$ by using (\ref{rho-n}) and (\ref{rho-star}) we
conclude that $\Pi_{\{|k\rangle\}}(\rho_{i}^{k})=\rho_{n(k)}$ for
all $i$ and $k$. Thus the states $\{\rho_{i}^{k}\}$ are not
contained in $\mathcal{L}(\rho_{n_{*}})$ for all sufficiently
large $k$ and hence the "tail"  of the sequence $\{\{\pi_{i}^{k},
\rho_{i}^{k}\}\}_{k}$ is an approximating sequence for the set
$\mathcal{B}\backslash\mathcal{L}(\rho_{n_{*}})$. This implies
$\bar{C}(\mathcal{B})=\bar{C}(\mathcal{B}\backslash\mathcal{L}(\rho_{n_{*}}))$.

\subsection{The set $\mathcal{V}_{\sigma,c}$}

The set $\mathcal{V}_{\sigma,c}$ is introduced in section 3 as the
set defined by the inequality $H(\rho\|\sigma)\leq c$, where
$\sigma$ is a state and $c$ is a nonnegative number $c$. If
$\sigma$ is a state with infinite dimensional support then the
family of nonempty sets
$\{\mathcal{K}_{\sigma,c}\}_{c\in\mathbb{R}_{+}}$ is strictly
increasing and $\mathcal{K}_{\sigma,0}=\{\sigma\}$.

By theorem 1 every set $\mathcal{A}$ with finite $\chi$-capacity
is contained in the compact convex set
$\mathcal{V}_{\Omega(\mathcal{A}),\bar{C}(\mathcal{A})}$ such that
$$
\Omega(\mathcal{V}_{\Omega(\mathcal{A}),\bar{C}(\mathcal{A})})=
\Omega(\mathcal{A}) \quad \mathrm{and}\quad
\bar{C}(\mathcal{V}_{\Omega(\mathcal{A}),\bar{C}(\mathcal{A})})=\bar{C}(\mathcal{A}).
$$
Below we consider the $\chi$-capacity of the set
$\mathcal{V}_{\sigma,c}$ with arbitrary $\sigma$ and $c$.

Proposition 3a gives necessary and sufficient conditions of
boundedness and of continuity of the entropy on the set
$\mathcal{V}_{\sigma,c}$ in terms of the decrease coefficient
$\mathrm{dc}(\sigma)$ of the state $\sigma$. This proposition also
shows that existence of the Gibbs state of the set
$\mathcal{V}_{\sigma,c}$ is equivalent to the inequality $c\leq
c_{*}(\sigma)$.\footnote{The parameters $\mathrm{dc}(\sigma)$ and
$c_{*}(\sigma)$  are defined before proposition 3a.} The questions
concerning the $\chi$-capacity of the set $\mathcal{V}_{\sigma,c}$
are considered in the following proposition. Let
$c^{*}(\sigma)=\displaystyle\frac{\mathrm{Tr}\sigma^{\mathrm{dc}(\sigma)}(-\log\sigma)}
{\mathrm{Tr}\sigma^{\mathrm{dc}(\sigma)}}$ if
$\mathrm{Tr}\sigma^{\mathrm{dc}(\sigma)}<+\infty$ and
$c^{*}(\sigma)=+\infty$ otherwise. Note that
$c^{*}(\sigma)=\displaystyle\frac{c_{*}(\sigma)+\log\mathrm{Tr}\sigma^{\mathrm{dc}(\sigma)}}{1-\mathrm{dc}(\sigma)}
\geq c_{*}(\sigma)$ if $\mathrm{dc}(\sigma)<1$ and
$c^{*}(\sigma)=c_{*}(\sigma)=H(\sigma)$ if
$\mathrm{dc}(\sigma)=1$. \vspace{5pt}

\textbf{Proposition 3b.} \textit{Let $\sigma$ be an arbitrary
infinite dimensional state.}\vspace{5pt}

\textit{If $c\leq H(\sigma)\leq+\infty$ then
$$
\bar{C}(\mathcal{V}_{\sigma,c})=c\quad and\quad
\Omega(\mathcal{V}_{\sigma,c})=\sigma.
$$}\vspace{-10pt}

\textit{If $H(\sigma)<c\leq c^{*}(\sigma)$ then
$$
\bar{C}(\mathcal{V}_{\sigma,c})=\lambda^{*}c+\log\mathrm{Tr}\sigma^{\lambda^{*}}\quad
and\quad \Omega(\mathcal{V}_{\sigma,c})=
\left(\mathrm{Tr}\sigma^{\lambda^{*}}\right)^{-1}\sigma^{\lambda^{*}},
$$
where $\lambda^{*}$ is uniquely defined by the equation
$$
\mathrm{Tr}\sigma^{\lambda}(-\log
\sigma)=c\mathrm{Tr}\sigma^{\lambda}.
$$}\vspace{-10pt}

\textit{If $c^{*}(\sigma)<+\infty$ and $c\geq c^{*}(\sigma)$ then
$$
\bar{C}(\mathcal{V}_{\sigma,c})=
    \mathrm{dc}(\sigma)c+\log\mathrm{Tr}\sigma^{\mathrm{dc}(\sigma)}\quad
and\quad \Omega(\mathcal{V}_{\sigma,c})=
    \left(\mathrm{Tr}\sigma^{\mathrm{dc}(\sigma)}\right)^{-1}\sigma^{\mathrm{dc}(\sigma)}.
$$}\vspace{-10pt}

\textit{In the all cases
$\bar{C}(\mathcal{V}_{\sigma,c})=\inf_{\lambda\in(\mathrm{dc}(\sigma);1]}\left(\lambda
c+\log\mathrm{Tr}\sigma^{\lambda}\right)$.}\vspace{10pt}

\textit{The following statements are equivalent:}
\begin{enumerate}[i)]
  \item \textit{the inequality $c\leq c^{*}(\sigma)$ holds;}
  \item
  \textit{$\bar{C}(\mathcal{V}_{\sigma,c})\leq H(\Omega(\mathcal{V}_{\sigma,c}))$;}
  \item
  \textit{$\bar{C}(\mathcal{V}_{\sigma,c})=\bar{C}(\mathcal{V}_{\sigma,c}\cap
           \mathcal{L}(\Omega(\mathcal{V}_{\sigma,c})))$;}
  \item \textit{there exists optimal measure for the set
  $\mathcal{V}_{\sigma,c}$.}
\end{enumerate}

\textit{The set $\mathcal{V}_{\sigma,c}$ is regular if and only if
$\mathrm{dc}(\sigma)<1$ and $c<c^{*}(\sigma)$.}\vspace{5pt}

In fig.1 the result of numerical calculations of the
$\chi$-capacity of the set $\mathcal{V}_{\sigma,c}$ as a function
of $c$ for the state $\sigma$ with finite $c^{*}(\sigma)$ is
shown.

\textbf{Proof.} Let $\sigma=\sum_{k}\lambda_{k}|k\rangle\langle
k|$ be a full rank state so that $-\log\sigma$ is a
$\mathfrak{H}$-operator. The inequality
\begin{equation}\label{chi-cap-ineq-2}
\bar{C}(\mathcal{V}_{\sigma,c})\leq c
\end{equation}
follows from expression (\ref{chi-cap-exp}) in theorem 1.

Let $c\leq H(\sigma)\leq+\infty$. Consider the subset
$\mathcal{T}=\mathcal{V}_{\sigma,c}\cap\mathcal{L}(\sigma)$ of
$\mathcal{V}_{\sigma,c}$. By monotonicity of the $\chi$-capacity
and (\ref{chi-cap-ineq-2}) we have
$\bar{C}(\mathcal{T})\leq\bar{C}(\mathcal{V}_{\sigma,c})\leq
c<+\infty$. So, to prove that $\bar{C}(\mathcal{V}_{\sigma,c})=c$
it is sufficient to show that $\bar{C}(\mathcal{T})\geq c$.

Let $G$ be the group of all unitaries in
$\mathfrak{B}(\mathcal{H})$ diagonizable in the basis
$\{|k\rangle\}$. Since the set $\mathcal{T}$ is invariant under
the action of the automorphism $U(\cdot)U^{*}$ for each $U\in G$
corollary 9 implies $\Omega(\mathcal{T})=\sigma$. By expression
(\ref{chi-cap-exp}) in theorem 1 to show that
$\bar{C}(\mathcal{T})\geq c$ it is sufficient to find a state
$\sigma_{c}$ in the set $\mathcal{T}$ such that
$H(\sigma_{c}\|\Omega(\mathcal{T}))=H(\sigma_{c}\|\sigma)=c$.

By proposition 6a in the case $H(\sigma)<+\infty$ the relative
entropy $H(\rho\|\sigma)$ is a continuous function on
$\mathcal{L}(\sigma)$ with the range $[0;H(\sigma)]$. This implies
existence of the state $\sigma_{c}$ with the desired properties.

In the case $H(\sigma)=+\infty$ existence of the state
$\sigma_{c}$ follows from lemma 6 below (with $n=1$).

Thus $\bar{C}(\mathcal{V}_{\sigma,c})=\bar{C}(\mathcal{T})=c$ and
theorem 2C implies
$\Omega(\mathcal{V}_{\sigma,c})=\Omega(\mathcal{T})=\sigma$.

Let $c>H(\sigma)$. Since
$\mathcal{K}_{-\log\sigma,c}\subset\mathcal{V}_{\sigma,c}$
monotonicity of the $\chi$-capacity implies
\begin{equation}\label{chi-cap-ineq}
\bar{C}(\mathcal{K}_{-\log\sigma,c})\leq\bar{C}(\mathcal{V}_{\sigma,c})
\end{equation}

Note that $c^{*}(\sigma)=h_{*}(-\log\sigma)$. By proposition 1b to
prove the all assertions concerning the cases $H(\sigma)<c\leq
c^{*}(\sigma)$ and $c\geq c^{*}(\sigma)$ it is sufficient to show
that
\begin{equation}\label{coincidence}
\bar{C}(\mathcal{V}_{\sigma,c})=\bar{C}(\mathcal{K}_{-\log\sigma,c})
\end{equation}
since this equality and theorem 2C imply
$\Omega(\mathcal{V}_{\sigma,c})=\Omega(\mathcal{K}_{-\log\sigma,c})$.

Suppose $\mathrm{dc}(\sigma)=\mathrm{ic}(-\log\sigma)=1$. Then
$c^{*}(\sigma)=h_{*}(-\log\sigma)=H(\sigma)$. By proposition 1b
$\bar{C}(\mathcal{K}_{-\log\sigma,c})=c$ for all $c\geq
H(\sigma)$. Thus inequalities (\ref{chi-cap-ineq-2}) and
(\ref{chi-cap-ineq}) imply equality (\ref{coincidence}).

Suppose $\mathrm{dc}(\sigma)=\mathrm{ic}(-\log\sigma)<1$. Then
lemma 3 implies
$$
H(\rho\|(\mathrm{Tr}\sigma^{\lambda})^{-1}\sigma^{\lambda})\leq
\lambda H(\rho\|\sigma)+\log\mathrm{Tr}\sigma^{\lambda}\leq\lambda
c+\log\mathrm{Tr}\sigma^{\lambda}
$$
for all $\rho$ in $\mathcal{V}_{\sigma,c}$ and for all
$\lambda\in(\mathrm{dc}(\sigma);1]$. By the second part of theorem
1 we have
$$
\bar{C}(\mathcal{V}_{\sigma,c})\leq\inf_{\lambda\in(\mathrm{dc}(\sigma);1]}\sup_{\rho\in\mathcal{V}_{\sigma,c}}
H(\rho\|(\mathrm{Tr}\sigma^{\lambda})^{-1}\sigma^{\lambda})\leq\inf_{\lambda\in(\mathrm{dc}(\sigma);1]}\left(\lambda
c+\log\mathrm{Tr}\sigma^{\lambda}\right).
$$
By proposition 1b
$\bar{C}(\mathcal{K}_{-\log\sigma,c})=\inf_{\lambda\in(\mathrm{dc}(\sigma);+\infty)}\left(\lambda
c+\log\mathrm{Tr}\sigma^{\lambda}\right)$ and it is easy to see
that the condition $c> H(\sigma)$ implies that the last infinum is
achieved at some $\lambda^{*}\leq 1$. Thus this infinum coincides
with the previous one and hence (\ref{coincidence}) holds in this
case.

Equivalence of statements $(i)-(iv)$ will be shown by proving the
following implications
$(i)\Rightarrow(iv)\Rightarrow(ii)\Rightarrow(i)$ and
$(i)\Rightarrow(iii)\Rightarrow(i)$.

$(i)\Rightarrow(iv)$. In the case $H(\sigma)<+\infty$ existence of
an optimal measure for the set $\mathcal{V}_{\sigma,c}$ under the
condition $c\leq c^{*}(\sigma)$ is proved by considering the
following subcases separately
\begin{itemize}
  \item $c\leq H(\sigma)$;
  \item $\mathrm{dc}(\sigma)<1$ and $H(\sigma)<c\leq c^{*}(\sigma)$.
\end{itemize}

If $c\leq H(\sigma)$ then by the above observation
$\bar{C}(\mathcal{V}_{\sigma,c})=\bar{C}(\mathcal{T})$, where
$\mathcal{T}=\mathcal{V}_{\sigma,c}\cap\mathcal{L}(\sigma)$. By
proposition 6a the entropy is continuous on the set $\mathcal{T}$,
which implies its regularity. It follows from this and theorem 3
that there exists an optimal measure for the set $\mathcal{T}$.
Since $\bar{C}(\mathcal{V}_{\sigma,c})=\bar{C}(\mathcal{T})$ and
$\mathcal{T}\subset\mathcal{V}_{\sigma,c}$ this measure is an
optimal measure for the set $\mathcal{V}_{\sigma,c}$.

If $\mathrm{dc}(\sigma)<1$ and $H(\sigma)<c\leq c^{*}(\sigma)$
then by the above observation
$\bar{C}(\mathcal{V}_{\sigma,c})=\bar{C}(\mathcal{K}_{-\log\sigma,c})$
and $c^{*}(\sigma)=h_{*}(-\log\sigma)$. By proposition 1b there
exists an optimal measure for the set
$\mathcal{K}_{-\log\sigma,c}$. Since
$\bar{C}(\mathcal{V}_{\sigma,c})=\bar{C}(\mathcal{K}_{-\log\sigma,c})$
and $\mathcal{K}_{-\log\sigma,c}\subset\mathcal{V}_{\sigma,c}$
this measure is an optimal measure for the set
$\mathcal{V}_{\sigma,c}$.

In the case $H(\sigma)=+\infty$, in which $c^{*}(\sigma)=+\infty$,
existence of an optimal measure is verified by the following
direct construction.

For given $c$ let $m$ and $\rho_{c,1,m}$ be a natural number and a
state provided by lemma 6.  Let
$P_{m}=\sum_{k=1}^{m}|k\rangle\langle k|$ and $G_{m}$ be the
\textit{compact} group of all unitaries in
$\mathfrak{B}(P_{m}(\mathcal{H}))$ diagonizable in the basis
$\{|k\rangle\}_{k=1}^{m}$ in $P_{m}(\mathcal{H})$. For arbitrary
$U$ in $G_{m}$ denote by $\hat{U}$ the unitary operator $U\oplus
I_{\mathcal{H}\ominus P_{m}(\mathcal{H})}$ in
$\mathfrak{B}(\mathcal{H})$. By using the construction of the
state $\rho_{c,1,m}$ it is easy to see that
$$
\int_{G_{m}}\hat{U}\rho_{c,1,m}\hat{U}^{*}\mu_{H}(dU)=\sigma,
$$
where $\mu_{H}$ is the Haar measure on $G_{m}$. Since
$$
H(\hat{U}\rho_{c,1,m}\hat{U}^{*}\|\sigma)=H(\rho_{c,1,m}\|\hat{U}^{*}\sigma\hat{U})=H(\rho_{c,1,m}\|\sigma)=c
$$
the image of the measure $\mu_{H}$ under the mapping $U\mapsto
\hat{U}\rho_{c,1,m}\hat{U}^{*}$ is an optimal measure for the set
$\mathcal{V}_{\sigma,c}$, which is supported by the set
$\mathcal{L}(\sigma)$ by the construction.

$(iv)\Rightarrow(ii)$. This directly follows from corollary 11.

$(ii)\Rightarrow(i)$. If $c^{*}(\sigma)$ is finite and
$c>c^{*}(\sigma)$ then the proof of the previous part of the
proposition, propositions 1a and 1b imply
\begin{equation}\label{t-ineq}
\bar{C}(\mathcal{V}_{\sigma,c})=\bar{C}(\mathcal{K}_{-\log\sigma,c})>
H(\Omega(\mathcal{V}_{\sigma,c})=\Omega(\mathcal{K}_{-\log\sigma,c})).
\end{equation}

$(i)\Rightarrow(iii)$. If $c\leq H(\sigma)$ then by the proof of
the previous part of the proposition
$\bar{C}(\mathcal{V}_{\sigma,c})=\bar{C}(\mathcal{T})$ and
$\Omega(\mathcal{V}_{\sigma,c})=\sigma$, where
$\mathcal{T}=\mathcal{V}_{\sigma,c}\cap\mathcal{L}(\sigma)$. If
$H(\sigma)< c\leq c^{*}(\sigma)$ then by the proof of the previous
part of the proposition, propositions 1b and 6b we have
\begin{equation}\label{tt-ineq}
\bar{C}(\mathcal{V}_{\sigma,c})=\bar{C}(\mathcal{K}_{-\log\sigma,c})=
H(\Omega(\mathcal{K}_{-\log\sigma,c}))=
\bar{C}(\mathcal{L}(\Omega(\mathcal{K}_{-\log\sigma,c}))).
\end{equation}

Since $\mathcal{L}(\Omega(\mathcal{K}_{-\log\sigma,c}))\subset
\mathcal{K}_{-\log\sigma,c}\subset\mathcal{V}_{\sigma,c}$ and
$\Omega(\mathcal{V}_{\sigma,c})=\Omega(\mathcal{K}_{-\log\sigma,c})$
in this case we obtain $(iii)$.

$(iii)\Rightarrow(i)$. If $c^{*}(\sigma)$ is finite and
$c>c^{*}(\sigma)$ then inequality (\ref{t-ineq}) holds which
contradicts to $(iii)$ by proposition 6b.

If $\mathrm{dc}(\sigma)<1$ and $c<c^{*}(\sigma)$ then by the above
observation $\mathrm{dc}(\Omega(\mathcal{V}_{\sigma,c}))<1$ and
regularity of the set $\mathcal{V}_{\sigma,c}$ follows from
theorem 2E.

To prove the converse assertion note that lemma 7 below and the
above observation imply that the second regularity condition does
not hold for the set $\mathcal{V}_{\sigma,c}$ for arbitrary
infinite dimensional state $\sigma$ and $c>0$. Thus it is
sufficient to show the first regularity condition does not hold
for the set $\mathcal{V}_{\sigma,c}$ if either
$\mathrm{dc}(\sigma)=1$ or $c\geq c^{*}(\sigma)$.

If $\mathrm{dc}(\sigma)=1$ then by the above observation
$\Omega(\mathcal{V}_{\sigma,c})=\sigma$ for arbitrary $c$. In the
case $H(\sigma)<+\infty$ proposition 2 implies existence of the
sequence of states $\{\rho_{n}\}$ such that
$$
\lim_{n\rightarrow+\infty}H(\rho_{n}\|\sigma)=0\quad\mathrm{and}\quad
\lim_{n\rightarrow+\infty}H(\rho_{n})>H(\sigma).
$$
Thus the state $\rho_{n}$ lie in $\mathcal{V}_{\sigma,c}$ for
sufficiently large $n$ and hence the first regularity condition
does not hold. In the case $H(\sigma)=+\infty$ the first
regularity condition does not hold obviously.

If $\mathrm{dc}(\sigma)<1$ and $c\geq c^{*}(\sigma)$ then by the
above observation
$\Omega(\mathcal{V}_{\sigma,c})=(\mathrm{Tr}\sigma^{\mathrm{dc}(\sigma)})^{-1}\sigma^{\mathrm{dc}(\sigma)}$.
In the proof of proposition 3a it is shown that for arbitrary $m$
the states in the sequence $\{\rho_{n}^{m}\}_{n}$ for which
relations (\ref{s-lim-exp}) are valid lie in the set
$\mathcal{V}_{\sigma,c}$ for all sufficiently large $n$. Thus the
first regularity condition does not hold in this case. $\square$

The set $\mathcal{K}_{\sigma,c}$ with $H(\sigma)=+\infty$ is a
nontrivial example of a nonregular set containing states with
infinite entropy but having finite $\chi$-capacity and possessing
the optimal measure.

\textbf{Lemma 6.} \textit{Let
$\sigma=\sum_{k=1}^{\infty}\lambda_{k}|k\rangle\langle k|$ be a
state with infinite entropy. For arbitrary natural $n$ let
$\mathcal{L}_{n}(\sigma)$ be the convex closed subset of
$\mathcal{L}(\sigma)$ consisting of all states $\rho$ such that
$\langle i|\rho|j\rangle=0$ if $i\neq j$ and either $i<n$ or
$j<n$.} \textit{Then for arbitrary $c\geq 0$ and $n\in\mathbb{N}$
there exist a natural $m$ and a state $\rho_{c,n,m}$ in
$\mathcal{L}_{n}(\sigma)$ such that
$$
H(\rho_{c,n,m}\|\sigma)=c
$$
and $\langle i|\rho|j\rangle=0$ if $i\neq j$ and either $i>m$ or
$j>m$.}\vspace{5pt}

\textbf{Proof.} Let $c\geq 0$ and $n\in\mathbb{N}$ be arbitrary.
Consider the state
$$
\sigma_{n}=\mu_{n}^{-1}\sum_{k=n}^{+\infty}\lambda_{k}|k\rangle\langle
k|,
$$
where $\mu_{n}=\sum_{k=n}^{+\infty}\lambda_{k}$, and the sequence
of states
$$
\{\rho^{m}_{n}=\mu_{n}^{-1}\sum_{n\leq i,j\leq
m}\sqrt{\lambda_{i}}\sqrt{\lambda_{j}}|i\rangle\langle
j|+\mu_{n}^{-1}\sum_{k> m}\lambda_{k}|k\rangle\langle k|\}_{m},
$$
converging in the trace norm to the pure state
$\rho^{*}_{n}=\mu_{n}^{-1}\sum_{i,j\geq n
}\sqrt{\lambda_{i}}\sqrt{\lambda_{j}}|i\rangle\langle j|$ as
$m\rightarrow+\infty$. Since $H(\sigma_{n})=+\infty$ proposition
6a implies $H(\rho^{*}_{n}\|\sigma_{n})=+\infty$. By using this
and the general properties of the relative entropy we obtain
$$
H(\rho^{m}_{n}\|\sigma_{n})<+\infty,\quad\forall m\in\mathbb{N}
\quad\quad\mathrm{and}\quad\quad
\lim_{m\rightarrow+\infty}H(\rho^{m}_{n}\|\sigma_{n})=+\infty.
$$
Thus there exists natural $m(c)$ such that $c\mu_{n}^{-1}\leq
H(\rho^{m(c)}_{n}\|\sigma_{n})<+\infty$.  The convex lower
semicontinuous function
$f(\lambda)=H(\lambda\rho^{m(c)}_{n}+(1-\lambda)\sigma_{n}\|\sigma_{n})$
does not exceed $\lambda H(\rho^{m(c)}_{n}\|\sigma_{n})$ on
$[0;1]$ and hence it is continuous on $[0;1]$ \cite{J&T}. Since
$f(0)=0$ and $f(1)=H(\rho^{m(c)}_{n}\|\sigma_{n})\geq
c\mu_{n}^{-1}$ there exists $\lambda^{*}\in[0;1]$ such that
$f(\lambda^{*})=c\mu_{n}^{-1}$.

Let $m=m(c)$ and
$\rho_{c,n,m}=\sum_{k=1}^{n-1}\lambda_{k}|k\rangle\langle
k|+\mu_{n}(\lambda^{*}\rho^{m}_{n}+(1-\lambda^{*})\sigma_{n})$. It
is easy to see that $H(\rho_{c,n,m}\|\sigma)=\mu_{n}
H(\lambda^{*}\rho^{m(c)}_{n}+(1-\lambda^{*})\sigma_{n}\|\sigma_{n})=c$
and that $\rho_{c,n,m}\in\mathcal{L}_{n}(\sigma)$. By the
construction $\langle i|\rho|j\rangle=0$ if $i\neq j$ and either
$i>m$ or $j>m$. $\square$

\textbf{Lemma 7.} \textit{Let $\sigma$ be a state with infinite
dimensional support. Then the relative entropy
$H(\rho\|(\mathrm{Tr}\sigma^{\lambda})^{-1}\sigma^{\lambda})$ is
not a continuous function of the state $\rho$ on the set
$\mathcal{V}_{\sigma,c}$ for arbitrary $c>0$ and for arbitrary
$\lambda$ such that $\mathrm{Tr}\sigma^{\lambda}<+\infty$.}

\textbf{Proof.} Without loss of generality we may assume that
$\sigma$ is a full rank state. Let $\varrho$ be a pure state such
that $H(\varrho\|\sigma)=+\infty$ and $P_{n}$ be the spectral
projector of the state $\sigma$, corresponding to its maximal $n$
eigenvalues. Then the sequence of pure states
$\{\varrho_{n}=(\mathrm{Tr}P_{n}\varrho)^{-1}P_{n}\varrho P_{n}\}$
converges to the pure state $\varrho$ and by using general
properties of the relative entropy we have
$$
H(\varrho_{n}\|\sigma)<+\infty\quad
\mathrm{for}\;\mathrm{all}\;n\quad \mathrm{and}
\quad\lim_{n\rightarrow+\infty}H(\varrho_{n}\|\sigma)=+\infty.
$$
Consider the sequence
$\{\eta_{n}=c(H(\varrho_{n}\|\sigma))^{-1}\}_{n\geq n_{0}}$, where
$n_{0}$ is chosen to be so large that $H(\varrho_{n}\|\sigma)>c$
for all $n\geq n_{0}$. Let
$\rho_{n}=\eta_{n}\varrho_{n}+(1-\eta_{n})\sigma$ for all $n\geq
n_{0}$. Then by using general properties of the relative entropy
we obtain
$$
c-h_{2}(\eta_{n})=\eta_{n}H(\varrho_{n}\|\sigma)-h_{2}(\eta_{n})\leq
H(\rho_{n}\|\sigma)\leq \eta_{n}H(\varrho_{n}\|\sigma)=c,
$$
where $h_{2}(x)=-x\log x-(1-x)\log(1-x)$.

Since $\eta_{n}\rightarrow 0$ as $n\rightarrow 0$ this inequality
implies that
\begin{equation}\label{re-lim-exp}
\rho_{n}\in\mathcal{V}_{\sigma,c}\quad
\mathrm{for}\;\mathrm{all}\;n\quad \mathrm{and}
\quad\lim_{n\rightarrow+\infty}H(\rho_{n}\|\sigma)=c.
\end{equation}
Let $\lambda$ be an arbitrary positive number such that
$\mathrm{Tr}\sigma^{\lambda}<+\infty$. By lemma 3 we have
\begin{equation}\label{e-re-exp-n}
H(\rho_{n}\|(\mathrm{Tr}\sigma^{\lambda})^{-1}\sigma^{\lambda})=
\lambda
H(\rho_{n}\|\sigma)+\log\mathrm{Tr}\sigma^{\lambda}-(1-\lambda)H(\rho_{n}).
\end{equation}

By using general properties of the entropy we obtain
$$
(1-\eta_{n})H(\sigma)\leq H(\rho_{n})\leq
(1-\eta_{n})H(\sigma)+h_{2}(\eta_{n}),
$$
for all $n\geq n_{0}$ and hence
$\lim_{n\rightarrow+\infty}H(\rho_{n})=H(\sigma)$.

Thus (\ref{re-lim-exp}) and (\ref{e-re-exp-n})  implies
$$
\lim_{n\rightarrow+\infty}H(\rho_{n}\|(\mathrm{Tr}\sigma^{\lambda})^{-1}\sigma^{\lambda})=
c+\log\mathrm{Tr}\sigma^{\lambda}-(1-\lambda)H(\sigma).
$$
By the construction the sequence $\{\rho_{n}\}$ of states in
$\mathcal{V}_{\sigma,c}$ tends to the state $\sigma$. Since
$H(\sigma\|(\mathrm{Tr}\sigma^{\lambda})^{-1}\sigma^{\lambda})=
\log\mathrm{Tr}\sigma^{\lambda}-(1-\lambda)H(\sigma)$ the previous
expression means
$$
\lim_{n\rightarrow+\infty}H(\rho_{n}\|(\mathrm{Tr}\sigma^{\lambda})^{-1}\sigma^{\lambda})=
H(\sigma\|(\mathrm{Tr}\sigma^{\lambda})^{-1}\sigma^{\lambda})+c,
$$
which implies discontinuity of the function
$H(\rho\|(\mathrm{Tr}\sigma^{\lambda})^{-1}\sigma^{\lambda})$ on
the set $\mathcal{V}_{\sigma,c}$. $\square$

Nonregularity of the set $\mathcal{V}_{\sigma,c}$ for arbitrary
state with infinite entropy are illustrated by the following
example.

\textbf{Example of a decreasing sequence of closed sets with the
same positive $\chi$-capacity, having the intersection with zero
$\chi$-capacity.}

For arbitrary natural $n$ let $\mathcal{L}_{n}(\sigma)$ be the
convex closed subset of $\mathfrak{S}(\mathcal{H})$ introduced in
lemma 6. For given $c>0$ consider the monotonously decreasing
sequence $\{\mathcal{A}_{n}=\mathcal{L}_{n}(\sigma)\cap
\mathcal{V}_{\sigma,c}\}$ of closed convex sets. Corollary 9
implies that $\Omega(\mathcal{A}_{n})=\sigma$ - the only state in
$\mathcal{A}_{n}$ invariant under the action of all automorphism
from $\mathfrak{F}(\mathcal{A}_{n})$. Lemma 6 provides existence
of the state $\rho_{c,n,m}$ in $\mathcal{A}_{n}$ such that
$H(\rho_{c,n,m}\|\Omega(\mathcal{A}_{n}))=H(\rho_{c,n,m}\|\sigma)=c$,
which by theorem 1 implies $\bar{C}(\mathcal{A}_{n})\geq c$. By
theorem 2C we have $\bar{C}(\mathcal{A}_{n})\leq
\bar{C}(\mathcal{V}_{\sigma,c})=c$ and hence
$$
\bar{C}(\mathcal{A}_{n})=c\;\,\mathrm{for}\;\,\mathrm{all}\;\,
n\;\,\mathrm{while}\;\,\bar{C}\left(\bigcap_{n}\mathcal{A}_{n}\right)=0\;\,
\mathrm{since}\;\,\bigcap_{n}\mathcal{A}_{n}=\{\sigma\}.
$$

\subsection{The set $\mathcal{A}\otimes\mathcal{B}$}

Let $\mathcal{H}$ and $\mathcal{K}$ be separable Hilbert spaces.
For arbitrary sets $\mathcal{A}\subseteq\mathfrak{S}(\mathcal{H})$
and $\mathcal{B}\subseteq\mathfrak{S}(\mathcal{K})$ consider the
set
$$
\mathcal{A}\otimes\mathcal{B}=
\{\omega\in\mathfrak{S}(\mathcal{H}\otimes\mathcal{K})|\,
\omega^{\mathcal{H}}\in\mathcal{A},\omega^{\mathcal{K}}\in\mathcal{B}\},
$$
where $\omega^{\mathcal{H}}=\mathrm{Tr}_{\mathcal{K}}\omega$ and
$\omega^{\mathcal{K}}=\mathrm{Tr}_{\mathcal{H}}\omega$.

In \cite{Sh-2} the following lemma was proved.

\textbf{Lemma 8.} \textit{The set $\mathcal{A}\otimes \mathcal{B}$
is a convex subset of $\mathfrak{S}(\mathcal{H}\otimes
\mathcal{K})$ if and only if the sets $\mathcal{A}$ and
$\mathcal{B}$ are convex subsets of $\mathfrak{S}(\mathcal{H})$
and of $\mathfrak{S}(\mathcal{K})$ correspondingly.}

\textit{The set $\mathcal{A}\otimes \mathcal{B}$ is a compact
subset of $\mathfrak{S}(\mathcal{H}\otimes \mathcal{K})$ if and
only if the sets $\mathcal{A}$ and $\mathcal{B}$ are compact
subsets of $\mathfrak{S}(\mathcal{H})$ and of
$\mathfrak{S}(\mathcal{K})$ correspondingly.}

The properties of the restriction of the entropy to the set
$\mathcal{A}\otimes \mathcal{B}$ are also determined by the
properties of the restrictions of the entropy to the sets
$\mathcal{A}$ and $\mathcal{B}$.

\textbf{Proposition 10.} \textit{Let $\mathcal{A}$ and
$\mathcal{B}$ be an arbitrary subsets of
$\mathfrak{S}(\mathcal{H})$ and of $\mathfrak{S}(\mathcal{K})$
correspondingly.}

\textit{The entropy is bounded on the set $\mathcal{A}\otimes
\mathcal{B}$ if and only if the entropy is bounded on the sets
$\mathcal{A}$ and $\mathcal{B}$.}

\textit{The entropy is continuous on the set $\mathcal{A}\otimes
\mathcal{B}$ if and only if the entropy is continuous on the sets
$\mathcal{A}$ and $\mathcal{B}$.}

\textbf{Proof.} If the entropy is bounded (continuous) on the set
$\mathcal{A}\otimes \mathcal{B}$ then it is bounded (continuous)
on the sets $\mathcal{A}$ and $\mathcal{B}$ since for every state
$\rho$ in $\mathcal{A}$ and for every state $\sigma$ in
$\mathcal{B}$ the state $\rho\otimes\sigma$ lies in
$\mathcal{A}\otimes \mathcal{B}$ and
$H(\rho\otimes\sigma)=H(\rho)+H(\sigma)$.

If the entropy is bounded on the sets $\mathcal{A}$ and
$\mathcal{B}$ then the entropy is bounded on the set
$\mathcal{A}\otimes \mathcal{B}$ due to its subadditivity.

Suppose, the entropy is continuous on the sets $\mathcal{A}$ and
$\mathcal{B}$. Let $\omega_{0}$ be a state in $\mathcal{A}\otimes
\mathcal{B}$ and $\{\omega_{n}\}$ be a sequence of states in
$\mathcal{A}\otimes \mathcal{B}$ converging to the state
$\omega_{0}$. Since
$$
H(\omega_{n})=H(\omega_{n}^{\mathcal{H}})+H(\omega_{n}^{\mathcal{K}})-
H(\omega_{n}\|\,\omega_{n}^{\mathcal{H}}\otimes\omega_{n}^{\mathcal{K}})
$$
the assumption and lower semicontinuity of the relative entropy
imply
$$
\begin{array}{c}
\limsup\limits_{n\rightarrow+\infty}H(\omega_{n})=\lim\limits_{n\rightarrow+\infty}H(\omega_{n}^{\mathcal{H}})
+\lim\limits_{n\rightarrow+\infty}H(\omega_{n}^{\mathcal{K}})-
\liminf\limits_{n\rightarrow+\infty}H(\omega_{n}\|\,\omega_{n}^{\mathcal{H}}\otimes\omega_{n}^{\mathcal{K}})\\\\\leq
H(\omega_{0}^{\mathcal{H}})+H(\omega_{0}^{\mathcal{K}})-
H(\omega_{0}\|\,\omega_{0}^{\mathcal{H}}\otimes\omega_{0}^{\mathcal{K}})=H(\omega_{0}).
\end{array}
$$
This and lower semicontinuity of the entropy implies
$\lim\limits_{n\rightarrow+\infty}H(\omega_{n})=H(\omega_{0})$.
$\square$

The important example of the set $\mathcal{A}\otimes \mathcal{B}$
is the set consisting of all states $\omega$ in
$\mathfrak{S}(\mathcal{H}\otimes \mathcal{K})$ with given partial
traces $\omega^{\mathcal{H}}=\rho$ and
$\omega^{\mathcal{K}}=\sigma$. Following \cite{Par-2} we denote
this set $\mathcal{C}(\rho,\sigma)$. By lemma 8 the set
$\mathcal{C}(\rho,\sigma)$ is convex and compact for arbitrary
$\rho$ and $\sigma$. By subaddditivity of the entropy
$\sup_{\omega\in\mathcal{C}(\rho,\sigma)}H(\omega)=H(\rho)+H(\sigma)$.
Similarly to the case of the set $\mathcal{L}(\sigma)$ finiteness
of the entropy on the set $\mathcal{C}(\rho,\sigma)$ implies its
continuity.

\textbf{Corollary 12.} \textit{The entropy is continuous on the
set $\mathcal{C}(\rho,\sigma)$ if and only if the entropies
$H(\rho)$ and $H(\sigma)$ are finite.}

For two arbitrary ensembles $\{\pi_{i}, \rho_{i}\}$ and
$\{\lambda_{j}, \sigma_{j}\}$ of states in $\mathcal{A}$ and in
$\mathcal{B}$ correspondingly the ensemble $\{\pi_{i}\lambda_{j},
\rho_{i}\otimes\sigma_{j}\}$ of states in
$\mathcal{A}\otimes\mathcal{B}$ is called the tensor product of
the two above ensembles. By considering such tensor products of
all possible ensembles of states in $\mathcal{A}$ and
$\mathcal{B}$ it is easy to deduce from the definition that
\begin{equation}\label{superadditivity}
\bar{C}(\mathcal{A}\otimes\mathcal{B})\geq\bar{C}(\mathcal{A})+\bar{C}(\mathcal{B}).
\end{equation}

There exist nontrivial examples of sets $\mathcal{A}$ and
$\mathcal{B}$, for which equality holds in
(\ref{superadditivity}). This takes place if $\mathcal{A}$ and
$\mathcal{B}$ are sets of the types considered in subsection 5.2.
But there exist examples of sets $\mathcal{A}$ and $\mathcal{B}$,
for which strict inequality holds in (\ref{superadditivity}).
Moreover, if $\mathcal{A}=\{\rho\}$ and $\mathcal{B}=\{\sigma\}$
where $\rho$ and $\sigma$ are isomorphic states in
$\mathfrak{S}(\mathcal{H})$ and in $\mathfrak{S}(\mathcal{K})$
with infinite entropy then by proposition 11 below the left side
of (\ref{superadditivity}) is equal to the infinity while the
right side is obviously equal to zero.\footnote{The strict
inequality in (\ref{superadditivity}) does not contradict to the
additivity conjecture for the $\chi$-capacity of quantum channels.
Indeed, if $\mathcal{A}$ and $\mathcal{B}$ are the output sets of
particular channels $\Phi$ and $\Psi$ correspondingly then the
output set of the channel $\Phi\otimes\Psi$ is a proper subset of
the set $\mathcal{A}\otimes\mathcal{B}$.}

Note that the equality in (\ref{superadditivity}) implies
\begin{equation}\label{a-s-t-p}
\Omega(\mathcal{A}\otimes\mathcal{B})=\Omega(\mathcal{A})\otimes\Omega(\mathcal{B}).
\end{equation}
Indeed, if $\{\{\pi_{i}^{n}, \rho_{i}^{n}\}\}_{n}$ and
$\{\{\lambda_{j}^{n}, \sigma_{j}^{n}\}\}_{n}$ are some
approximating sequences of ensembles for the sets $\mathcal{A}$
and $\mathcal{B}$ correspondingly then by the assumed equality in
(\ref{superadditivity}) the sequence of ensembles
$\{\{\pi_{i}^{n}\lambda_{j}^{n},
\rho_{i}^{n}\otimes\sigma_{j}^{n}\}\}_{n}$ will be approximating
sequence for the set $\mathcal{A}\otimes\mathcal{B}$. By theorem 1
the sequences $\{\bar{\rho}_{n}\}$ and $\{\bar{\sigma}_{n}\}$
converges to the optimal average states $\Omega(\mathcal{A})$ and
$\Omega(\mathcal{B})$ correspondingly. So, the sequence
$\{\bar{\rho}_{n}\otimes\bar{\sigma}_{n}\}$ converges to the state
$\Omega(\mathcal{A})\otimes\Omega(\mathcal{B})$ and, hence, by
theorem 1 this state is the optimal average state
$\Omega(\mathcal{A}\otimes\mathcal{B})$ of the set
$\mathcal{A}\otimes\mathcal{B}$. The below proposition 11 shows,
in particular, that (\ref{a-s-t-p}) does not imply
(\ref{superadditivity}).

In the rest of this section we restrict out attention on the set
$\mathcal{C}(\rho,\sigma)$. Let
$\rho=\sum_{i}\pi_{i}|e_{i}\rangle\langle e_{i}|$ and
$\sigma=\sum_{j}\lambda_{j}|f_{j}\rangle\langle f_{j}|$, where
$\{|e_{i}\rangle\}$ and $\{|f_{j}\rangle\}$ are orthonormal
systems of vectors in $\mathcal{H}$ and in $\mathcal{K}$
correspondingly. Let $E_{ij}=|e_{i}\rangle \langle e_{j}|$ and
$F_{kl}=|f_{k}\rangle \langle f_{l}|$ be one rank operators in
$\mathfrak{B}(\mathcal{H})$ and in $\mathfrak{B}(\mathcal{K})$
correspondingly. For arbitrary probability distributions
$\{\pi_{i}\}$ and $\{\lambda_{j}\}$ let
$\mathcal{C}(\{\pi_{i}\},\{\lambda_{j}\})$ be the set of all
probability distribution $\{\omega_{ij}\}$ such that
$\sum_{j}\omega_{ij}=\pi_{i}$ and
$\sum_{i}\omega_{ij}=\lambda_{j}$, so that
$\mathcal{C}(\{\pi_{i}\},\{\lambda_{j}\})$ is the classical analog
of the set $\mathcal{C}(\rho,\sigma)$. Denote by
$\mathcal{C}_{s}(\rho,\sigma)$ the closed convex subset of
$\mathcal{C}(\rho,\sigma)$ consisting of all states of the form
$\sum_{ij}\omega_{ij}E_{ii}\otimes F_{jj}$, where
$\{\omega_{ij}\}\in\mathcal{C}(\{\pi_{i}\},\{\lambda_{j}\})$. The
set $\mathcal{C}_{s}(\rho,\sigma)$ can be identified with the
classical analog $\mathcal{C}(\{\pi_{i}\},\{\lambda_{j}\})$ of the
set $\mathcal{C}(\rho,\sigma)$.

Let $G$ be the group of all unitaries in
$\mathfrak{B}(\mathcal{H}\otimes\mathcal{K})$, diagonizable in the
basis $\{|e_{i}\otimes f_{j}\rangle\}$. We will use the following
simple observation.

\textbf{Lemma 9.} \textit{Let $\rho=\sum_{i}\pi_{i}|e_{i}\rangle
\langle e_{i}|$ and $\sigma=\sum_{j}\lambda_{j}|f_{j}\rangle
\langle f_{j}|$ be two states in $\mathfrak{S}(\mathcal{H})$ and
in $\mathfrak{S}(\mathcal{K})$ correspondingly. An arbitrary state
$\omega$ in $\mathcal{C}(\rho,\sigma)$ can be represented by
$$
\omega=\sum_{ij}\omega_{ij}E_{ii}\otimes F_{jj}+\sum_{i\neq j,
k\neq l }\eta_{ijkl}E_{ij}\otimes F_{kl},
$$
where
$\{\omega_{ij}\}\in\mathcal{C}(\{\pi_{i}\},\{\lambda_{j}\})$.}

\textit{The set $\mathcal{C}(\rho,\sigma)$ is invariant under the
action of the automorphism $U(\cdot)U^{*}$ for arbitrary $U\in G$
while $\mathcal{C}_{s}(\rho,\sigma)$ is the set of all invariant
states in $\mathcal{C}(\rho,\sigma)$ for the group of the above
automorphisms.}

\textbf{Proof.} An arbitrary state $\omega$ in
$\mathcal{C}(\rho,\sigma)$ can be represented by
$$
\omega=\sum_{ijkl}\eta_{ijkl}E_{ij}\otimes F_{kl}.
$$
The requirements
$\mathrm{Tr}_{\mathcal{K}}\omega=\rho=\sum_{i}\pi_{i}E_{ii}$ and
$\mathrm{Tr}_{\mathcal{H}}\omega=\sigma=\sum_{j}\lambda_{j}F_{jj}$
provides the first statement of the lemma.

Since an arbitrary $U$ in $G$ is defined by the set
$\{\varphi_{ij}(U)\}_{ij}$ of numbers in $[0;2\pi)$ via the
expression
$$
U=\sum_{i,j}\exp(\mathrm{i}\varphi_{ij}(U))E_{ii}\otimes F_{jj},
$$
we have $U E_{ii}\otimes F_{jj}U^{*}=E_{ii}\otimes F_{jj}$ and $U
E_{ij}\otimes
F_{kl}U^{*}=\exp(\mathrm{i}(\varphi_{ik}-\varphi_{jl}))E_{ij}\otimes
F_{kl}$ for this $U$. By this  for the above
$\omega\in\mathcal{C}(\rho,\sigma)$ and $U$ we obtain
$$
U\omega U^{*}=\sum_{ij}\omega_{ij}E_{ii}\otimes F_{jj}+\sum_{i\neq
j, k\neq l
}\eta_{ijkl}\exp(\mathrm{i}(\varphi_{ik}-\varphi_{jl}))E_{ij}\otimes
F_{kl},
$$
which provides the second statement of the lemma. $\square$

The following proposition shows that the problems of calculation
of the $\chi$-capacity and of finding the optimal average state of
the set $\mathcal{C}(\rho,\sigma)$ are nontrivial even in the
symmetrical case $\rho\cong\sigma$.

\textbf{Proposition 11.} \textit{Let
$\rho=\sum_{i}\lambda_{i}|e_{i}\rangle \langle e_{i}|$ and
$\sigma=\sum_{j}\lambda_{j}|f_{j}\rangle \langle f_{j}|$ be two
isomorphic states supported by the subspaces
$\mathcal{H}_{\rho}\subseteq\mathcal{H}$ and
$\mathcal{K}_{\sigma}\subseteq\mathcal{K}$ correspondingly such
that
$H(\rho)=H(\sigma)=-\sum_{i}\lambda_{i}\log\lambda_{i}=h\leq+\infty$.
Then
$$
h\leq\bar{C}(\mathcal{C}(\rho,\sigma))\leq 2h,
$$
where the equality in the left side holds if and only if $\rho$
and $\sigma$ are pure states.}

\textit{In the case $h<+\infty$ there exists an optimal measure
$\mu_{*}(\rho,\sigma)$ with the barycenter
$\Omega(\mathcal{C}(\rho,\sigma))$ in
$\mathcal{C}_{s}(\rho,\sigma)$ having the support
$\mathcal{H}_{\rho}\otimes\mathcal{K}_{\sigma}$ and the following
statements are equivalent:}
\begin{enumerate}[(i)]
  \item \textit{$\bar{C}(\mathcal{C}(\rho,\sigma))=2h$;}
  \item \textit{$\Omega(\mathcal{C}(\rho,\sigma))=\rho\otimes\sigma$;}
  \item \textit{$\rho$ and $\sigma$ are multiples of projectors of the
  same finite rank;}
  \item \textit{$\mu_{*}(\rho,\sigma)$ is supported by pure states.}
\end{enumerate}

\textbf{Proof.} By subadditivity of the entropy $H(\omega)\leq
H(\rho)+H(\sigma)=2h$ for all $\omega$ in
$\mathcal{C}(\rho,\sigma)$. This implies the upper bound for
$\bar{C}(\mathcal{C}(\rho,\sigma))$.

Suppose $\bar{C}(\mathcal{C}(\rho,\sigma))$ is finite. By theorem
1 there exists the unique state $\Omega(\mathcal{C}(\rho,\sigma))$
in $\mathcal{C}(\rho,\sigma)$ such that
\begin{equation}\label{chi-center-ineq}
H(\omega\|\Omega(\mathcal{C}(\rho,\sigma)))\leq
\bar{C}(\mathcal{C}(\rho,\sigma)),\quad
\forall\omega\in\mathcal{C}(\rho,\sigma).
\end{equation}

By lemma 9 and corollary 8 this state
$\Omega(\mathcal{C}(\rho,\sigma))$ is invariant under automorphism
$U(\cdot)U^{*}$ for arbitrary $U$ in $G$ and hence
\begin{equation}\label{chi-center-rep}
\Omega(\mathcal{C}(\rho,\sigma))=\sum_{ij}\omega_{ij}E_{ii}\otimes
F_{jj}
\end{equation}
for some probability distribution $\{\omega_{ij}\}$ from
$\mathcal{C}(\{\lambda_{i}\},\{\lambda_{j}\})$. All elements
$\omega_{ij}$ of this distribution must be positive since
otherwise it is easy to find $\omega$ in
$\mathcal{C}(\rho,\sigma)$ such that
$H(\omega\|\Omega(\mathcal{C}(\rho,\sigma)))=+\infty$
contradicting to (\ref{chi-center-ineq}).

Let
$\omega=\sum_{ij}\sqrt{\lambda_{i}}\sqrt{\lambda_{j}}E_{ij}\otimes
F_{ij}$ be a pure state in $\mathcal{C}(\rho,\sigma)$. By
(\ref{chi-center-ineq}) and (\ref{chi-center-rep}) we have

\begin{equation}\label{H-exp}
\begin{array}{c}
\bar{C}(\mathcal{C}(\rho,\sigma))\geq
H(\omega\|\Omega(\mathcal{C}(\rho,\sigma)))=-\mathrm{Tr}\omega\log(\Omega(\mathcal{C}(\rho,\sigma))\\\\=
-\mathrm{Tr}\sum_{ij}\sqrt{\lambda_{i}}\sqrt{\lambda_{j}}\log\omega_{jj}E_{ij}\otimes
F_{ij}=-\sum\limits_{i}\lambda_{i}\log\omega_{ii}.
\end{array}
\end{equation}
If $\rho$ and $\sigma$ are not pure states then the right side of
this expression is greater than
$-\sum\limits_{i}\lambda_{i}\log\lambda_{i}=h$ since
$\omega_{ii}+\sum_{j\neq i}\omega_{ij}=\lambda_{i}$ and
$\omega_{ij}>0$ for all $i$ and $j$.

The existence of optimal measure in the case $h<+\infty$ follows
from corollary 12 and  theorem 3.

The asserted equivalence of statements $(i)-(iv)$ will be proved
in the following order
$(ii)\Rightarrow(i)\Rightarrow(iv)\Rightarrow(iii)\Rightarrow(ii)$.

$(ii)\Rightarrow(i)$ Suppose,
$\Omega(\mathcal{C}(\rho,\sigma))=\rho\otimes\sigma=\sum_{ij}\lambda_{i}\lambda_{j}E_{ii}\otimes
F_{jj}$. Let $\omega$ be the above pure state. By using expression
(\ref{H-exp}) with $\omega_{ij}=\lambda_{i}\lambda_{j}$ we have
$$
\bar{C}(\mathcal{C}(\rho,\sigma))\geq
H(\omega\|\Omega(\mathcal{C}(\rho,\sigma)))=-\sum\limits_{i}\lambda_{i}\log\lambda_{i}^{2}=2h.
$$
Since the converse inequality is already proved we obtain
$\bar{C}(\mathcal{C}(\rho,\sigma))=2h$.

$(i)\Rightarrow(iv)$ Suppose
$\bar{C}(\mathcal{C}(\rho,\sigma))=2h=H(\rho\otimes\sigma)$. Let
$\mu_{*}$ be an arbitrary optimal measure for the set
$\mathcal{C}(\rho,\sigma)$. Since $2h$ is the maximum of the
entropy on the set $\mathcal{C}(\rho,\sigma)$ we necessarily have
$\int H(\omega)\mu_{*}(d\omega)=0$ and hence $\mu_{*}$ is
supported by pure states in $\mathcal{C}(\rho,\sigma)$.

$(iv)\Rightarrow(iii)$ Let $\mu_{*}$ be an optimal measure for the
set $\mathcal{C}(\rho,\sigma)$ supported by pure states. This
implies that its barycenter $\Omega(\mathcal{C}(\rho,\sigma))$
lies in the convex closure of pure states in
$\mathcal{C}(\rho,\sigma)$. Since by the above observation
$\Omega(\mathcal{C}(\rho,\sigma))$ is a state in
$\mathcal{C}_{s}(\rho,\sigma)$ supported by
$\mathcal{H}_{\rho}\otimes\mathcal{K}_{\sigma}$ lemma 10 below
implies that $\rho$ and $\sigma$ are multiples of projectors of
the same finite rank.

$(iii)\Rightarrow(ii)$ Suppose $\rho$ and $\sigma$ are multiples
of projectors. By lemma 10 below there exists an ensemble of pure
states in $\mathcal{C}(\rho,\sigma)$ with the average state
$\rho\otimes\sigma$. Since this ensemble is obviously optimal for
the set $\mathcal{C}(\rho,\sigma)$ its the average state coincides
with $\Omega(\mathcal{C}(\rho,\sigma))$. $\square$

\textbf{Lemma 10.} \textit{Let $\rho$ and $\sigma$ be two states
supported by the subspaces
$\mathcal{H}_{\rho}\subseteq\mathcal{H}$ and
$\mathcal{K}_{\sigma}\subseteq\mathcal{K}$ correspondingly.}

\textit{The following statements are equivalent:}
\begin{enumerate}[i)]
  \item \textit{the set $\mathcal{C}_{s}(\rho,\sigma)$ contains a
state with the support
$\mathcal{H}_{\rho}\otimes\mathcal{K}_{\sigma}$, which lies the
convex closure of the set of all pure states in
$\mathcal{C}(\rho,\sigma)$;}
  \item \textit{the states $\rho$ and $\sigma$ are multiples of
projectors of the same finite rank;}
  \item \textit{the state
$\rho\otimes\sigma$ in $\mathcal{C}_{s}(\rho,\sigma)$ can be
represented as a finite convex combination of pure states in
$\mathcal{C}(\rho,\sigma)$.}
\end{enumerate}

\textbf{Proof.} The all statements of the lemma imply that the
states $\rho$ and $\sigma$ are isomorphic. Otherwise there exist
no pure states in $\mathcal{C}(\rho,\sigma)$.

It is sufficient to show $(i)\Rightarrow(ii)$ and
$(ii)\Rightarrow(iii)$.

$(i)\Rightarrow(ii)\,$ Let
$\hat{\omega}=\sum_{ij}\omega_{ij}E_{ii}\otimes F_{jj}$ be a state
in $\mathcal{C}_{s}(\rho,\sigma)$, contained in the convex closure
of the set of all pure states in $\mathcal{C}(\rho,\sigma)$. By
lemma l in \cite{H-Sh-W} there exists a measure $\mu$ supported by
pure states in $\mathcal{C}(\rho,\sigma)$ such that
$$
\hat{\omega}=\int_{\mathcal{C}(\rho,\sigma)}\omega\mu(d\omega)
$$

It is sufficient to prove that the state $\rho$ has no different
positive eigenvalues. Suppose $\lambda_{i}$ and $\lambda _{j}$ are
such eigenvalues. By using the Schmidt decomposition for any pure
state $\omega$ in $\mathcal{C}(\rho,\sigma)$ it is easy to see
that $E_{ii}\otimes F_{jj}\,\omega=0$. Hence
$$
\omega_{ij}E_{ii}\otimes F_{jj}=E_{ii}\otimes
F_{jj}\hat{\omega}=\int_{\mathcal{C}(\rho,\sigma)}E_{ii}\otimes
F_{jj}\omega \mu(d\omega)=0,
$$
which implies that the support of the state $\hat{\omega}$ does
not coincide with $\mathcal{H}_{\rho}\otimes\mathcal{K}_{\sigma}$.

$(ii)\Rightarrow(iii)\,$ Let $\rho=d^{-1}P$ and $\sigma=d^{-1}Q$,
where $P$ and $Q$ are $d$-dimensional projectors in
$\mathfrak{B}(\mathcal{H})$ and in $\mathfrak{B}(\mathcal{K})$
correspondingly. Let $\{|\varphi_{i}\rangle\}$ is a particular
basis of maximally entangled vectors in $P(\mathcal{H})\otimes
Q(\mathcal{K})$. Then
$\rho\otimes\sigma=d^{-2}\sum_{i}|\varphi_{i}\rangle\langle\varphi_{i}|$.$\square$

\textbf{Remark 10.} It is interesting to compare the
$\chi$-capacity of the set $\mathcal{C}(\rho,\sigma)$ with the
$\chi$-capacity of the set $\mathcal{C}_{s}(\rho,\sigma)$ which
can be identified with the classical analog
$\mathcal{C}(\{\pi_{i}\},\{\lambda_{j}\})$ of the set
$\mathcal{C}(\rho,\sigma)$. Let $\rho$ and $\sigma$ are multiples
of $d$-dimensional projectors. In this case the set
$\mathcal{C}(\{\pi_{i}\},\{\lambda_{j}\})$ consists of all
probability distribution $\{\omega_{ij}\}_{i,j=1}^{d}$  such that
$\sum_{i=1}^{d}\omega_{ij}=d^{-1}=\sum_{j=1}^{d}\omega_{ij}$. It
is easy to see that the optimal ensemble for the set
$\mathcal{C}_{s}(\rho,\sigma)\cong\mathcal{C}(\{\pi_{i}\},\{\lambda_{j}\})$
consists of $d$ states, having one nonzero element $d^{-1}$ in
each row and in each column, with equal probabilities, so that the
average state is the uniform distribution
$\{\omega_{ij}=d^{-2}\}$. Thus
$$
\bar{C}(\mathcal{C}_{s}(\rho,\sigma))=\log d^{2}-\log d=\log
d=h=\textstyle\frac{1}{2}\,\bar{C}(\mathcal{C}(\rho,\sigma)),
$$
where the last equality follows from proposition 11. So, using
entangled states in $\mathcal{C}(\rho,\sigma)$ leads to twice
increasing of the $\chi$-capacity.

\subsection{An orbit of a compact group of
automorphisms}

Let $G$ be a compact group and $\{U_{g}\}_{g\in G}$ be its unitary
(projective) representation on the Hilbert space $\mathcal{H}$.
Let $\sigma$ be an arbitrary state in $\mathfrak{S}(\mathcal{H})$.
Consider the set $\mathcal{O}_{G,U_{g},\sigma}=\left\{ U_{g}\sigma
U_{g}^{*};g\in G\right\}$. This set is compact as the image of the
compact set $G$ under the continuous mapping $g\mapsto U_{g}\sigma
U_{g}^{*}$. This and separability of the space
$\mathfrak{S}(\mathcal{H})$ implies compactness of its convex
closure $\overline{\mathrm{co}}(\mathcal{O}_{G,U_{g},\sigma})$.
Let $\omega(G,U_{g},\sigma)=\int_{G}U_{g}\sigma
U_{g}^{*}\mu_{H}(dg)$ be a state in
$\overline{\mathrm{co}}(\mathcal{O}_{G,U_{g},\sigma})$, where
$\mu_{H}$ is the Haar measure on $G$.

\textbf{Proposition 12.} \textit{The entropy is bounded on the set
$\overline{\mathrm{co}}(\mathcal{O}_{G,U_{g},\sigma})$ if and only
if $H(\omega(G,U_{g},\sigma))<+\infty$. In this case the entropy
is continuous on the set
$\overline{\mathrm{co}}(\mathcal{O}_{G,U_{g},\sigma})$ and
achieves its maximum at the Gibbs state
$$
\Gamma(\overline{\mathrm{co}}(\mathcal{O}_{G,U_{g},\sigma}))=\omega(G,U_{g},\sigma).
$$}\vspace{-10pt}

\textit{The $\chi$-capacity
$\bar{C}(\mathcal{O}_{G,U_{g},\sigma})$ of the set
$\mathcal{O}_{G,U_{g},\sigma}$ is equal to
$H(\sigma\,\|\,\omega(G,U_{g},\sigma))$. If the $\chi$-capacity is
finite then the image of the Haar measure $\mu_{H}$ corresponding
to the mapping $g \mapsto U_{g}\sigma U_{g}^{*}$ is the optimal
measure for the set $\mathcal{O}_{G,U_{g},\sigma}$ and
$$
\Omega(\mathcal{O}_{G,U_{g},\sigma})=\omega(G,U_{g},\sigma).
$$}\vspace{-10pt}

\textit{The set $\mathcal{O}_{G,U_{g},\sigma}$ is regular of and
only if it has finite $\chi$-capacity.}

\textbf{Proof.} Since $\int_{G}U_{g}\rho
U_{g}^{*}\mu_{H}(dg)=\omega(G,U_{g},\sigma)$ for arbitrary state
$\rho$ in $\overline{\mathrm{co}}(\mathcal{O}_{G,U_{g},\sigma})$
the boundedness assertion of the proposition easily follows from
concavity of the entropy and Jensen's
inequality.\footnote{Application of Jensen's inequality in this
case is valid since the entropy can be represented as a pointwise
limit of a monotonously increasing sequence of continuous concave
functions \cite{L}.} The continuity assertion follows from
corollary 3 since
$\mathrm{Tr}\sigma(-\log\omega(G,U_{g},\sigma))=H(\omega(G,U_{g},\sigma))$.

The set $\mathcal{O}_{G,U_{g},\sigma}$ is invariant under the
action of the family of automorphisms
$\{U_{g}(\cdot)U_{g}^{*}\}_{g\in G}$  and $\omega(G,U_{g},\sigma)$
is the only invariant state in
$\overline{\mathrm{co}}(\mathcal{O}_{G,U_{g},\sigma})$ for this
family. It follows from corollary 9 that
$\bar{C}(\mathcal{O}_{G,U_{g},\sigma})=H(\sigma\|\omega(G,U_{g},\sigma))$
and that
$\Omega(\mathcal{O}_{G,U_{g},\sigma})=\omega(G,U_{g},\sigma)$.

The assertion concerning existence of optimal measure for the set
$\mathcal{O}_{G,U_{g},\sigma}$ is obvious.

The regularity assertion follows from the above observation since
it is easy to see that
$H(\rho\|\omega(G,U_{g},\sigma))=H(\sigma\|\omega(G,U_{g},\sigma))$
for all $\rho$ in $\mathcal{O}_{G,U_{g},\sigma}$. $\square$

\textbf{Example of a closed set having optimal measure, but having
no atomic optimal measure.} Let $G=\mathbb{T}$ - one dimensional
rotation group represented as the interval $[\pi,\pi)$. In this
case the Haar measure is the normalized Lebesgue measure
$\frac{dx}{2\pi }$. Let $\mathcal{H}=\mathcal{L}^{2}$ $[\pi,\pi)$.
We may consider elements of $\mathcal{L}_{2}([\pi,\pi))$ as
$2\pi$-periodic functions on $\mathbb{R}$. Let
$\{U_{\lambda}\}_{\lambda\in\mathbb{T}}$ be unitary representation
of the group $\mathbb{T}$ defined by
$$
U_{\lambda}(\psi(x))=\psi(x-\lambda),\quad \psi(x)\in
\mathcal{L}_{2}([\pi,\pi)).
$$
For given $|\varphi_{0}\rangle$ in $\mathcal{L}_{2}([\pi,\pi))$
consider the set
$\mathcal{O}_{\mathbb{T},U_{\lambda},|\varphi_{0}\rangle\langle\varphi_{0}|}$.
It this case
$$
\omega(\mathbb{T},U_{\lambda},|\varphi_{0}\rangle\langle\varphi_{0}|)=
\frac{1}{2\pi}\int\limits_{-\pi}^{+\pi}|\varphi_{\lambda}\rangle\langle
\varphi_{\lambda}|d\lambda,
$$
where $|\varphi_{\lambda}\rangle=U_{\lambda}|\varphi_{0}\rangle$.
Note that
$\overline{\mathrm{co}}(\mathcal{O}_{\mathbb{T},U_{\lambda},|\varphi_{0}\rangle\langle\varphi_{0}|})$
is the closure of the output set of the channel $\Phi$ considered
in \cite{H-Sh-W}. In the proof of theorem 4 in \cite{H-Sh-W} it
was shown that
\begin{equation}
\bar{C}\left(\mathcal{O}_{\mathbb{T},U_{\lambda},|\varphi_{0}\rangle\langle\varphi_{0}|}\right)=
H\left(\omega(\mathbb{T},U_{\lambda},|\varphi_{0}\rangle\langle\varphi_{0}|)\right)=-\sum_{n=-\infty
}^{\infty }c^{2}_{n}(\varphi_{0})\log c^{2}_{n}(\varphi_{0}),
\label{orbit-cap}
\end{equation}
where $\{c_{n}(\varphi_{0})\}_{n\in\mathbb{Z}}$ are the set of the
Fourier coefficients of the function $\varphi_{0}$ with respect to
trigonometric orthomormal system
$\{\exp(\mathrm{i}nx)\}_{n\in\mathbb{Z}}$. By proposition 12
finiteness of the above series means continuity of the entropy on
the set
$\overline{\mathrm{co}}(\mathcal{O}_{\mathbb{T},U_{\lambda},|\varphi_{0}\rangle\langle\varphi_{0}|})$.
Proposition 12 also implies that the image of the normalized
Lebesgue measure $\frac{dx}{2\pi }$ corresponding to the mapping
$\lambda\mapsto
U_{\lambda}|\varphi_{0}\rangle\langle\varphi_{0}|U_{\lambda}^{*}$
is an optimal measure for the set
$\mathcal{O}_{\mathbb{T},U_{\lambda},|\varphi_{0}\rangle\langle\varphi_{0}|}$.
This measure is nonatomic, but its existence does not mean that
there is no purely atomic optimal measure in this case. We will
show that for a particular function $\varphi_{0}$ there is no
purely atomic optimal measure for the set
$\mathcal{O}_{\mathbb{T},U_{\lambda},|\varphi_{0}\rangle\langle\varphi_{0}|}$.

Let
$$
   \varphi_{0}(x)=\left\{
   \begin{array}{ll}
    0,& x\in [-\pi;0)\\
    \sqrt{2},& x\in [0;+\pi).
   \end{array} \right.
$$

It this case $c_{n}(\varphi_{0})\sim n^{-1}$ so that the series in
(\ref{orbit-cap}) is finite.

By proposition 8 to prove nonexistence of an atomic optimal
measure it is sufficient to show that the state
$\Omega(\mathcal{O}_{\mathbb{T},U_{\lambda},|\varphi_{0}\rangle\langle\varphi_{0}|})=
\omega(\mathbb{T},U_{\lambda},|\varphi_{0}\rangle\langle\varphi_{0}|)$
can not be represented as a countable convex combination of states
in
$\mathcal{O}_{\mathbb{T},U_{\lambda},|\varphi_{0}\rangle\langle\varphi_{0}|}$.
For this aim it is possible to apply the method used in
\cite{H-Sh-W}, but we will consider another approach based on the
theory of generalized functions (distributions).

Suppose
$\omega(\mathbb{T},U_{\lambda},|\varphi_{0}\rangle\langle\varphi_{0}|)
=\sum_{i=1}^{+\infty}\pi_{i}|\varphi_{\lambda_{i}}\rangle\langle
\varphi_{\lambda_{i}}|$. Without loss of generality we may assume
that $\pi_{1}\geq\pi_{i}$ for all $i>1$ and that $\lambda_{1}=0$.
For arbitrary $\eta$ we have
\begin{equation}\label{const}
\begin{array}{c}
 \displaystyle\sum_{i=1}^{+\infty}\pi_{i}\langle
\varphi_{\eta}|\varphi_{\lambda_{i}}\rangle^{2}=\langle
\varphi_{\eta}|\omega(\mathbb{T},U_{\lambda},|\varphi_{0}\rangle\langle\varphi_{0}|)|\varphi_{\eta}\rangle\\\\
\displaystyle=\frac{1}{2\pi}\int\limits_{-\pi}^{+\pi}\langle
\varphi_{\eta}|\varphi_{\lambda}\rangle^{2}d\lambda=\mathrm{Const}(\eta).
\end{array}
\end{equation}

Let $\theta(x)$ be the $2\pi$-periodical function equal to
$(1-\pi^{-1}|x|)^{2}$ on $[-\pi;+\pi]$. Then $\langle
\varphi_{\eta}|\varphi_{\lambda}\rangle^{2}=\theta(\eta-\lambda)$
for all $\lambda$ and $\eta$. Since for each $\lambda$ the
function $\theta_{\lambda}(x)=\theta_{0}(x-\lambda)$ is locally
integrable it generates elements $\tilde{\theta}_{\lambda}$ of the
space $\mathfrak{D}'$ of generalized functions.\footnote{The space
$\mathfrak{D}'$ is the linear space of continuous linear
functional on the space $\mathfrak{D}$ of smooth functions with
finite support \cite{Courant}.} Let $\tilde{\theta}'_{\lambda}\in
\mathfrak{D}'$ be the (generalized) derivative of the generalized
function $\tilde{\theta}_{\lambda}\in \mathfrak{D}'$. Then
(\ref{const}) implies
\begin{equation}\label{const+}
\mathfrak{D}'-\lim_{n\rightarrow+\infty}\sum_{i=1}^{n}\pi_{i}\tilde{\theta}'_{\lambda_{i}}=0.
\end{equation}

Let
$$
   \omega_{\delta}(x)=\left\{
   \begin{array}{lr}
    \exp(-(1-(x/\delta)^{2})^{-1}),& x\in [-\delta;+\delta]\\
    0, & x\in \mathbb{R}\backslash[-\delta;+\delta]
   \end{array} \right.
$$
be a function from the space $\mathfrak{D}$ for each $\delta>0$.
By direct integration it is easy to see that
$$
\begin{array}{c}
\displaystyle\tilde{\theta}'_{\lambda}(\omega'_{\delta})=
\int\limits_{-\infty}^{+\infty}\theta'_{\lambda}(x)\omega'_{\delta}(x)dx
=\frac{2}{\pi}\int\limits_{-\delta}^{\lambda}\left(1+\frac{x-\lambda}{\pi}\right)
\,\omega'_{\delta}(x)dx\\\\\displaystyle+
\frac{2}{\pi}\int\limits_{\lambda}^{+\delta}\left(\frac{x-\lambda}{\pi}-1\right)\,\omega'_{\delta}(x)dx
=\frac{4\omega_{\delta}(\lambda)}{\pi}-\frac{2\delta I}{\pi^{2}}
\end{array}
$$
if $\lambda\in[-\delta;+\delta]$ and
$$
\tilde{\theta}'_{\lambda}(\omega'_{\delta})=\int\limits_{-\infty}^{+\infty}\theta'_{\lambda}(x)\omega'_{\delta}(x)dx
=\frac{2}{\pi^{2}}\int\limits_{-\delta}^{+\delta}x\omega'_{\delta}(x)dx=-\frac{2\delta
I}{\pi^{2}}
$$
if $\lambda\in\mathbb{R}\backslash[-\delta;+\delta]$, where
$I=\delta^{-1}\int\limits_{-\delta}^{+\delta}\omega_{\delta}(x)dx=\int\limits_{-1}^{+1}\exp(-(1-x^{2})^{-1})dx$
is a positive number.

Let
$\mathcal{N}(\delta)=\{i\in\mathbb{N}\,|\,\lambda_{i}\in[-\delta;+\delta]\}$
and $\mathcal{K}_{n}=\{2,3,...,n\}$. By using the above
expressions we obtain
$$
\begin{array}{c}
\displaystyle\sum\limits_{i=1}^{n}\pi_{i}\tilde{\theta}'_{\lambda_{i}}(\omega'_{\delta})
=\pi_{1}\tilde{\theta}'_{0}(\omega'_{\delta}) +\sum\limits_{i\in
\mathcal{N}(\delta)\cap\mathcal{K}_{n}}\pi_{i}
\tilde{\theta}'_{\lambda_{i}}(\omega'_{\delta})
+\sum\limits_{i\in(\mathbb{N}\backslash\mathcal{N}(\delta))\cap\mathcal{K}_{n}}
\pi_{i}\tilde{\theta}'_{\lambda_{i}}(\omega'_{\delta})\\\\
\displaystyle\geq \pi_{1}\left(\frac{4}{e\pi}-\frac{2\delta
I}{\pi^{2}}\right) -\left(\frac{4}{e\pi}-\frac{2\delta
I}{\pi^{2}}\right)\sum\limits_{i\in\mathcal{N}(\delta),i>1}\pi_{i}-\frac{2\delta
I}{\pi^{2}},\quad \forall n.
\end{array}
$$
Since $\sum\limits_{i\in\mathcal{N}(\delta), i>1}\pi_{i}$
obviously tends to zero as $\delta$ tends to zero the above
inequality implies $\liminf_{n\rightarrow+\infty}
\sum\limits_{i=1}^{n}\pi_{i}\tilde{\theta}'_{\lambda_{i}}(\omega'_{\delta})>0$
for all sufficiently small $\delta$, which contradicts to
(\ref{const+}).

\section{On another definition of $\bar{C}(\mathcal{A})$ and of $\Omega(\mathcal{A})$}

It is known that the entropy and the relative entropy for general
quantum states can be introduced via finite dimensional definition
and a limiting procedure. To show this consider the nonlinear
mapping
$$
\Theta_{P}(\rho)=(\mathrm{Tr}P\rho)^{-1}P\rho P
$$
corresponding to arbitrary finite rank projector $P$ and having
the domain
$\mathfrak{D}(\Theta_{P})=\{\rho\in\mathfrak{S}(\mathcal{H})\,|\,P\rho\neq
0\}$. By the results in \cite{L} the entropy $H(\rho)$ of an
arbitrary state $\rho$ can be defined by
$$
H(\rho)=\lim_{n\rightarrow+\infty}H(\Theta_{P_{n}}(\rho)),
$$
while the relative entropy $H(\rho\|\sigma)$ for arbitrary states
$\rho$ and $\sigma$ - by
$$
H(\rho\|\sigma)=\lim_{n\rightarrow+\infty}H(\Theta_{P_{n}}(\rho)\|\Theta_{P_{n}}(\sigma)),
$$
where $\{P_{n}\}$ is an arbitrary increasing sequence of finite
rank projectors strongly converging to the identity operator
$I_{\mathcal{H}}$.\footnote{It is assumed that $n$ is sufficiently
large so that $\rho$ and $\sigma$ lie in
$\mathfrak{D}(\Theta_{P_{n}})$} This implies that both above
limits exist (finite or infinite) and do not depend on the choice
of the sequence $\{P_{n}\}$. Since the states
$\Theta_{P_{n}}(\rho)$ and $\Theta_{P_{n}}(\sigma)$ are supported
by finite dimensional subspaces $P_{n}(\mathcal{H})$ for all $n$
this observation reduces the definition of the entropy and of the
relative entropy to the finite dimensional case.

In this section we obtain the analogous results for the
$\chi$-capacity and for the optimal average state of an arbitrary
set of states. Since for any closed subset of states in the
$d$-dimensional Hilbert space the supremum in the definition of
the $\chi$-capacity can be over all ensembles of $d^{2}$ states
the $\chi$-capacity and the optimal average state of this subset
can be defined by linear programming procedure \cite{Shor}. So,
the results of this section provides the definition of the
$\chi$-capacity and of the optimal average state for an arbitrary
set of the infinite dimensional states, which can be used (in
principal) for their numerical approximations.

It is clear that for arbitrary projector $P$ the corresponding
mapping $\Theta_{P}(\sigma)$ is continuous in each point of its
domain. Despite nonlinearity of this mapping the following result is
valid.

\textbf{Lemma 11.} \textit{For arbitrary convex subset
$\mathcal{A}$ of $\mathfrak{D}(\Theta_{P})$ its image
$\Theta_{P}(\mathcal{A})$ under the mapping $\Theta_{P}$ is a
convex subset of $\mathfrak{S}(\mathcal{H})$.}

\textit{For arbitrary ensemble $\{\pi_{i}, \rho_{i}\}_{i=1}^{m}$
of states in $\Theta_{P}(\mathcal{A})$ there exists ensemble
$\{\lambda_{i}, \sigma_{i}\}_{i=1}^{m}$ of states in $\mathcal{A}$
such that}
$$
\Theta_{P}(\sigma_{i})=\rho_{i}\quad and \quad
\lambda_{i}\mathrm{Tr}P\sigma_{i}=\pi_{i}\sum_{j=1}^{m}\lambda_{j}\mathrm{Tr}P\sigma_{j}\quad
for\;\;i=\overline{1,m}.
$$

\textbf{Proof.} It is sufficient to prove the second statement of
the lemma since it implies
$$
\Theta_{P}\left(\sum_{i}\lambda_{i}\sigma_{i}\right)=\sum_{i}\pi_{i}
\rho_{i}.
$$

For each $i$ the state $\rho_{i}$ in $\Theta_{P}(\mathcal{A})$ is
an image of a particular state $\sigma_{i}$ in $\mathcal{A}$. Let
$\eta_{i}=\pi_{i}(\mathrm{Tr}P\sigma_{i})^{-1}$ be a positive
number for each $i=\overline{1,m}$ and
$\left\{\lambda_{i}=\eta_{i}\left(\sum_{j=1}^{m}\eta_{j}\right)^{-1}\right\}$
be a probability distribution. By summing the equalities
$\lambda_{i}\mathrm{Tr}P\sigma_{i}=\pi_{i}\left(\sum_{j=1}^{m}\eta_{j}\right)^{-1}$
we obtain
$\sum_{i=1}^{m}\lambda_{i}\mathrm{Tr}P\sigma_{i}=\left(\sum_{j=1}^{m}\eta_{j}\right)^{-1}$.$\square$

\textbf{Lemma 12.} \textit{Let $\mathcal{A}$ be a set with finite
$\chi$-capacity and $P$ be a projector such that
$\eta(\mathcal{A},
P)=\inf_{\rho\in\mathcal{A}}\mathrm{Tr}P\rho>0$. Then
$$
\eta(\mathcal{A}, P)\bar{C}(\Theta_{P}(\mathcal{A}))\leq
\bar{C}(\mathcal{A}).
$$
} \textbf{Proof.} For arbitrary ensemble $\{\pi_{i}, \rho_{i}\}$
of states in $\Theta_{P}(\mathcal{A})$ let $\{\lambda_{i},
\sigma_{i}\}$ be the corresponding ensemble of states in
$\mathcal{A}$ provided by lemma 11. It follows that
$\eta=\sum_{i}\lambda_{i}\eta_{i}$, where
$\eta_{i}=\mathrm{Tr}P\sigma_{i}$ and
$\eta=\mathrm{Tr}P\bar{\sigma}$.

Consider the channel
$$
\Phi(\rho)=P\rho P+(\mathrm{Tr}(I-P)\rho)\tau,
$$
where $\tau$ is a pure state corresponding to arbitrary unit
vector in $\mathcal{H}\ominus P(\mathcal{H})$. By general
properties of the relative entropy we obtain
$$
\begin{array}{c}
\chi(\{\lambda_{i},
\Phi(\sigma_{i})\})=\sum\limits_{i}\lambda_{i}H(P\sigma_{i}P
\|P\bar{\sigma}P)\\\\+\sum\limits_{i}\lambda_{i}H((\mathrm{Tr}(I-P)\sigma_{i})\tau\|(\mathrm{Tr}(I-P)\bar{\sigma})\tau)
\geq\sum\limits_{i}\lambda_{i}H(P\sigma_{i}P
\|P\bar{\sigma}P)\\\\=
\sum\limits_{i}\lambda_{i}H(\eta_{i}\rho_{i}\|\eta\bar{\rho})\geq
\sum\limits_{i}\lambda_{i}\eta_{i}H(\rho_{i}\|\bar{\rho})
=\eta\sum\limits_{i}\pi_{i}H(\rho_{i}\|\bar{\rho})\geq\eta(\mathcal{A},
P)\chi(\{\pi_{i},\rho_{i}\}).
\end{array}
$$

By monotonicity of the relative entropy we have
$$
\chi(\{\lambda_{i},
\Phi(\sigma_{i})\})\leq\chi(\{\lambda_{i},\sigma_{i}\}).
$$
The two above inequalities implies the statement of the lemma.
$\square$

\textbf{Remark 11.} The constant $\eta(\mathcal{A}, P)$ in lemma
12 cannot be replaced by $1$ (see the example in remark 12
below).$\square$

Now we can prove the following approximation result.

\textbf{Theorem 4.} \textit{Let $\mathcal{A}$ be an arbitrary
subset of $\mathfrak{S}(\mathcal{H})$.}

\textit{If the $\chi$-capacity of the set $\mathcal{A}$ is finite
then
$$
\lim_{n\rightarrow+\infty}\bar{C}(\Theta_{P_{n}}(\mathcal{A}))=\bar{C}(\mathcal{A})\quad
and \quad
\lim_{n\rightarrow+\infty}\Omega(\Theta_{P_{n}}(\mathcal{A}))=\Omega(\mathcal{A})
$$
for arbitrary sequence $\{P_{n}\}$ of projectors strongly
converging to $I_{\mathcal{H}}$.}

\textit{If there exists a sequence of projectors $\{P_{n}\}$
strongly converging to $I_{\mathcal{H}}$ such that the mappings in
the corresponding sequence $\{\Theta_{P_{n}}\}$ are well defined
on the set $\mathcal{A}$ and the sequence
$\{\bar{C}(\Theta_{P_{n}}(\mathcal{A}))\}$ is bounded then
$\bar{C}(\mathcal{A})$ is finite.}

\textbf{Proof.} Let $\bar{C}(\mathcal{A})<+\infty$ and $\{P_{n}\}$
be an arbitrary sequence of projectors strongly converging to
$I_{\mathcal{H}}$. By theorem 2D the set $\mathcal{A}$ is compact.
By compactness criterion
$\lim_{n\rightarrow+\infty}\eta(\mathcal{A},P_{n})=1$, where
$\eta(\mathcal{A},
P_{n})=\inf_{\rho\in\mathcal{A}}\mathrm{Tr}P_{n}\rho$. Thus
$\mathcal{A}\subseteq\mathfrak{D}(\Theta_{P_{n}})$ for all
sufficiently large $n$ and by lemma 12 we have
$$
\limsup_{n\rightarrow+\infty}\bar{C}(\Theta_{P_{n}}(\mathcal{A}))\leq\bar{C}(\mathcal{A}).
$$
Since $\Theta_{P_{n}}(\rho)\rightarrow \rho$ as
$n\rightarrow+\infty$ the first part of lemma 4 implies
$$
\liminf_{n\rightarrow+\infty}\bar{C}(\Theta_{P_{n}}(\mathcal{A}))\geq\bar{C}(\mathcal{A}).
$$
By the two above inequalities we obtain the first limit expression
in the theorem, the second follows from the first and the second
part of lemma 4.

If $\bar{C}(\mathcal{A})=+\infty$ and $\{P_{n}\}$ be a sequence of
finite dimensional projectors strongly converging to
$I_{\mathcal{H}}$ such that
$\mathcal{A}\subseteq\mathfrak{D}(\Theta_{P_{n}})$ for all
sufficiently large $n$ then the first part of lemma 4 implies
$$
\lim_{n\rightarrow+\infty}\bar{C}(\Theta_{P_{n}}(\mathcal{A}))=+\infty.\square
$$
\textbf{Remark 12.} The convergence of the sequence
$\{\bar{C}(\Theta_{P_{n}}(\mathcal{A}))\}$ to
$\bar{C}(\mathcal{A})$ has different nature depending on the
choice of the sequence $\{P_{n}\}$. It may seem surprising that
for a particular set $\mathcal{A}$ and a  sequence $\{P_{n}\}$ the
sequence $\{\bar{C}(\Theta_{P_{n}}(\mathcal{A}))\}$ converges to
$\bar{C}(\mathcal{A})$ strongly decreasing. Indeed, let
$\mathcal{A}$ be the set consisting of two states
$\{\frac{1}{2}\rho+\frac{1}{2}\sigma_{i}\}_{i=1,2}$, where $\rho$
is a state with infinite dimensional support $\mathcal{H}_{\rho}$
such that $\mathcal{H}\ominus\mathcal{H}_{\rho}$ is a two
dimensional subspace and $\sigma_{1}, \sigma_{2}$ are the states
corresponding to orthonormal unit vectors in
$\mathcal{H}\ominus\mathcal{H}_{\rho}$. Let $\{P_{n}\}$ be such
sequence of finite rank projectors that
$P_{n}(\mathcal{H})\supseteq \mathcal{H}\ominus\mathcal{H}_{\rho}$
and the sequence $\{\eta_{n}=\mathrm{Tr}P_{n}\rho\}$ is strongly
increasing to $1$. It is easy to obtain that
$$
\bar{C}(\Theta_{P_{n}}(\mathcal{A}))=\frac{1}{1+\eta_{n}}\log
2\;\searrow\; \frac{1}{2}\log 2=\bar{C}(\mathcal{A})
\quad\mathrm{as}\quad n\rightarrow+\infty.
$$

\section{Appendix}

In this section the detailed investigation of the properties of
the function $F_{H}(h)=\sup_{\rho\in\mathcal{K}_{H,h}}H(\rho)$
described in proposition 1a is presented.

Note first that by lower semicontinuity of the entropy
$\lim_{n\rightarrow+\infty}F_{H}(h)=\sup_{\rho\in\mathfrak{S}(\mathcal{H})}H(\rho)=+\infty$
for arbitrary value of $\mathrm{ic}(H)$ since
$\overline{\bigcup_{h\in\mathbb{R}}\mathcal{K}_{H,h}}=\mathfrak{S}(\mathcal{H})$.

Consider the function
$$
g(\lambda, h)=\sum_{k=1}^{+\infty}(h_{k}-h)\exp(-\lambda h_{k}).
$$
By using the theorem about series depending on parameters it is
easy to see that this function is differentiable at any point
$(\lambda, h)$ with $\lambda>\mathrm{ic}(H)$ and
\begin{equation}\label{p-der}
\frac{\partial g(\lambda, h)}{\partial
\lambda}=\sum_{k=1}^{+\infty}h_{k}(h-h_{k})\exp(-\lambda
h_{k}),\quad \frac{\partial g(\lambda, h)}{\partial
h}=-\sum_{k=1}^{+\infty}\exp(-\lambda h_{k}).
\end{equation}
By the observation in the proof of proposition 1a for each $h$ in
$(h_{\mathrm{m}}(H); h_{*}(H))$ there exists the unique
$\lambda^{*}=\lambda^{*}(h)>\mathrm{ic}(H)$ such that
$g(\lambda^{*}(h), h)=0$. It follows from (\ref{p-der}) that
$$
\left.\frac{\partial g(\lambda, h)}{\partial
\lambda}\right|_{\lambda=\lambda^{*}(h)}=\sum_{k=1}^{+\infty}(h_{k}-h)^{2}\exp(-\lambda^{*}(h)
h_{k})<0.
$$

By the implicit function theorem the function $\lambda^{*}(h)$ is
differentiable on $(h_{\mathrm{m}}(H); h_{*}(H))$ and
\begin{equation}\label{der-exp}
\begin{array}{c}
\displaystyle\frac{d \lambda^{*}(h)}{d h}=-\left[\frac{\partial
g(\lambda, h)}{\partial \lambda}\right]^{-1}\frac{\partial
g(\lambda, h)}{\partial
h}\\\\=-\displaystyle\left[\sum_{k=1}^{+\infty}(h_{k}-h)^{2}\exp(-\lambda^{*}(h)
h_{k})\right]^{-1}\sum_{k=1}^{+\infty}\exp(-\lambda^{*}(h)h_{k})<0
\end{array}
\end{equation}

Expression (\ref{h-exp-2}) implies
\begin{equation}\label{F-exp}
F_{H}(h)=\lambda^{*}(h)h+\log\sum_{k=1}^{+\infty}\exp(-\lambda^{*}(h)h_{k})
\end{equation}
for all $h$ in $(h_{\mathrm{m}}(H); h_{*}(H)]$.

By direct derivatives calculation we obtain
\begin{equation}\label{d-F-exp}
\frac{d F_{H}(h)}{d h}=\frac{d}{d
h}\left[\lambda^{*}(h)h+\log\sum_{k=1}^{+\infty}\exp(-\lambda^{*}(h)
h_{k})\right]=\lambda^{*}(h),
\end{equation}
where the equality $g(\lambda^{*}(h), h)=0$ was used. This and
(\ref{der-exp}) implies
$$
\frac{d^{2}F_{H}(h)}{d h^{2}}=\frac{d \lambda^{*}(h)}{d h}<0,
$$
which shows strict concavity of the function $F_{H}(h)$ on
$(h_{\mathrm{m}}(H); h_{*}(H))$.

Suppose $h_{*}(H)<+\infty$. If $h>h_{*}(H)$ then by the proved
part of the proposition 1a
\begin{equation}\label{F-exp-2}
F_{H}(h)=\mathrm{ic}(H)h+\log\sum_{k=1}^{+\infty}\exp(-\mathrm{ic}(H)h_{k})
\end{equation}
is a linear function and
\begin{equation}\label{d-F-exp-2}
\frac{d F_{H}(h)}{d h}=\mathrm{ic}(H).
\end{equation}

If $h=h_{*}(H)$ then by the observation in the proof of
proposition 1a $\lambda^{*}(h)=\mathrm{ic}(H)$ and hence
representations (\ref{F-exp})  and (\ref{F-exp-2}) coincides in
this case.

To show smoothness of the function $F_{H}(h)$ at the point
$h_{*}(H)$ note that $\lambda^{*}(h)\rightarrow\mathrm{ic}(H)$ as
$h\rightarrow h_{*}(H)-0$. Indeed, by (\ref{der-exp}) the function
$\lambda^{*}(h)$ is decreasing on $(h_{\mathrm{m}}(H); h_{*}(H))$
and for arbitrary $\lambda>\mathrm{ic}(H)$ there exists
$h_{\lambda}=\left[\sum_{k=1}^{+\infty}\exp(-\lambda
h_{k})\right]^{-1}\sum_{k=1}^{+\infty}h_{k}\exp(-\lambda h_{k})$
such that $\lambda=\lambda^{*}(h_{\lambda})$.

Thus (\ref{F-exp}),(\ref{d-F-exp}),(\ref{F-exp-2}) and
(\ref{d-F-exp-2}) imply
$$
\lim_{h\rightarrow
h_{*}(H)-0}F_{H}(h)=F_{H}(h_{*}(H))\quad\mathrm{and}\quad
\lim_{h\rightarrow h_{*}(H)-0}\frac{d F_{H}(h)}{d h}=\frac{d
F_{H}(h)}{d h}|_{h=h_{*}(H)+0}
$$
and hence the function $F_{H}(h)$ has a continuous derivative at
the point $h_{*}(H)$.

To prove right continuity of the function $F_{H}(h)$ at the point
$h_{\mathrm{m}}(H)$ note first that
\begin{equation}\label{lim-h-min}
\lambda^{*}(h)\rightarrow+\infty\quad \mathrm{as} \quad
h\rightarrow h_{\mathrm{m}}+0.
\end{equation}
Indeed, by (\ref{der-exp}) the function $\lambda^{*}(h)$ is
decreasing on $(h_{\mathrm{m}}(H); h_{*}(H))$ and hence there
exists $\lambda^{\mathrm{m}}=\lim_{h\rightarrow
h_{\mathrm{m}}(H)+0}\lambda^{*}(h)$. If
$\lambda^{\mathrm{m}}<+\infty$ then by passing to the limit as
$h\rightarrow h_{\mathrm{m}}(H)+0$ in the identity
$$
\sum_{k=1}^{+\infty}h_{k}\exp(-\lambda^{*}(h) h_{k})\equiv
h\sum_{k=1}^{+\infty}\exp(-\lambda^{*}(h) h_{k}),
$$
valid for all $h$ in $(h_{\mathrm{m}}(H); h_{*}(H))$, we obtain a
contradiction.

Let $d=\dim\mathcal{H}_{\mathrm{m}}(H)$. It is easy to see that
\begin{equation}\label{P-Q-exp}
P(h)=\log\sum_{k=1}^{+\infty}\exp(-\lambda^{*}(h)
h_{k})=-\lambda^{*}(h)h_{\mathrm{m}}(H)+Q(h),
\end{equation}
where $Q(h)=\log(d+\sum_{k>d}^{+\infty}\exp(-\lambda^{*}(h)
(h_{k}-h_{\mathrm{m}}(H)))$ is a nondecreasing function on
$(h_{\mathrm{m}}(H); h_{*}(H))$ tending to $\log d$ as
$h\rightarrow h_{\mathrm{m}}(H)+0$.

Since the function $F_{H}(h)$ is obviously nonnegative and
nondecreasing on $[h_{\mathrm{m}}(H); +\infty)$ there exists
$\lim_{h\rightarrow h_{\mathrm{m}}(H)+0}F_{H}(h)\geq
F_{H}(h_{\mathrm{m}}(H))$. This, (\ref{F-exp}) and (\ref{P-Q-exp})
imply that there exists $\lim_{h\rightarrow
h_{\mathrm{m}}(H)+0}\lambda^{*}(h)(h-h_{\mathrm{m}}(H))=C<+\infty$
and that
$$
\lim_{h\rightarrow h_{\mathrm{m}}(H)+0}F_{H}(h)=C+\log
d=C+F_{H}(h_{\mathrm{m}}(H)).
$$
Thus to prove right continuity of the function $F_{H}(h)$ at the
point $h_{\mathrm{m}}(H)$ it is sufficient to show that $C=0$.
This can be done by proving that
\begin{equation}\label{int-finiteness}
\int_{h_{\mathrm{m}}(H)}^{h''}\lambda^{*}(h)dh=\lim_{h'\rightarrow
h_{\mathrm{m}}(H)+0}\int_{h'}^{h''}\lambda^{*}(h)dh<+\infty,
\end{equation}
for some $h''>h_{\mathrm{m}}(H)$. Indeed, finiteness of this
integral and the assumption $C>0$ imply finiteness of the integral
$\int_{h_{\mathrm{m}}(H)}^{h''}(h-h_{\mathrm{m}}(H))^{-1}dh$.

It is easy to see that
$$
\frac{dP(h)}{dh}=-h\frac{d\lambda^{*}(h)}{dh}\quad \mathrm{and}\;
\mathrm{hence}\quad-\frac{d\lambda^{*}(h)}{dh}(h-h_{\mathrm{m}}(H))=\frac{dQ(h)}{dh}.
$$
By direct integration we obtain
$$
Q(h'')-Q(h')=\lambda^{*}(h')(h'-h_{\mathrm{m}}(H))-\lambda^{*}(h'')(h''-h_{\mathrm{m}}(H))
+\int_{h'}^{h''}\lambda^{*}(h)dh.
$$
This and the mentioned before existence of $\lim_{h'\rightarrow
h_{\mathrm{m}}(H)+0}Q(h')=\log d$ and of $\lim_{h'\rightarrow
h_{\mathrm{m}}(H)+0}\lambda^{*}(h')(h'-h_{\mathrm{m}}(H))=C<+\infty$
imply (\ref{int-finiteness}).

By the above observation
$$
\frac{F_{H}(h)-F_{H}(h_{\mathrm{m}}(H))}{h-h_{\mathrm{m}}(H)}\geq\lambda^{*}(h),\quad\forall
h>h_{\mathrm{m}}(H),
$$
and hence (\ref{lim-h-min}) implies $\frac{d F_{H}(h)}{d
h}|_{h=h_{\mathrm{m}}(H)+0}=+\infty$.

\vspace{10pt}

\textbf{Acknowledgments.} The author is grateful to A. S. Holevo
for the permanent help and the useful discussion. The work was
partially supported by the program "Modern problems of theoretical
mathematics" of Russian Academy of Sciences.

\end{document}